\documentclass[preprint,3p,times]{elsarticle}

\usepackage{amssymb}
\usepackage{amsmath}
\usepackage{gensymb}
\usepackage{empheq}
\usepackage{tabularx}
\usepackage{float}
\usepackage{svg}
\usepackage[labelfont=bf]{caption}
\usepackage{multirow}

\usepackage{eurosym}

\usepackage{mathtools}
\usepackage{commath}
\usepackage{todonotes}
\biboptions{numbers,sort&compress}
\usepackage[hyphens]{url}
\usepackage{multicol}
\usepackage{lscape}
\usepackage{tablefootnote}

\usepackage[title]{appendix}

\usepackage{soul}

\usepackage{scrextend}
\usepackage[colorlinks]{hyperref}
\usepackage[noabbrev,nameinlink]{cleveref} 

\usepackage{vcell}

\makeatletter
\renewcommand*{\eqref}[1]{%
  \hyperref[{#1}]{\textup{\tagform@{\ref*{#1}}}}%
}
\makeatother

\usepackage{nomencl}
\makenomenclature

\usepackage{cleveref}
\Crefformat{figure}{#2Fig.~#1#3}
\Crefmultiformat{figure}{Figs.~#2#1#3}{ and~#2#1#3}{, #2#1#3}{ and~#2#1#3}
\crefname{paragraph}{section}{sections}
\Crefname{paragraph}{Section}{Sections}

\usepackage{lineno}
\usepackage{tikz}

\usepackage{etoolbox}
\renewcommand{\nomgroup}[1]{%
  \item[\textbf{%
  \ifthenelse{\equal{#1}{A}}{Acronyms}{}%
  \ifthenelse{\equal{#1}{B}}{Sets}{}%
  \ifthenelse{\equal{#1}{C}}{Index}{}%
  \ifthenelse{\equal{#1}{D}}{Parameters}{}%
  \ifthenelse{\equal{#1}{E}}{Variables}{}%
}]}

\usepackage{titlesec}

\titleformat{\paragraph}
{\normalfont\normalsize\itshape}{\theparagraph}{1em}{}
\titlespacing*{\paragraph}
{0pt}{3.25ex plus 1ex minus .2ex}{1.5ex plus .2ex}

\usepackage{adjustbox}
\usepackage[skins,breakable]{tcolorbox}

\usepackage{graphicx,stackengine}
\newcommand\underlay[4]{%
  \stackengine{0pt}%
  {\kern#2\includegraphics[height=#1]{#4}}%
  {\includegraphics[height=#1]{#3}}%
  {O}{l}{F}{F}{L}%
}
\newcommand\addunderlay[4]{%
  \stackengine{0pt}%
  {\kern#2\includegraphics[height=#1]{#4}}%
  {#3}%
  {O}{l}{F}{F}{L}%
}

\journal{Journal of Cleaner Production}

\begin{document}

\begin{frontmatter}



\title{Decarbonisation of industry and the energy system: exploring mutual impacts and investment planning}


\author[inst1]{Quentin Raillard{-}{-}Cazanove\corref{cor1}}
\ead{quentin.raillard-cazanove@minesparis.psl.eu}

\affiliation[inst1]{organization={Mines Paris, PSL University, Centre for processes, renewable energy and energy systems (PERSEE)},
            city={Sophia Antipolis},
            postcode={06904}, 
            country={France}}

\author[inst1]{Thibaut Knibiehly}
\author[inst1]{Robin Girard\corref{cor1}}
\ead{robin.girard@minesparis.psl.eu}

\cortext[cor1]{Corresponding authors}

\begin{abstract}
The decarbonisation of the energy system is crucial for achieving climate goals and is inherently linked to the decarbonisation of industry. Despite this, few studies explore the simultaneous impacts of decarbonising both sectors. This paper aims to examine how industrial decarbonisation in Europe affects the energy system and vice versa. To address this, an industry model incorporating key heavy industry sectors across six European countries is combined with an energy system model for electricity and hydrogen covering fifteen European regions, refered to as the EU-15, divided into eleven zones. The study evaluates various policy scenarios under different conditions.

The results demonstrate that industrial decarbonisation leads to a significant increase in electricity and hydrogen demand. This additional demand for electricity is largely met through renewable energy sources, while hydrogen supply is predominantly addressed by blue hydrogen production when fossil fuels are authorized and the system lacks renewable energy. This increased demand results in higher prices with considerable regional disparities. Furthermore, the findings reveal that, regardless of the scenario, the electricity mix in the EU-15 remains predominantly renewable, exceeding 85\%.

A reduction in carbon taxes lowers the prices of electricity and hydrogen, but does not increase consumption, as the lower carbon tax makes the continued use of fossil fuels more attractive to industry. In scenarios that enforce a phase-out of fossil fuels, electricity prices rise, leading to a greater reliance on imports of low-carbon hydrogen and methanol. Results also suggest that domestic hydrogen production benefits from synergies between electrolytic hydrogen and blue hydrogen, helping to maintain competitive prices.
\end{abstract}

\begin{graphicalabstract}
\includegraphics[width=\columnwidth]{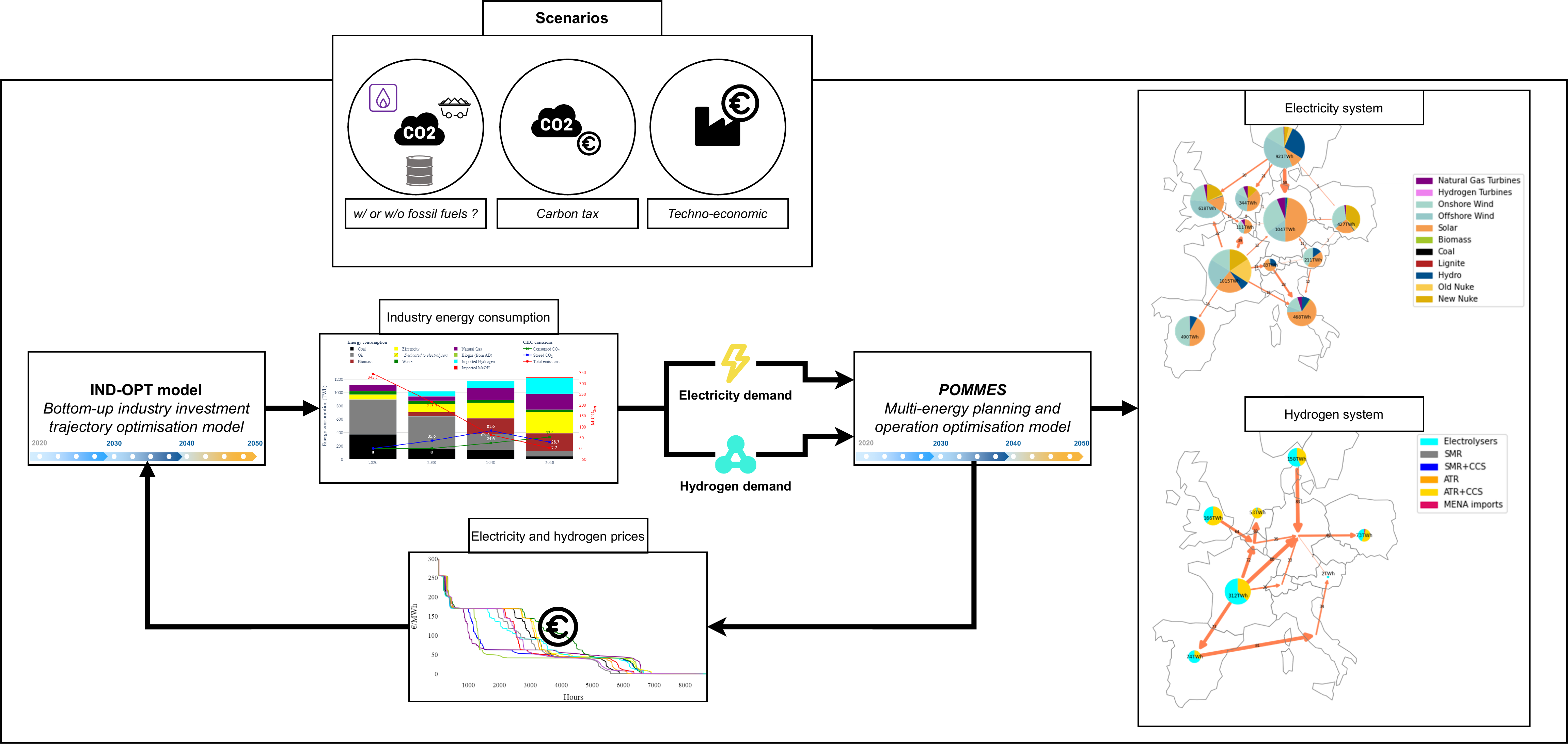}
\end{graphicalabstract}

\begin{highlights}
\item Industry decarbonisation significantly increase electricity and hydrogen demand
\item Higher electricity and hydrogen prices emerge with notable regional disparities, worsened by a phase-out of fossil fuels
\item Reducing carbon taxes lowers prices but makes fossil fuel usage more attractive, limiting decarbonisation progress
\item Synergies between electrolytic and blue hydrogen help maintain competitive hydrogen prices, supporting cost-effective production

\end{highlights}

\begin{keyword}
Decarbonisation \sep Industry \sep Energy System \sep Renewable Energy \sep Synergies


\end{keyword}

\end{frontmatter}


\printnomenclature

\section{Introduction}
\label{sec:introduction}

\subsection{Background}
\label{subsec:background}

Anthropogenic greenhouse gas emissions (GHG), the primary driver of global warming, are poised to inflict severe harm on ecosystems, compromise water availability, and jeopardise health, well-being, and economic stability \cite{IntergovernmentalPanelonClimateChangeIPCC2023ClimateVulnerability}. It is, therefore, imperative that we reduce our GHG emissions to mitigate these adverse effects.

\nomenclature[A]{GHG}{Greenhouse Gas}

In 2019, approximately 48\% of greenhouse gas emissions in Europe were attributable to industry and the production of electricity and heat \cite{Ritchie2024BreakdownSector}. It is therefore crucial to decarbonise these sectors to mitigate global warming, particularly as decarbonised electricity production can facilitate the decarbonisation of the residential and tertiary sectors, as well as transport (which accounted for 24\% of emissions in Europe in 2019 \cite{Ritchie2024BreakdownSector}).

Indeed, the direct or indirect use of low-carbon electricity is crucial for the decarbonisation of transport \cite{Shirizadeh2024Climate30years,Ruhnau2019Direct2050,EuropeanEnvironmentAgency2022DecarbonisingTransport,Jaramillo2023Transport,Tarvydas2022TheScenarios} and industry \cite{Tarvydas2022TheScenarios,Wei2019ElectrificationOutlook,Bashmakov2022Industry}. Similarly, the decarbonisation of buildings is largely reliant on electricity usage \cite{Cabeza2023Buildings,Santamouris2021PresentDecarbonisation,Tarvydas2022TheScenarios}.

\subsection{Industry decarbonisation}
\label{subsec:indus_decarb}

Industry, particularly heavy industry, exhibits a high energy demand and is characterised by significant emissions, especially for processes heavily reliant on fossil fuels. Decarbonising industry thus necessitates substantial technological advancements leading to the adoption of low-carbon energy sources.

It should be noted that many decarbonisation options are nearing commercial maturity \cite{Gailani2024AssessingSectors}, and both direct and indirect electrification present significant potential for reducing emissions \cite{Madeddu2020Thepower-to-heat,Wei2019ElectrificationOutlook}. Therefore, several prospective studies on industrial decarbonisation \cite{Raillard--Cazanove2024IndustryTrajectory,Fleiter2019IndustrialDecarbonisation,Sandberg2022TheSweden} found a profound shift in energy consumption patterns, which is expected to ultimately influence the dynamics of the energy system.

\subsection{Energy supply decarbonisation}
\label{subsec:elec_sys_plan}

The decarbonisation of the electricity system, a cornerstone of the European energy transition, necessitates the increased deployment of renewable energy sources alongside flexibility measures \cite{Shirizadeh2021Low-carbonStorage,Pastore2024Optimal2050,Luxembourg2024TIMES-Europe:Challenges,Colbertaldo2023ANeeds,ENTSOE2024TYNDPReport,RTE2024Bilan2023-2035,RTE2021Futurs2050}. Several studies and reports \cite{RTE2021Futurs2050,Shirizadeh2021Low-carbonStorage,Aszodi2023TheUnion} explore the potential role of nuclear energy, while highlighting the significant advantages of an interconnected system that promotes cooperation between nations \cite{Brown2018SynergiesSystem,Schlachtberger2017TheNetwork,RTE2021Futurs2050}.

The production of low-carbon hydrogen presents a significant challenge for decarbonising aviation, shipping, and industry \cite{Raillard--Cazanove2024IndustryTrajectory,Tarvydas2022TheScenarios}. In various prospective studies, hydrogen is predominantly produced via electrolysis \cite{Munster2024PerspectivesNeutrality,Kountouris2024AProduction,Tarvydas2022TheScenarios}, indicating that the electricity system must adapt to meet the corresponding demand. However, electrolysers can offer valuable flexibility for both hydrogen consumers \cite{Jodry2023IndustrialOptimisation} and the electricity system \cite{Tarvydas2022TheScenarios,RTE2021Futurs2050}. Furthermore, solutions such as Power-to-X-to-Power (PtXtP) are poised to address the seasonal intermittency of renewable energy sources in the long term \cite{RTE2021Futurs2050,Victoria2019TheSystem}. Consequently, the integrated modelling of an electricity/hydrogen system highlights systemic economical advantages \cite{Neumann2023TheEurope}.

\nomenclature[A]{PtXtP}{Power-to-X-to-Power}

\subsection{Interdependent dynamics}
\label{subsec:inter_dyn}

A cross-sectoral modelling approach allows for a better consideration of synergies and minimises trade-offs between sectors \cite{Babiker2023Cross-sectoralPerspectives}. The various stakeholders in the energy transition exhibit interconnected responses to climate policies \cite{BoaMorte2023ElectrificationGoals}. While urban and rural areas adopt a systemic approach with the automotive sector, the decarbonisation efforts of the energy and industrial sectors remain uncoordinated and lack a dynamic systemic perspective \cite{BoaMorte2023ElectrificationGoals}.

In most prospective studies modelling the European energy system, industry is treated merely as an exogenous demand parameter \cite{Luxembourg2024TIMES-Europe:Challenges,Fleiter2023METISSystem.,Neumann2023TheEurope,RTE2021Futurs2050,RTE2024Bilan2023-2035,Brown2018SynergiesSystem}. Although \citet{Fleiter2023METISSystem.} models industrial consumption within a dedicated model, it remains an exogenous parameter to the energy system under study. \citet{Papadaskalopoulos2018QuantifyingSystem} also treats industrial consumption as an exogenous parameter but allows the energy system to make it flexible, demonstrating the associated economic and renewable energy deployment benefits.

To the authors' knowledge, only one study stands out, \citet{Manuel2022HighDecarbonisation}, optimising the energy system of the Netherlands jointly with the technological deployment in the industrial sector. It then examines the impact of four industrial energy policies (Bio-based, CCUS-based, Electrification, Hydrogen-based) on the initial optimal system.

Nonetheless, changes in the electricity system, such as the availability of low-cost renewable energy can significantly impact industrial strategies, influencing decisions related to energy sourcing and investment in new technologies.

\subsection{Research gap and objectives}
\label{subsec:research_gap_obj}

Despite the evident interdependencies between the energy system and industry, limited research considers the bidirectional and simultaneous impact of decarbonisation in these sectors.

While \citet{Manuel2022HighDecarbonisation} investigates the impact of enforced technological choices in industry on the overall energy mix, we propose an approach, detailed in \Cref{sec:methods}, that focuses on the implications of an economically optimised industrial decarbonisation pathway within different political and macroeconomic contexts, focusing on its effects on the electricity system and the resulting synergies.

Thus, this paper aims to investigate how the decarbonisation of industry in Europe affects the electricity system's planning and operation, and vice versa. To achieve this, we have coupled an industrial model, based on previous work \cite{Raillard--Cazanove2024IndustryTrajectory}, with an energy system model both incorporating investment planning frameworks.

By exploring the interactions between these two models, this study provides insight into the synergistic planning required to achieve our decarbonisation goals.

\section{Methodology}
\label{sec:methods}

\subsection{Industry decarbonisation modelling}
\label{subsec:ind-opt}

\begin{figure}[H]
    \centering
    \includegraphics[width=0.8\textwidth]{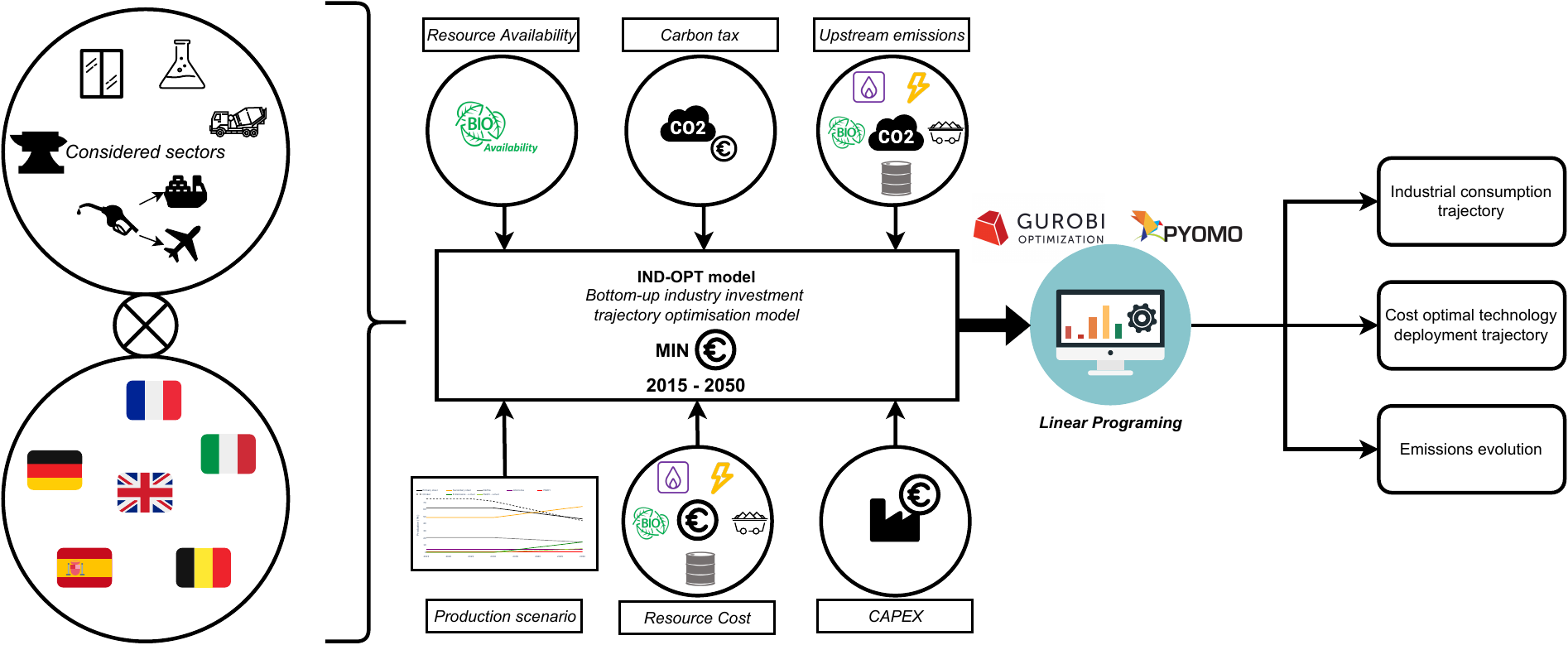}
    \caption{IND-OPT model simplified scheme}\label{fig:ind-opt}
\end{figure}

The industry modelling builds upon prior research \cite{Raillard--Cazanove2024IndustryTrajectory} that focuses on decarbonising the steel, chemical, cement, and glass sectors, along with e-fuel production in six European countries\footnote{France, Germany, Great Britain, Italy, Spain and Belgium}. The IND-OPT model employs a bottom-up technological approach, taking into account numerous technologies detailed in \Cref{tab:tec_sec}.

\begin{table}[H]
\centering
\begin{minipage}{\textwidth}
\resizebox{\textwidth}{!}{%

\begin{tabular}{cllllllll}
\hline
\multicolumn{1}{l}{Resource} & Hydrogen & Ammonia & Olefins & MeOH & E-Kerosene & Clinker & Steel & Glass \\ \hline
\multirow{7}{*}{Technologies} & SMR/eSMR\footnote{\label{ccs}Carbon capture compatible} & Haber-Bosch & NC/eNC\footref{ccs} & SMR-MSR & Fischer-Tropsch & Reference & BF-BOF\footref{ccs} & Regenerative SP\footref{ccs} \\
 & ATR\footref{ccs} &  & MTO & Gasification-MSR &  & Oxy-Ref\footref{ccs} & DRI-EAF\footref{ccs} & Regenerative EP\footref{ccs} \\
 & Gasification\footref{ccs} &  & OCM & CO\textsubscript{2} to MeOH &  & eC-pK\footref{ccs} & Electrowinning & Recuperative\footref{ccs} \\
 & Electrolysis &  &  &  &  & OC-HK\footref{ccs} & EAF (recycling) & Oxyfuel\footref{ccs} \\
 & Pyrolysis &  &  &  &  & eC-OK\footref{ccs} &  & TCR\footref{ccs} \\
 &  &  &  &  &  & eC-HK\footref{ccs} &  & Hybrid\footref{ccs} \\
 &  &  &  &  &  &  &  & Electric\footref{ccs} \\ \hline
\end{tabular}%

}
\end{minipage}
\caption{Production technologies considered in IND-OPT for the study based on \cite{Raillard--Cazanove2024IndustryTrajectory}}
\label{tab:tec_sec}
\end{table}

\nomenclature[A]{SMR}{Steam Methane Reforming}
\nomenclature[A]{ATR}{Autothermal Reforming}
\nomenclature[A]{NC}{Naphtha Cracking}
\nomenclature[A]{OCM}{Oxidative Coupling of Methane}
\nomenclature[A]{MTO}{Methanol to Olefins}
\nomenclature[A]{DRI}{Direct Reduced Iron}
\nomenclature[A]{EAF}{Electric Arc Furnace}
\nomenclature[A]{MSR}{Methanol Synthesis Reactor}
\nomenclature[A]{BF}{Blast Furnace}
\nomenclature[A]{BOF}{Basic Oxygen Furnace}
\nomenclature[A]{TCR}{Thermo-Chemical heat Recovery}

\subsection{Energy system modelling}
\label{subsec:pommes}

\begin{figure}[H]
    \centering
    \includegraphics[width=0.8\textwidth]{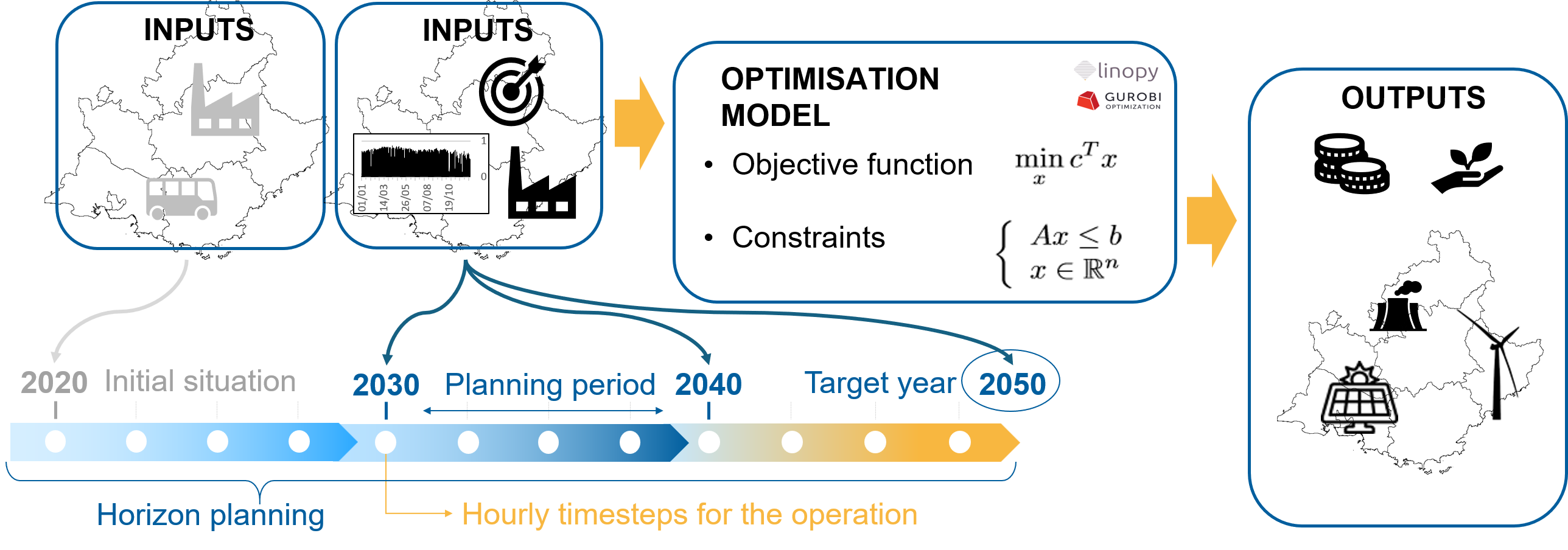}
    \caption{POMMES framework scheme}\label{fig:pommes}
\end{figure}
POMMES (Planning and Operation Model for Multi-Energy Systems) is a framework adapted from the model developed by \citet{Jodry2023IndustrialOptimisation}, specifically designed for multi-horizon modelling of energy systems, with the primary objective of cost minimisation \cite{Knibiehly2024PlanningGitLab}. In this study, POMMES has been employed to model the electricity and hydrogen system of 15 countries (grouped into 11 nodes) up to the year 2050. The modelling process includes the optimisation of system operations on an hourly basis, alongside a planning horizon—also optimised—set at 10-year intervals. Interconnections between countries are incorporated into the model, with POMMES capable of optimising these as well. The equations underlying the model are detailed in the Supplementary Materials.

\subsection{Models Coupling}
\label{subsec:model_coupling}

\subsubsection{Coupling method}
\label{subsubsec:coupling_method}

\begin{figure}[H]
    \centering
    \includegraphics[width=1.0\textwidth]{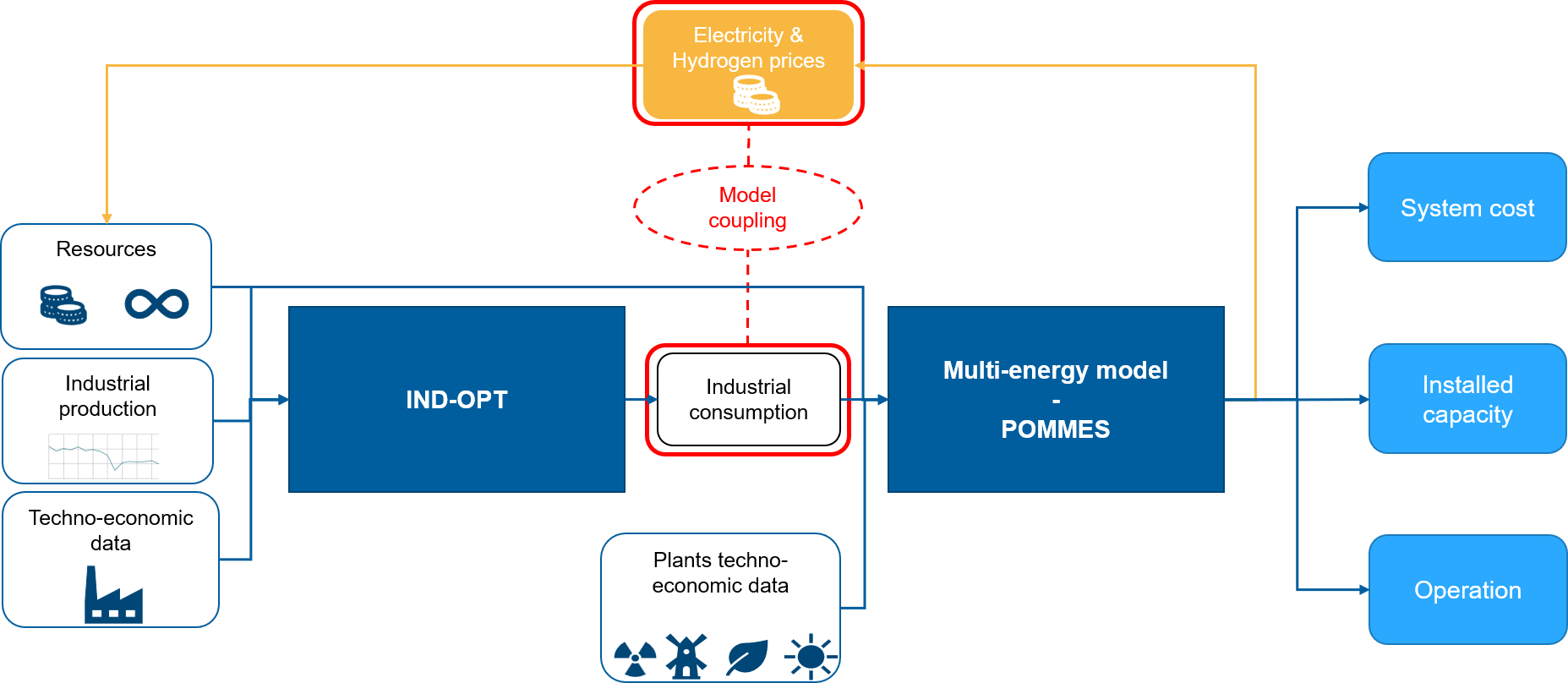}
    \caption{Models coupling scheme}\label{fig:model_coupling}
\end{figure}

The models were integrated as depicted in \Cref{fig:model_coupling}, based on industrial consumption data from IND-OPT and hydrogen and electricity prices from POMMES. The iterative process involves alternating calculations between IND-OPT and POMMES until convergence of consumption and prices is achieved over two successive iterations. The stopping criteria are defined as follows:
\begin{equation*}
    \text{\textit{Stopping criteria}}=
    \begin{cases}
        \abs{\overline{P}_{\text{\euro}}^{elec,i}-\overline{P}_{\text{\euro}}^{elec,i-1}}<0.1\text{\euro}/MWh \\
        \abs{\overline{P}_{\text{\euro}}^{hyrogen,i}-\overline{P}_{\text{\euro}}^{hyrogen,i-1}}<0.01\text{\euro}/kg \\
        \sum_{y \in Years}\abs{C_{TWh}^{elec,y,i}-C_{TWh}^{elec,y,i-1}}< 1 TWh * \text{\textit{Nb of regions}} \\
        \sum_{y \in Years}\abs{C_{TWh}^{hydrogen,y,i}-C_{TWh}^{hydrogen,y,i-1}}< 1 TWh * \text{\textit{Nb of regions}}
        
    \end{cases}
\end{equation*}

with, at iteration $i$, $\overline{P}_{\text{\euro}}^{elec,i}$/$\overline{P}_{\text{\euro}}^{hydrogen,i}$ the average price over the modelled horizons (2030, 2040 and 2050) for respectively electricity/hydrogen, and $C_{TWh}^{elec,y,i}$/$C_{TWh}^{hydrogen,y,i}$ the industrial consumption form IND-OPT at year $y$.

\subsubsection{Considered countries and data}
\label{subsubsec:countries_and_data}

\begin{figure}[H]
    \centering
    \includegraphics[width=0.8\textwidth]{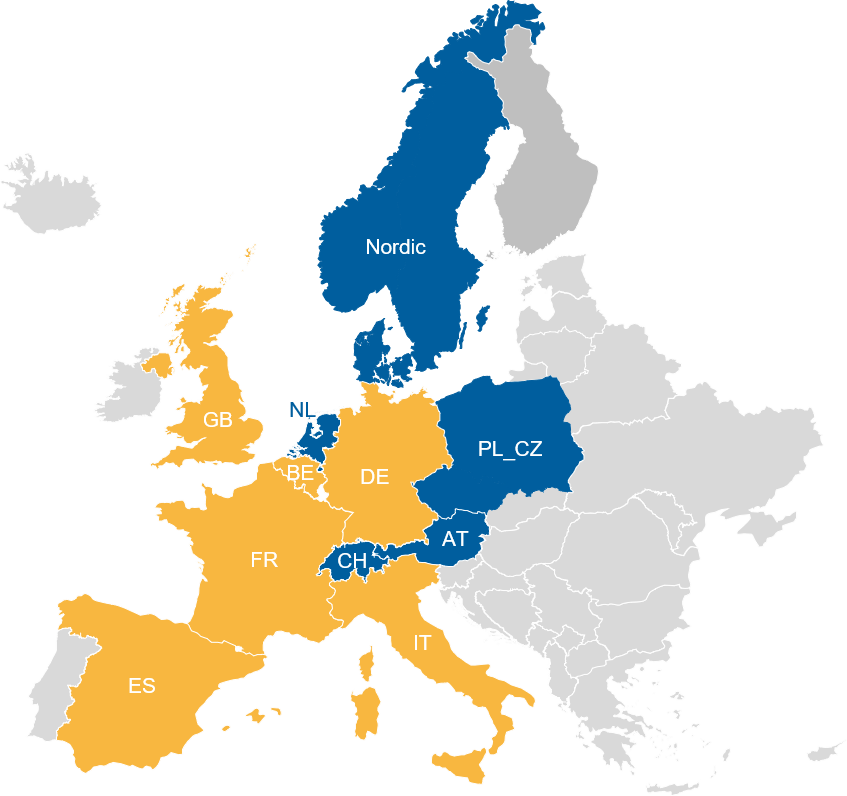}
    \caption{Considered area nodes | Areas in blue are only considered in POMMES while areas in yellow are included in both models}\label{fig:areas}
\end{figure}

The regions modelled are displayed in \Cref{fig:areas}. As a reminder, IND-OPT models industry based on \citet{Raillard--Cazanove2024IndustryTrajectory} and therefore only includes the countries shown in yellow in \Cref{fig:areas}, as well as the steel, cement, chemicals, glass, and e-fuel production sectors for air and sea transport. The associated consumption of electricity and hydrogen is thus used as an input parameter for POMMES. Consumption in other sectors (those not included in IND-OPT, such as residential, transport, etc.) is an exogenous parameter largely based on the Distributed Energy scenario in the TYNDP 2024 report \cite{ENTSOE2024TYNDPReport}. Detailed consumption assumptions are provided in the Supplementary Materials.

The technologies for generating and storing electricity and hydrogen are detailed in \ref{appendix:techno_economic_pommes}, along with their economic characteristics. For renewable and nuclear energy, the maximum deployment capacities are determined based on data from TYNDP 2024\cite{ENTSOE2024TYNDPReport}, as well as the FES \cite{ESO2024Future2024} and BP2050\cite{RTE2021Futurs2050} reports from ESO and RTE concerning nuclear energy in the UK and France.

The time series used for demand and availability of generation resources are based on 2018 data from the ENTSOE Transparency Plateform\footnote{\url{https://transparency.entsoe.eu/}} and Renewables.ninja\footnote{\url{https://www.renewables.ninja/}} (for renewables in particular). 

Electricity market prices are obtained from POMMES, derived from the Lagrangians of the adequacy constraint between electricity production and consumption. Country-specific network costs are incorporated during the post-processing stage before implementation in IND-OPT. These network costs, sourced from Eurostat \footnote{\url{https://ec.europa.eu/eurostat/databrowser/view/nrg_pc_205_c__custom_12113170/default/table?lang=en}} data, are detailed in \Cref{tab:network_cost}.
This electricity network tax is also applied within POMMES to hydrogen production technologies in the form of variable costs weighted by the amount of electricity consumed.

The models can also import hydrogen and methanol (MeOH) from the MENA region, which are treated as exogenous parameters. 
Building on the work of \citet{Lux2021SupplyRegion}, it is assumed that hydrogen imported from MENA is produced through electrolysis powered by renewable energy. The corresponding import prices are provided in \Cref{tab:mena_imports}. Methanol is assumed to be produced using DAC units, with the technical and economic data drawn from \citet{Raillard--Cazanove2024IndustryTrajectory}, in conjunction with electrolysers.

\nomenclature[A]{DAC}{Direct Air Capture}
\nomenclature[A]{MeOH}{Methanol}

\begin{table}[H]
\centering
\begin{tabular}{llll}
\hline
 & 2030 & 2040 & 2050 \\ \hline
Hydrogen (€/kg) & 5.1 & 4.2 & 3.3 \\
MeOH (€/kg) & 0.86 & 0.72 & 0.59 \\ \hline
\end{tabular}
\caption{Hydrogen and MeOH imports from MENA cost assumptions}
\label{tab:mena_imports}
\end{table}

From the perspective of allocating limited resources, it is assumed that bioenergy is reserved for industry and transport. Consequently, only IND-OPT can utilise it, within the limits defined in \citet{Raillard--Cazanove2024IndustryTrajectory}, while the SMR/ATR and CCGT/TAC in POMMES rely on natural gas.

\subsection{Studied frameworks}
\label{subsec:studied_frameworks }

\subsubsection{Policy scenarios}
\label{subsubsec:policy_scenarios}

Three policy scenarios were developed for this study to assess their impact on the industry and the energy system synergies:
\begin{itemize}
\item \textbf{Reference:} This scenario is based on the \citet{Raillard--Cazanove2024IndustryTrajectory} reference scenario, specifically the 300€/tCO\textsubscript{2} in 2050 carbon tax case, which was established as a zero direct emission threshold for the modelled industries.
\item \textbf{No Fossil 2050:} A variation of the Reference scenario where the models are constrained to achieve a fossil-free system by 2050. This scenario imposes a reduction in fossil fuel consumption starting in 2045, with the requirement that fossil fuel use shall not exceed 10\% of 2015 levels. \citet{Raillard--Cazanove2024IndustryTrajectory} demonstrated that a complete fossil phase-out driven solely by economic factors is challenging for the industry without the enforcement of additional policies.
\item \textbf{Low Carbon Tax:} This scenario modifies the Reference scenario by applying a 150€/tCO\textsubscript{2} carbon tax, aligning with the \citet{Raillard--Cazanove2024IndustryTrajectory} reference scenario.
\end{itemize}

\subsubsection{Technical and economical variations}
\label{subsubsec:tech_and_eco_variations}
In addition to the policy scenarios, we explored techno-economic variations within the energy system. Our analysis begins with a reference case, termed \textbf{Central}, which is comprehensively detailed in \Cref{subsubsec:countries_and_data} and \ref{appendix:techno_economic_pommes}. A crucial assumption in the Central case is that hydrogen production is treated as a domestic activity for each country, with no cross-border hydrogen exchange. Several other techno-economic scenarios are explored to assess different potential futures:
\begin{itemize}
    \item \textbf{Nuke\_Plus:} This scenario assumes that governments increase their equity stakes in nuclear projects, resulting in a reduction of the weighted average cost of capital (WACC) to 4\%. Additionally, it considers, as installable by POMMES, 8 GW of new nuclear capacity in Italy and 6 GW in Belgium by 2050. The United Kingdom is allowed to maintain its current nuclear capacity until 2050. The assumptions for Italy and the UK are based on recent political commitments, whereas the assumptions for Belgium are derived from the Electrification scenario of the PATHS2050 project by EnergyVille.

    \item \textbf{Exch\_Plus:} This scenario involves doubling the electric interconnection capacities that can be installed by 2050, enhancing cross-border electricity exchanges.

    \item \textbf{ENR\_Plus:} In this scenario, the renewable energy capacities that can be installed by 2050 are adjusted to reflect the highest deployment rates observed between 2030 and 2050. This adjustment addresses the observed decline in deployment rates in many countries between 2040 and 2050, as documented in the TYNDP 2024 dataset.

    \item \textbf{ENR\_Plus\_Plus:} Building on the ENR\_Plus scenario, this scenario assumes a WACC of 2\% for renewable energy projects, compared to the standard 4\%.

    \item \textbf{ENR\_Exch\_Plus:} This scenario combines the assumptions of ENR\_Plus with the enhanced electric interconnection capacities from the Exch\_Plus scenario, allowing for both increased renewable energy deployment and greater cross-border electricity exchange.

    \item \textbf{H\textsubscript{2}\_Exch\_Plus:} In this scenario, hydrogen interconnections between countries are allowed, following the projects outlined in the TYNDP 2024 report, facilitating cross-border hydrogen trade.
\end{itemize}
\nomenclature[A]{WACC}{Weighted Average Cost of Capital}

\section{Results}
\label{sec:results}
\subsection{Energy System Standalone Analysis}\label{subsec:energy_system_standalone}
The energy system was initially modelled without coupling, meaning industrial consumption from IND-OPT was not considered. The results indicate substantial differences when varying policy scenarios, though less so when considering technical and economic variations. Therefore, in this section, only the "Central" cases are considered across various policy scenarios.

\begin{figure}[H]
    \centering
    \includegraphics[width=\textwidth]{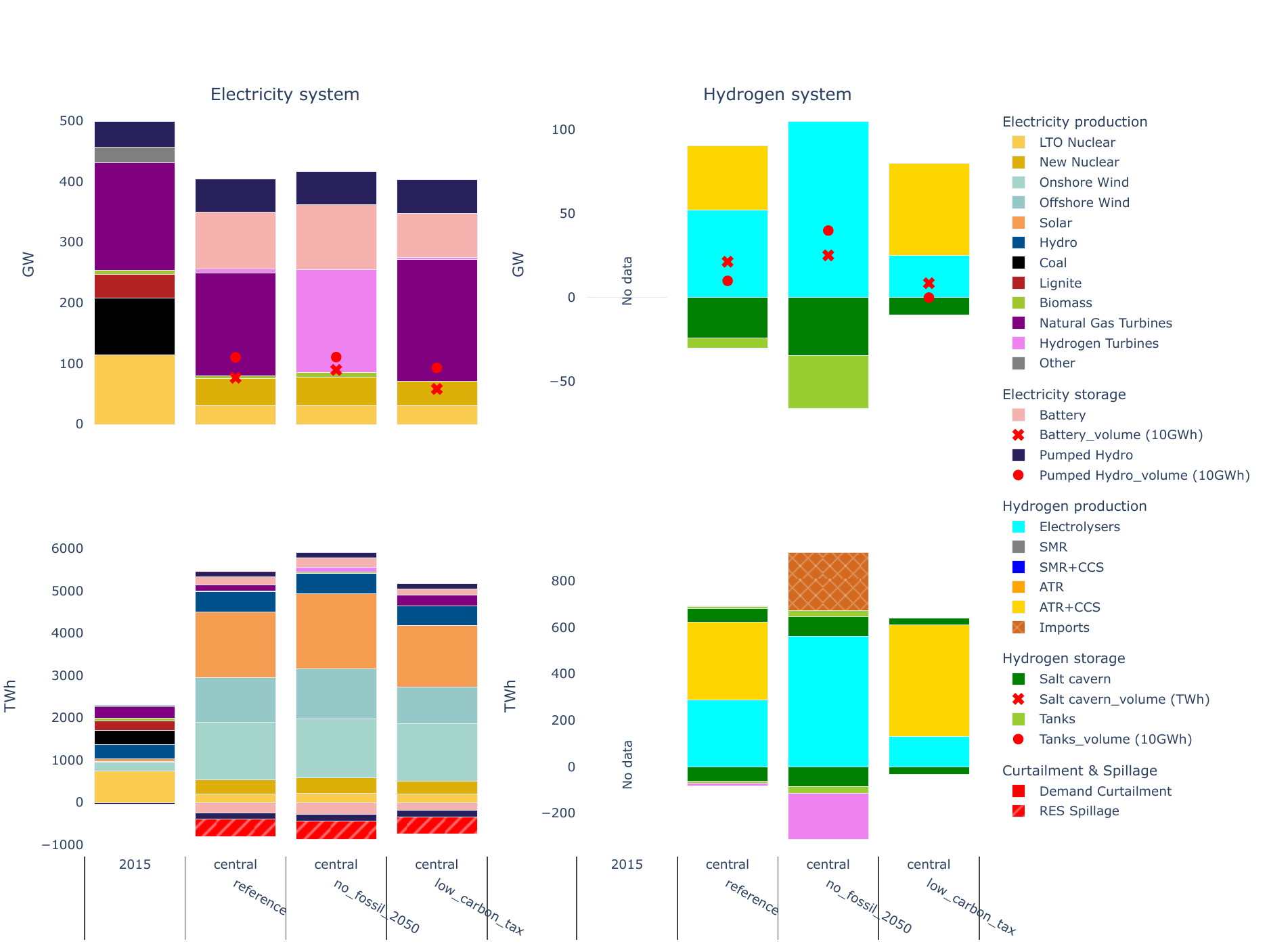}
    \caption{POMMES Standalone - Electricity and hydrogen system installed capacity (only distpatchables for electricity) and production per technology in 2050}\label{fig:partial_coupling}
\end{figure}

The results (\Cref{fig:partial_coupling}) reveal an electricity system predominantly powered by renewable energy, accounting for at least 85\% of total production (storage not included), with a relatively stable nuclear share at 10-11\%. Fossil fuel contributions, specifically from gas-fired power plants, account for 3\% and 5\% of electricity generation in the Reference and Low Carbon Tax scenarios, respectively. In contrast, fossil fuels constituted approximately 37\% of electricity production in 2015.

For hydrogen production (\Cref{fig:partial_coupling}), in the Reference scenario the distribution between electrolysis and ATR+CCS technologies is relatively balanced. A transition away from fossil fuels assigns a significant role to imports in ensuring supply. Naturally, a fossil-free approach implies the exclusion of ATR and SMR processes in POMMES results, which would necessitate a greater emphasis on storage capacity to manage the variable output from electrolysis. Similarly, in a fossil-free scenario, the electricity system replaces natural gas power plants with hydrogen-powered ones, significantly increasing the required hydrogen production to meet approximately 2\% of Europe’s electricity demand.

\begin{figure}[H]
    \centering
    \includegraphics[width=0.8\textwidth]{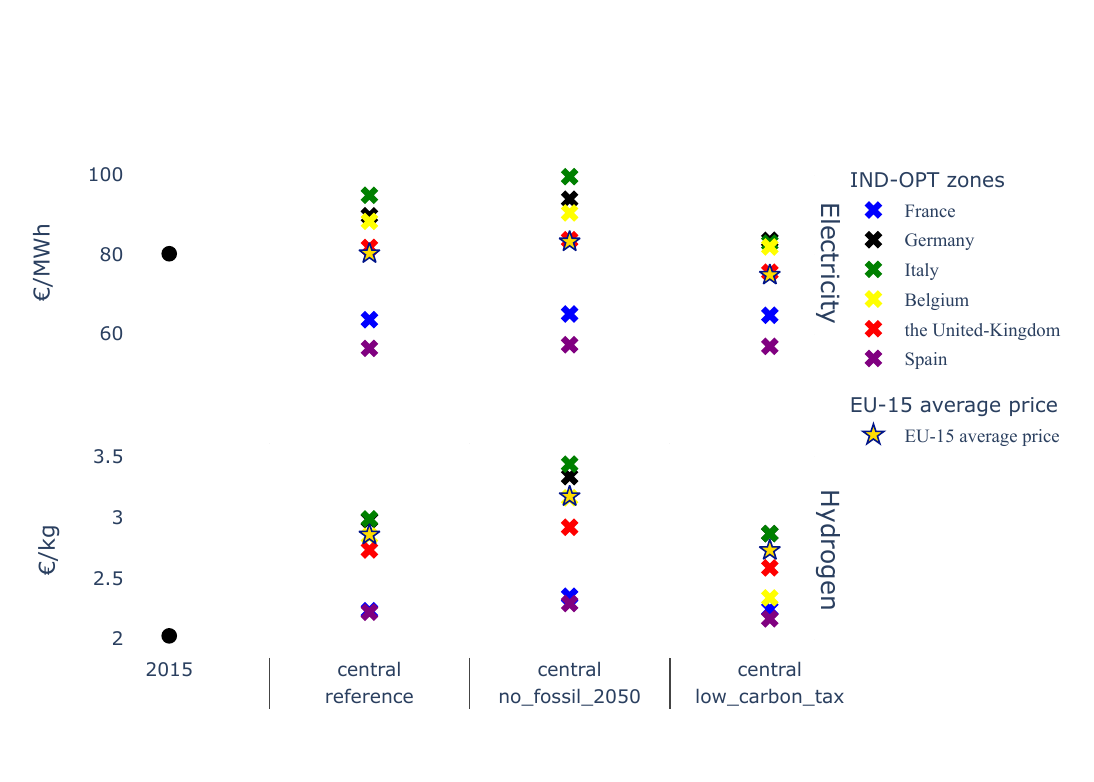}
    \caption{POMMES Standalone - Average electricity and hydrogen price in 2050}\label{fig:partial_coupling_price}
\end{figure}

Concerning electricity and hydrogen prices, the results presented in \Cref{fig:partial_coupling_price} reveal significant territorial disparities, with Spain and France consistently benefiting from lower prices. In contrast, Italy and Germany exhibit prices above the EU-15 average.

However, these prices are notably affected by carbon taxes. For most countries, a reduction by 50\% in the carbon tax leads to a decrease in electricity prices of around 5-6€/MWh. It is also noteworthy that a system devoid of fossil fuels results in higher prices—approximately 3€/MWh for electricity and 0.32€/kg for hydrogen at the EU-15 level. This increase can be attributed to the No Fossil 2050 scenario leading to a replacement of gas-fired power plants by hydrogen-fired power plants and an increased reliance on hydrogen imports, thereby driving up costs.

\subsection{Impact of additional industrial energy consumption}\label{subsec:consum_add_impact}

\begin{figure}[H]
    \centering
    \includegraphics[width=\textwidth]{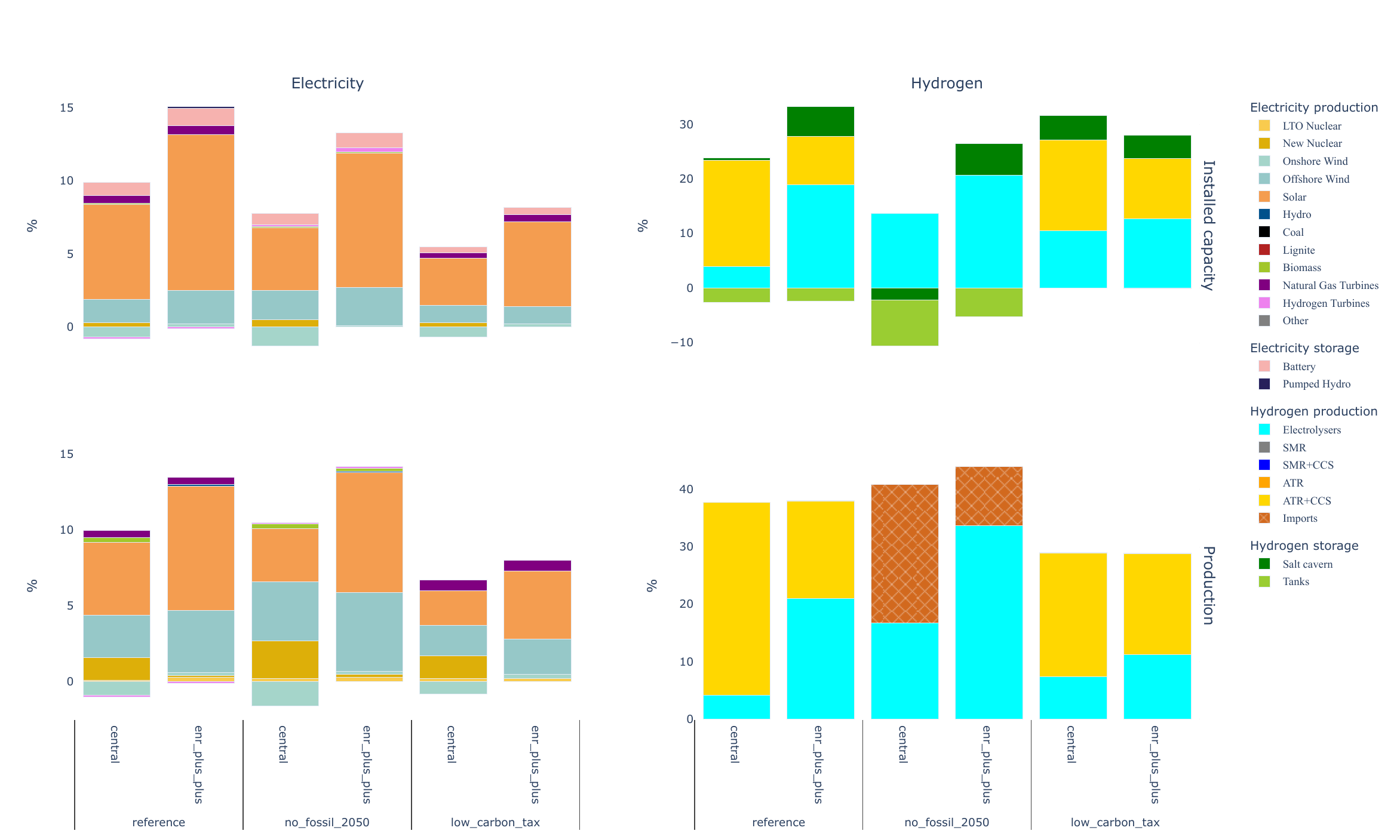}
    \caption{Impact of POMMES-IND-OPT coupling on POMMES Standalone installed capacity and production results in 2050}\label{fig:comp_coupling}
\end{figure}

When IND-OPT is integrated with POMMES, it results in additional electricity and hydrogen consumption that POMMES must address. As shown in \Cref{fig:comp_coupling}, the increased electricity demand requires a 5-8\% rise in electricity generation, primarily met through wind, solar, and nuclear energy sources.

Regarding hydrogen production for Central scenarios, the demand driven by IND-OPT necessitates a 29-41\% increase in hydrogen generation. In scenarios excluding "No Fossil 2050", more than 74\% of this additional production is supplied by ATR+CCS technologies. If we look at how the additional hydrogen demand is met when more renewable energies are authorised, we can see that electrolysis plays a much greater role in meeting this additional demand.

\begin{figure}[H]
    \centering
    \includegraphics[width=0.9\textwidth]{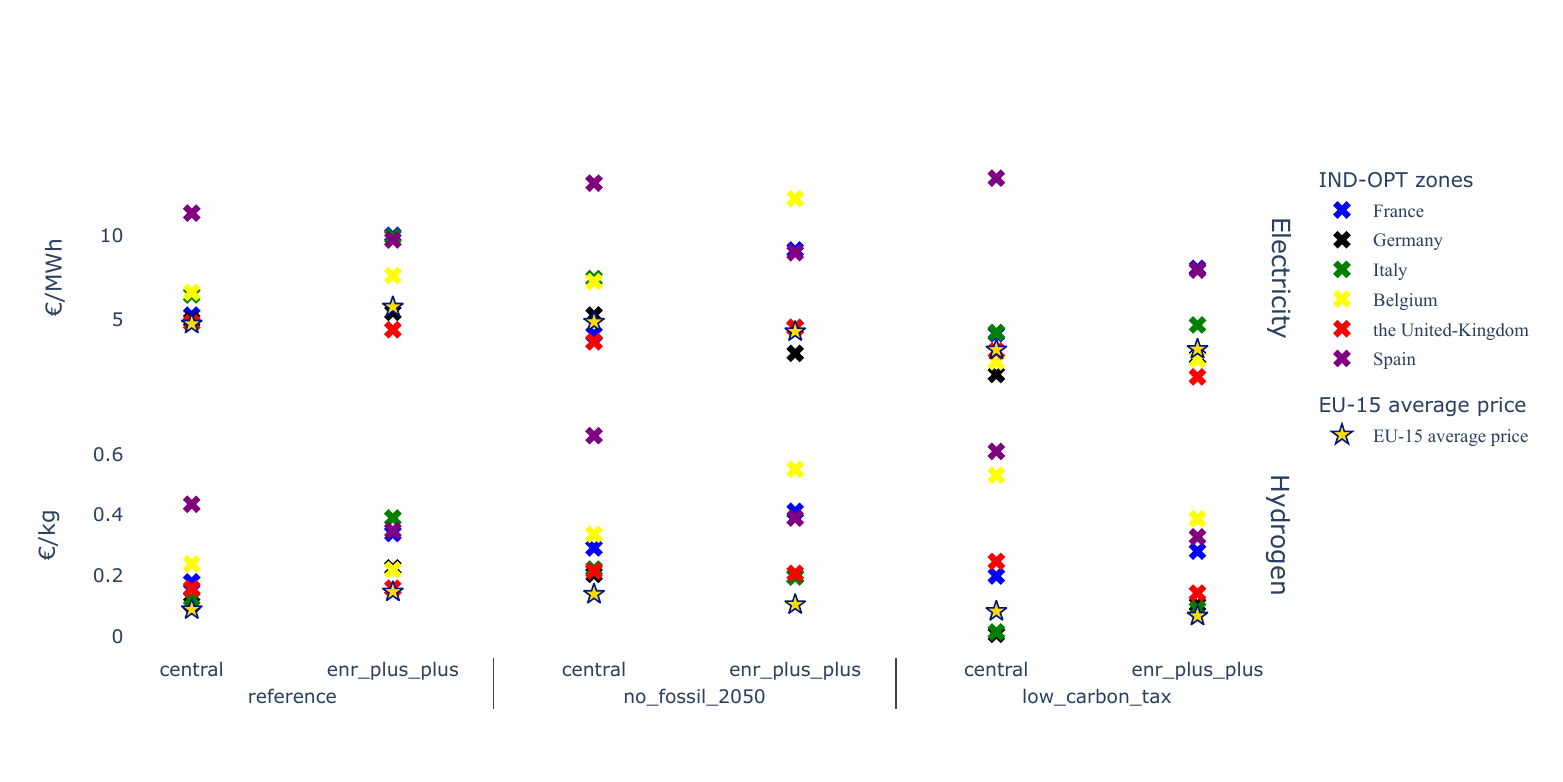}
    \caption{Impact of POMMES-IND-OPT coupling on POMMES Standalone electricity and hydrogen prices in 2050}\label{fig:comp_coupling_price}
\end{figure}

Regarding prices (\Cref{fig:comp_coupling_price}), the additional demand from the countries modelled in IND-OPT leads to an increase in both electricity and hydrogen prices in these countries, with varying impacts depending on the nation. Consequently, the average price in the EU-15 rises by 3-5€/MWh for electricity and 0.08-0.14€/kg for hydrogen. Spain appears as the most affected, experiencing an increase of up to 14€/MWh for electricity and 0.66€/kg for hydrogen. However, it seems that Spain's ability to install more low-cost renewable energy is significantly limiting the rise in prices.

This greater increase in costs observed in Spain can be explained by several factors. Firstly, the onshore wind potential had already been fully exploited before the coupling of the models. As a result (see \Cref{fig:comp_spain}), it is primarily solar energy, combined with batteries and a higher reliance on imports from France, that meets the additional electricity demand. This leads to a rise in electricity prices, which causes the additional hydrogen demand to be met by ATR+CCS, provided fossil fuels are allowed by the scenario. However, if Spain has the opportunity to install additional low-cost renewable energy (mainly solar, as the potential had not been fully reached), the increase in electricity prices is less significant. Consequently, the share of hydrogen production from electrolysers increases, which helps limit the rise in costs within the country.

However, these increases in prices should be viewed in context, as Spain, along with France, continues to have the lowest prices among the countries modelled by IND-OPT, as illustrated in \Cref{fig:full_coupling_elec_h2_prices}.

\subsection{Coupled Models: Integrated Results}\label{subsec:coupled_model_int_res}
\begin{figure}[H]
    \centering
    \includegraphics[width=0.8\textwidth]{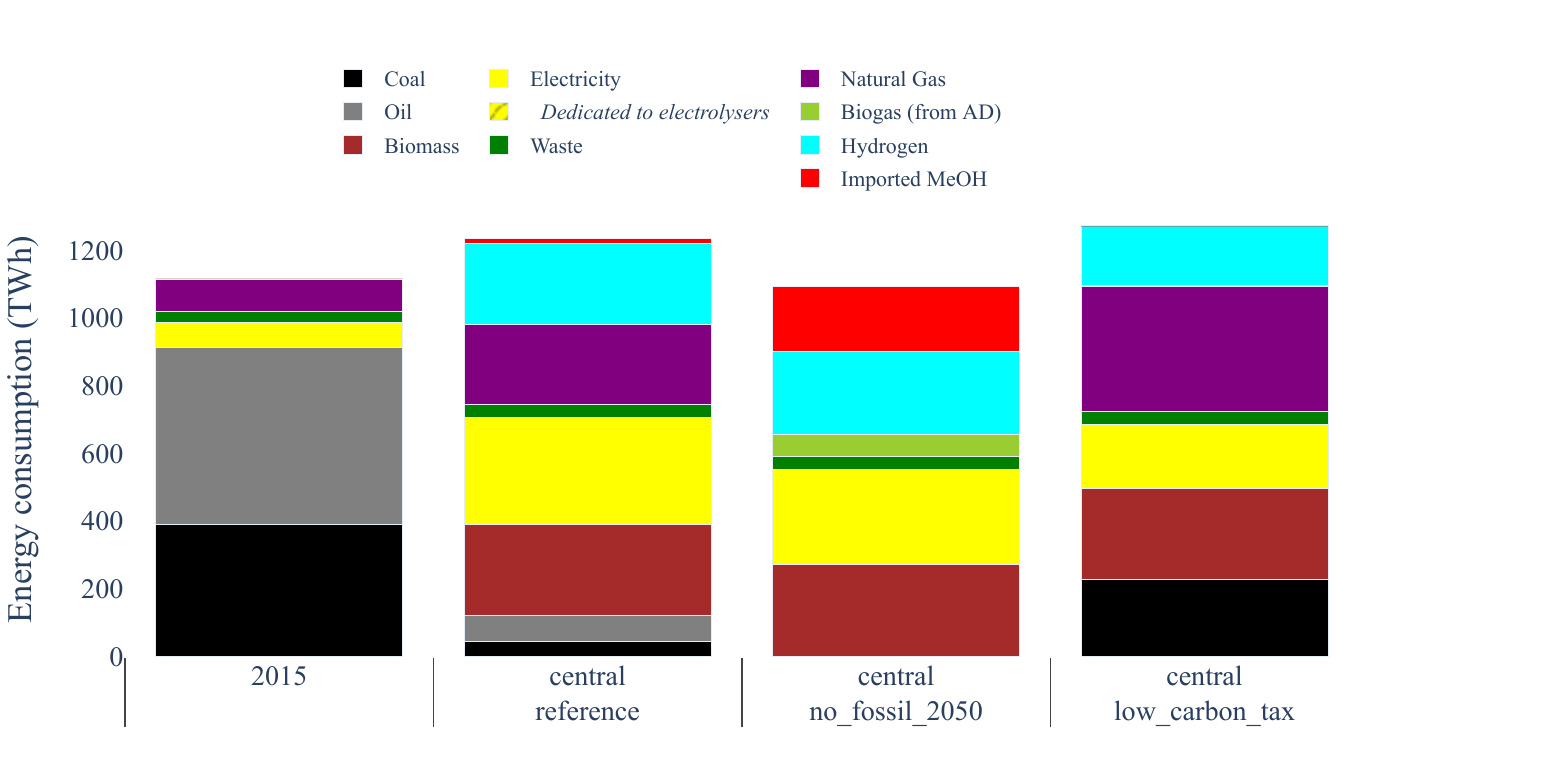}
    \caption{Heavy industry consumption in EU-5+1 from IND-OPT in 2050 | Electricity and hydrogen demands are met by POMMES}\label{fig:full_coupling_indus_consum}
\end{figure}

The impact of policy scenarios on industrial consumption is more significant than that of the techno-economic scenarios, which lead to only marginal changes. Therefore, \Cref{fig:full_coupling_indus_consum} presents the heavy industry consumption modelled by IND-OPT for the central case only across the three policy scenarios. Detailed information regarding the techno-economic scenarios impact is provided in \Cref{fig:full_coupling_indus_consum_appendix}.

A general increase in natural gas consumption is observed in scenarios permitting fossil fuel use, particularly for the production of MeOH in the chemical sector. Naturally, a reduction in the carbon tax leads to a higher share of fossil fuels in industrial consumption. However, it is noted that phasing out fossil fuels, as in the "No Fossil 2050" scenario, requires an increased reliance on methanol imports.

\begin{figure}[H]
    \centering
    \includegraphics[width=0.8\textwidth]{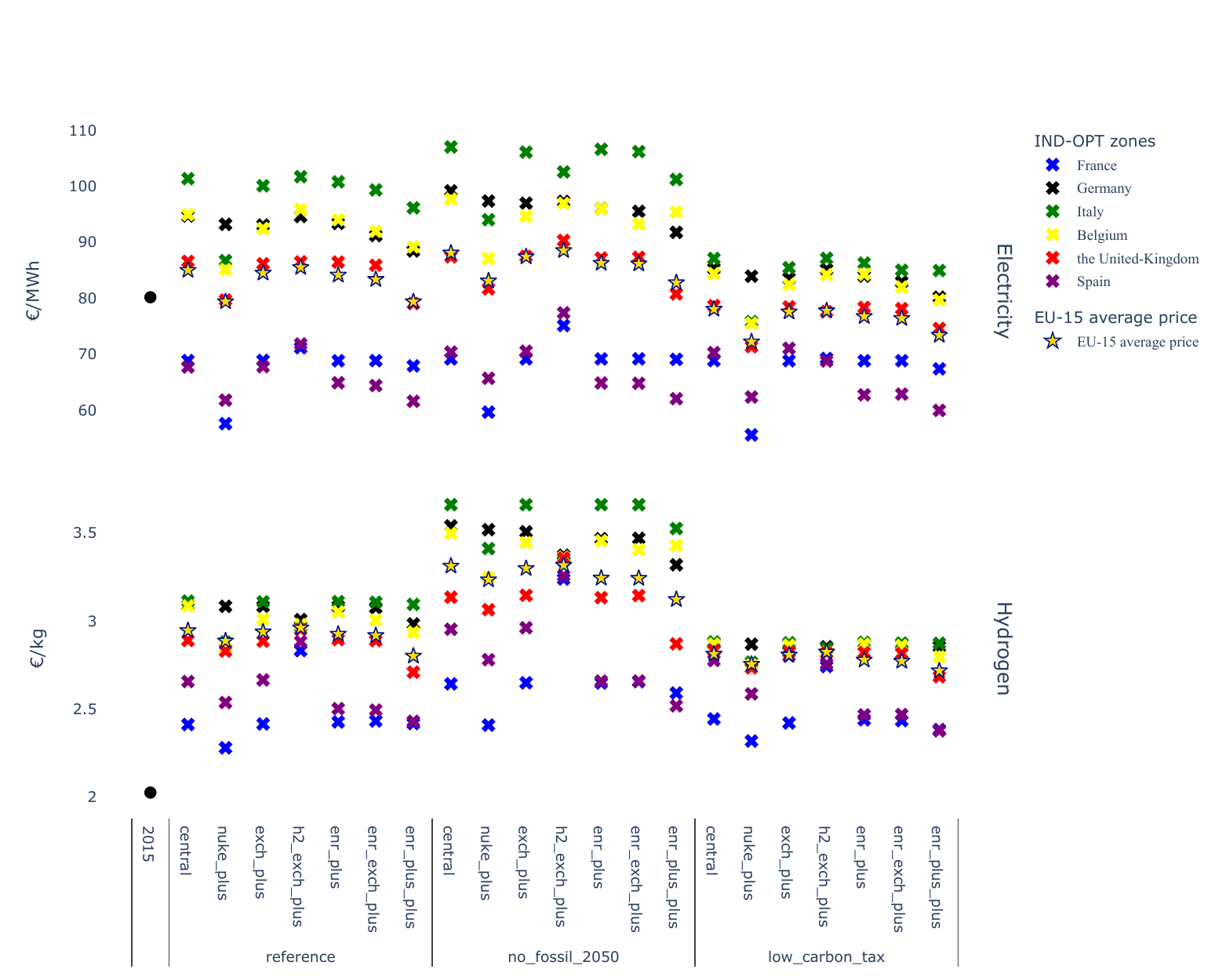}
    \caption{Average electricity and hydrogen price in 2050}\label{fig:full_coupling_elec_h2_prices}
\end{figure}

While \Cref{fig:full_coupling_elec_h2_prices} shows that a lower carbon tax results in electricity and hydrogen prices being reduced by 5-7 €/MWh and approximately 0.1 €/kg, respectively, these reductions are insufficient to drive an increase in electricity and hydrogen consumption compared to the Reference scenario. Indeed, the carbon tax of 150 €/tCO2, as opposed to 300 €/tCO2 in the Reference scenario, has a more significant impact on consumption.

\begin{figure}[H]
    \centering
    \includegraphics[width=\textwidth]{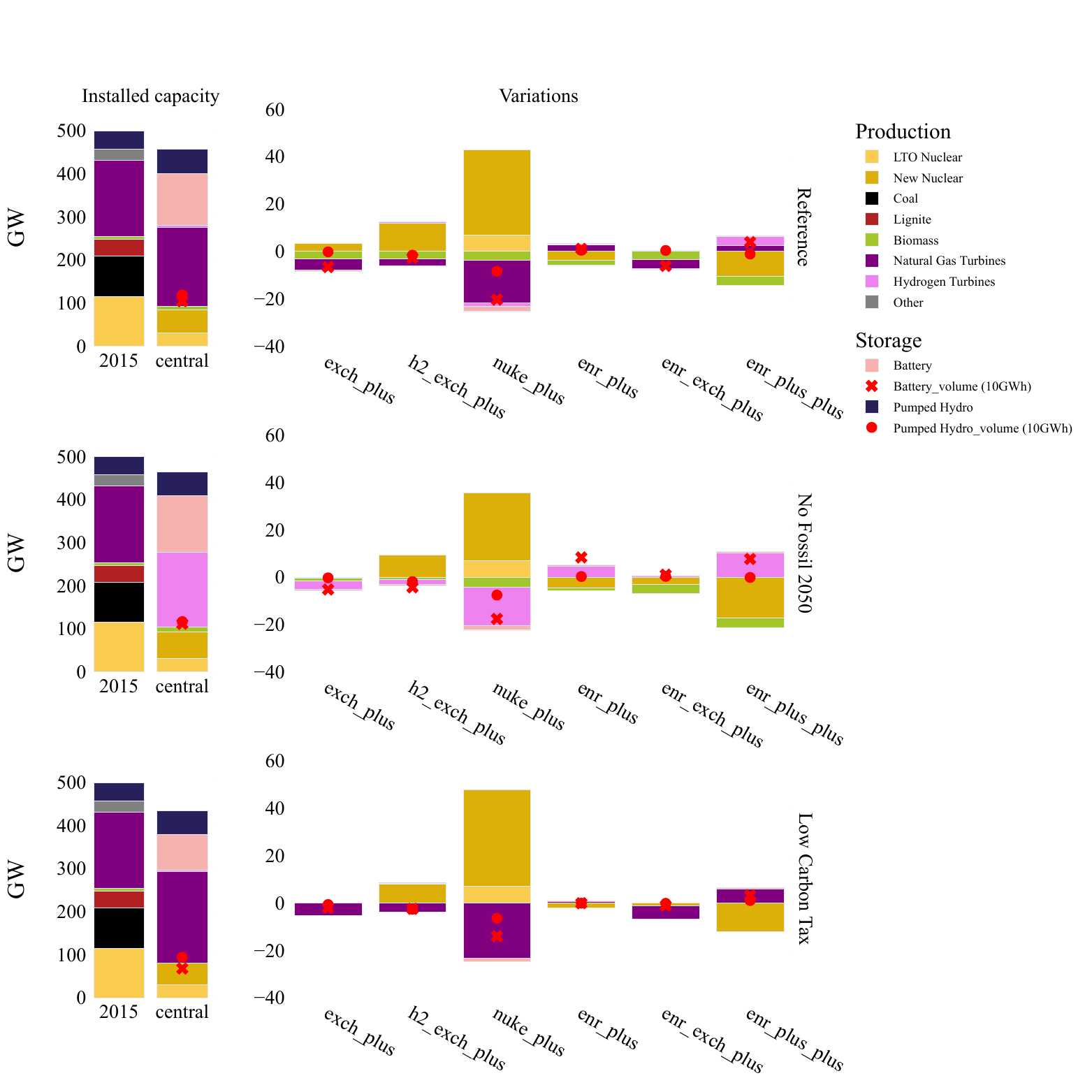}
    \caption{EU-15 electricity dispatchable installed capacity and variations in 2050}\label{fig:full_coupling_elec_dispatch_capa_var}
\end{figure}

\begin{figure}[H]
    \centering
    \includegraphics[width=\textwidth]{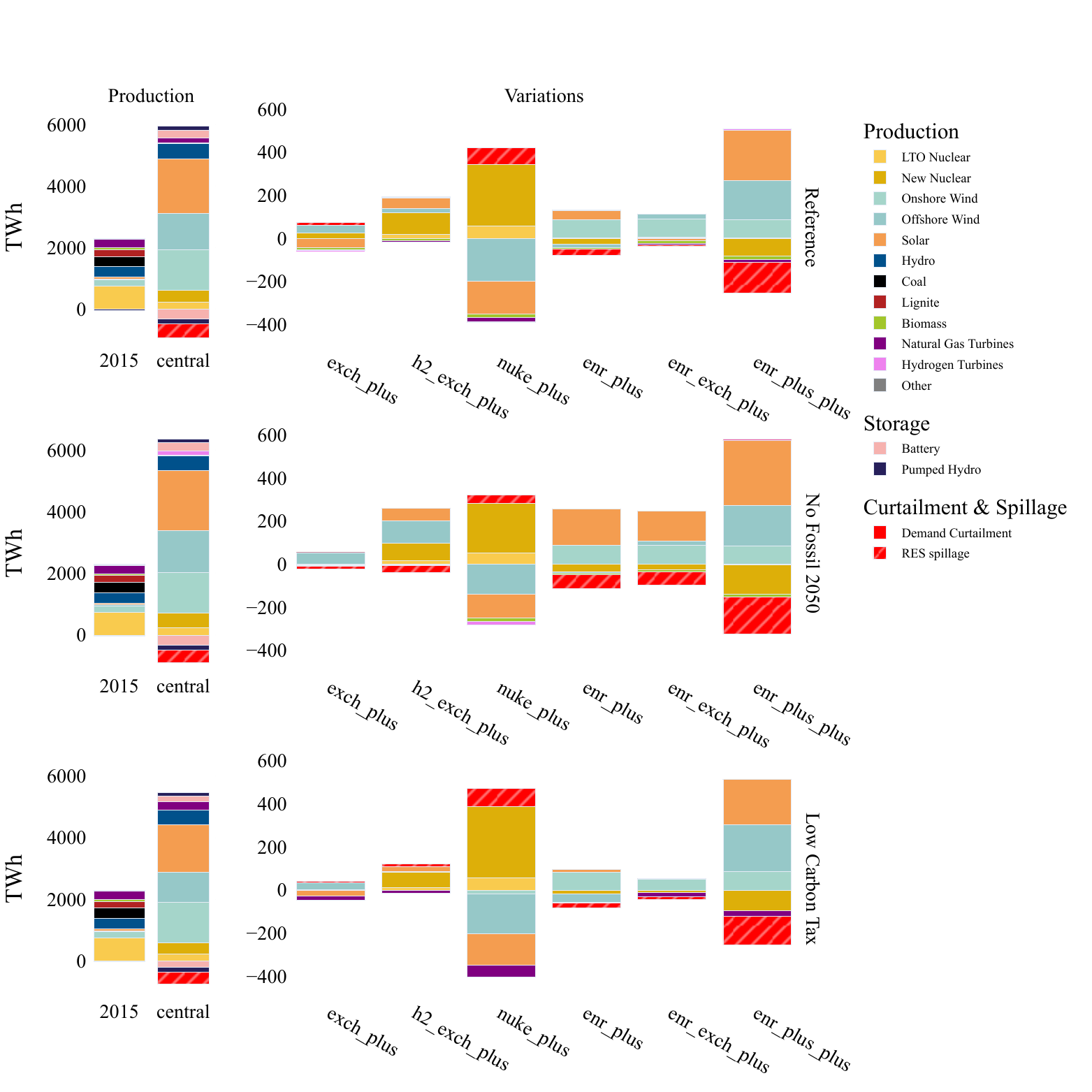}
    \caption{EU-15 electricity production and variations in 2050}\label{fig:full_coupling_elec_prod_var}
\end{figure}

Regarding the electricity system, the results reveal, as in \Cref{subsec:energy_system_standalone}, an electricity system predominantly driven by renewable energy sources and, to a lesser extent, nuclear power. The total installed capacity of dispatchable power and storage, shown in \Cref{fig:full_coupling_elec_dispatch_capa_var}, decreases slightly compared to 2015, despite electricity production being 2 to 3 times higher (\Cref{fig:full_coupling_elec_prod_var}). While scenarios with increased interconnections or nuclear energy show a reduction in thermal capacities (gas or hydrogen), scenarios with higher shares of renewable energy lead to a decline in nuclear power, with thermal plants used to compensate for intermittency. Naturally, the Nuke\_Plus scenarios result in an additional 35 to 48 GW of installed nuclear capacity.

As depicted in \Cref{fig:full_coupling_elec_prod_var}, these outcomes also impact electricity generation, where nuclear energy substitutes renewable energy generation (as well as their spillage) and vice versa. In the ENR\_Plus\_Plus scenario (with a lower capital cost for renewable energy), new nuclear power becomes significantly less competitive, reducing both its installation and production. This, however, leads to an increase in spillage, although a slight reduction in fossil fuel production is also observed. Nonetheless, the H\textsubscript{2}\_Exch\_Plus scenario, which facilitate hydrogen interconnections, benefit both nuclear and renewable energy sources.

\begin{figure}[H]
    \centering
    \includegraphics[width=\textwidth]{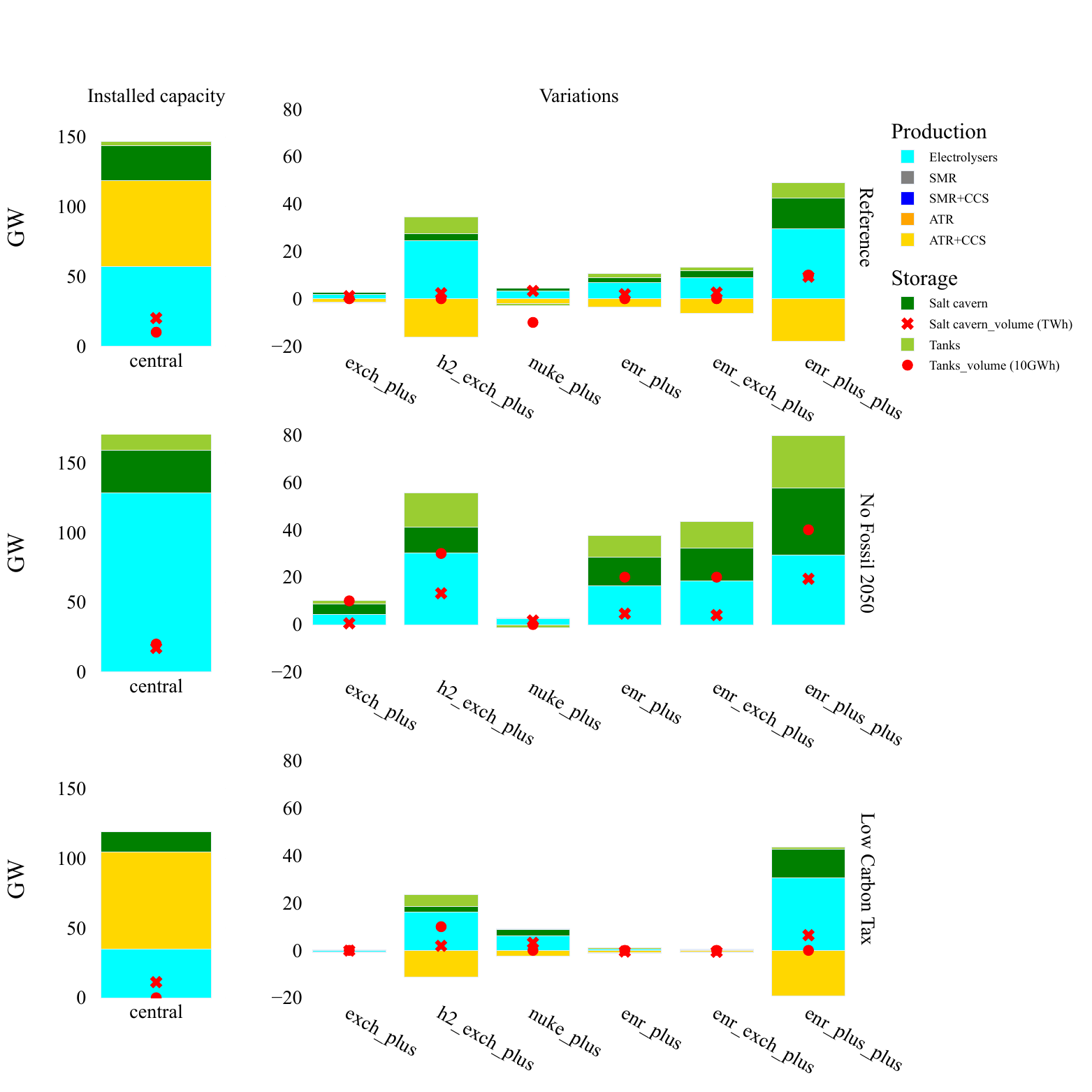}
    \caption{EU-15 hydrogen installed capacity and variations in 2050}\label{fig:full_coupling_h2_capa_var}
\end{figure}

\begin{figure}[H]
    \centering
    \includegraphics[width=\textwidth]{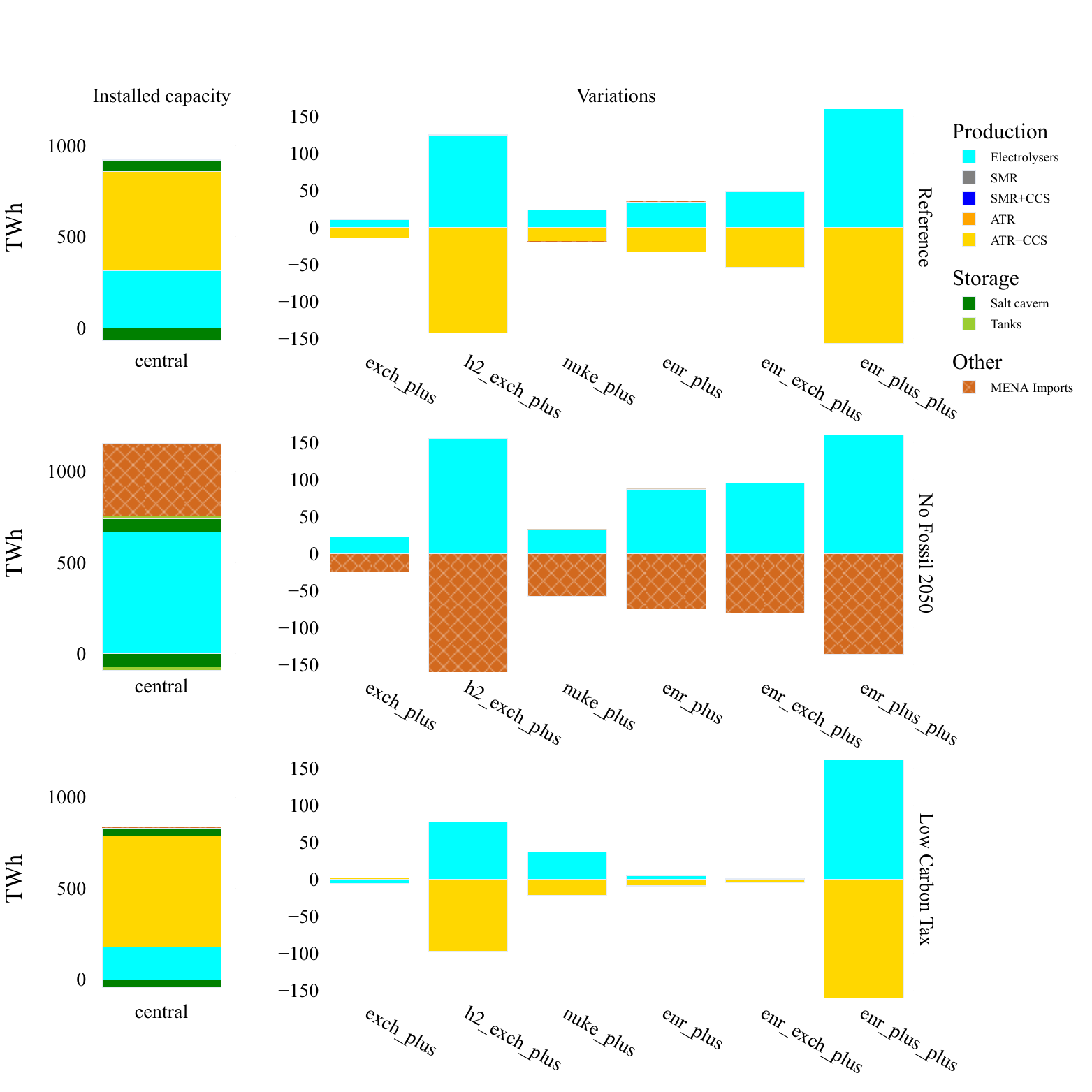}
    \caption{EU-15 hydrogen production and variations in 2050}\label{fig:full_coupling_h2_prod_var}
\end{figure}

For hydrogen production, \Cref{fig:full_coupling_h2_capa_var} shows that electrolyser capacities in the "Central" cases range between 35 and 128 GW. The installation of electrolysers appears to be correlated with the deployment of storage facilities. \Cref{fig:full_coupling_h2_prod_var} highlights a substantial reliance on hydrogen imports in the No Fossil 2050 scenarios. However, most techno-economic scenarios help reduce both fossil-based hydrogen production and imports.

Similar to \Cref{subsec:energy_system_standalone}, electricity and hydrogen prices, illustrated in \Cref{fig:full_coupling_elec_h2_prices}, exhibit significant variation depending on the country. The Nuke\_Plus and ENR\_Plus\_Plus scenarios result in the lowest prices. 
The results indicate that, with the exception of the No Fossil 2050 scenario, blue hydrogen constitutes a substantial proportion of production (\Cref{fig:full_coupling_h2_prod_var}). Regional disparities in production (\Cref{fig:map_ref_h2_network}) are mirrored in hydrogen prices and are a direct consequence of differences in regional electricity prices (\Cref{fig:full_coupling_elec_h2_prices}).

The addition of a hydrogen network in the H\textsubscript{2}\_Exch\_Plus techno-economic scenario slightly reduces hydrogen prices but, more importantly, brings the prices across countries closer to a common value. For instance, in the Reference scenario, countries such as France and Spain, with hydrogen prices of 2.41€ and 2.66€/kg respectively, see prices increase to 2.83€ and 2.88€/kg.

While \Cref{fig:partial_coupling_price} indicates that, in the Central cases, a system without fossil fuels results in an increase of approximately 3€/MWh for electricity and 0.32€/kg for hydrogen, \Cref{fig:full_coupling_elec_h2_prices} demonstrates that these values remain unchanged when the additional industrial consumption is accounted for.

\begin{figure}[H]
    \centering
    \includegraphics[width=0.9\textwidth]{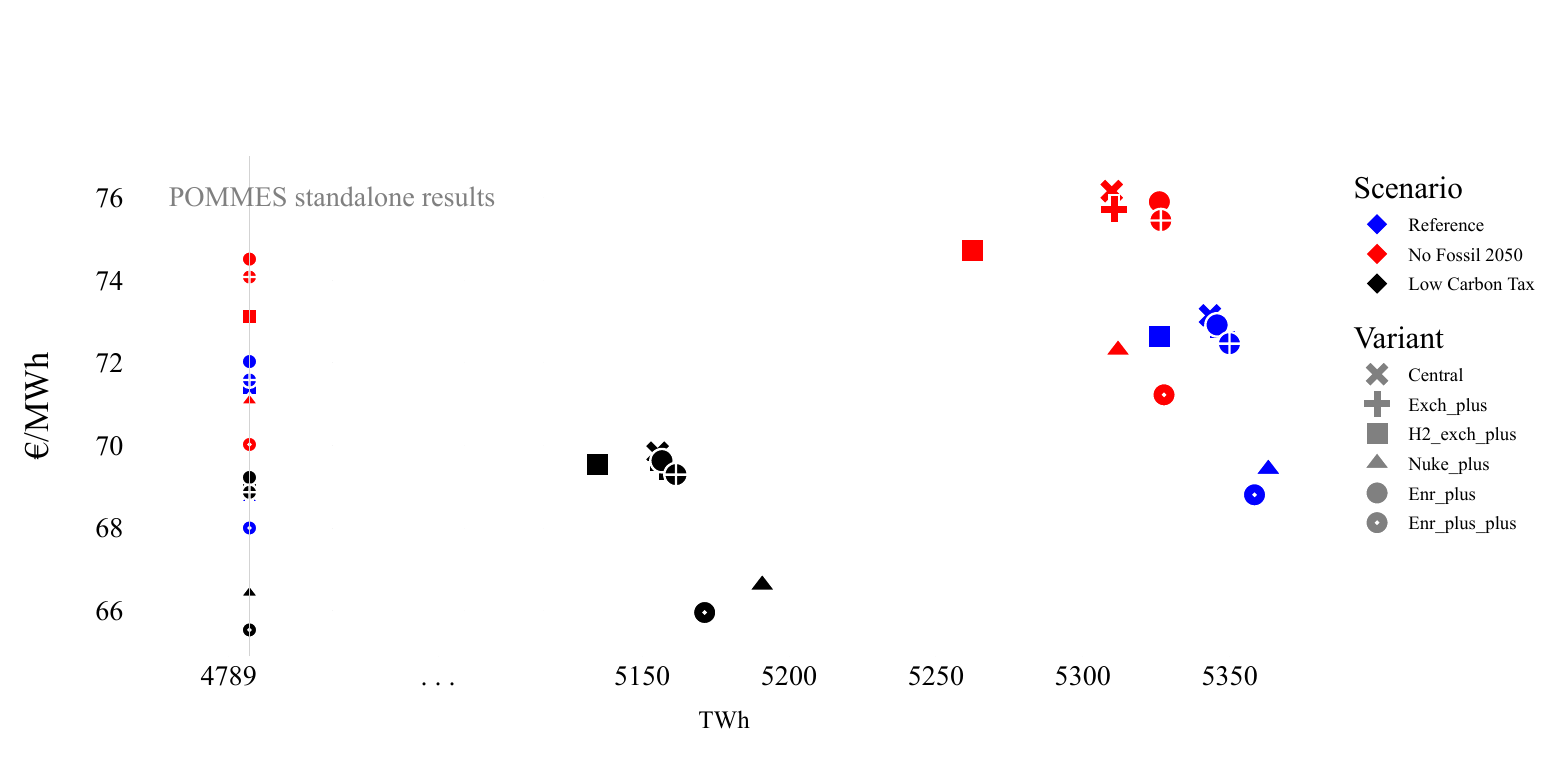}
    \caption{Electricity and hydrogen system overall cost in EU-15 in 2050 versus demand per scenario}\label{fig:system_cost_energy}
\end{figure}

\Cref{fig:system_cost_energy} illustrates the costs of the electricity and hydrogen system modelled by POMMES for 2050, according to the different scenarios and their impact on demand. It is evident that demand for electricity and hydrogen from IND-OPT varies significantly across both the policy scenarios and their respective techno-economic variation scenarios. The techno-economic scenarios that achieve the greatest cost reductions are those that lower the weighted average cost of capital (WACC) for nuclear energy (Nuke\_Plus) and renewable energy (ENR\_Plus\_Plus) to 4\% and 2\%, respectively. Although the No Fossil 2050 scenarios tend to be more expensive than others, the Nuke\_Plus and ENR\_Plus\_Plus scenarios still result in system costs that are considerably lower than in the Reference Central scenario.

It is worth noting that, although \Cref{fig:full_coupling_elec_h2_prices} demonstrates that the addition of a hydrogen network using the H\textsubscript{2}\_Exch\_Plus scenario has no effect on average prices at the European level, \Cref{fig:system_cost_energy} shows a reduction in the weighted system cost of between 0.4\% and 1.9\% compared to the Central cases.

\section{Discussion}
\label{sec:discussion}
\subsection{Results implication}
\label{subsec:results_implication}

The results of this study provide important insights into potential future compositions of the European energy and hydrogen systems, particularly in the context of increased industrial low-carbon energy consumption. The analysis emphasises the substantial role of renewable energy, accounting for at least 85\% of electricity production in 2050, and nuclear energy providing 10-11\% of the supply. These findings reflect a significant shift from the 2015 energy mix, where fossil fuels made up 37\% of electricity generation in EU-15, highlighting the potential for a highly decarbonised system. However, the industrial sector's energy demands pose significant challenges to achieving this transition cost-effectively.

The integration of additional industrial energy consumption (from the IND-OPT model) represents a critical factors affecting future energy systems. When IND-OPT is incorporated, electricity demand in EU-15 increases by 5-8\%, and hydrogen demand rises dramatically by 29-41\%, depending on the scenario, although IND-OPT only considers 6 countries. This significant increase in consumption has major implications for both energy generation and infrastructure. To meet this growing demand, the system must rely heavily on renewable and nuclear generation, requiring further expansion of wind and solar capacities alongside hydrogen production technologies. The costs associated with these adjustments vary considerably across different scenarios, with some scenarios achieving notable cost reductions by lowering the WACC for nuclear and renewable energy, while others, particularly those phasing out fossil fuels, tend to be more expensive.

The impact of industrial consumption is not limited to energy generation but extends to energy prices, particularly in electricity and hydrogen markets. The increased demand driven by industries leads to higher prices, with notable variations across regions. For instance, Spain and France, while continuing to enjoy some of the lowest prices, experience sharp increases in electricity prices. In Spain, prices rise by as much as 14€/MWh, while hydrogen prices increase by 0.66€/kg due to the industrial sector’s growing demand. This shows that even countries traditionally benefiting from lower prices are not insulated from the pressures of increased industrial consumption.

These results underscore the vital role of industry in shaping energy system outcomes and highlight the challenges of balancing decarbonisation goals with the energy needs of a changing industrial sector. The findings suggest that policymakers must address the industrial sector’s future energy needs through targeted interventions, expanding renewable capacities, and ensuring sufficient hydrogen infrastructure to meet demand. Without careful planning, the energy-intensive demands of industry could threaten the cost-effectiveness of the transition and exacerbate regional price disparities. Indeed, the results also show that the policy scenarios have a much greater impact on the industry and the energy systems than the techno-economic scenarios. 

In scenarios such as No Fossil 2050, where fossil fuels are phased out entirely, industrial demand significantly increases the reliance on electrolytic hydrogen production and imports. The transition away from fossil fuels would necessitate a shift to electrolysis-based hydrogen production, raising concerns about the availability of sufficient storage and infrastructure to handle the variability in renewable electricity generation. Additionally, the required expansion in hydrogen production capacity, particularly from ATR+CCS technologies, underscores the technical and economic hurdles that must be addressed to meet industrial demand. 

The analysis of hydrogen networks, such as in the H\textsubscript{2}\_Exch\_Plus scenario, reveals important insights into how industrial consumption impacts regional price dynamics. While hydrogen networks can help reduce price disparities between countries, not all nations benefit equally. The price convergence observed in the H\textsubscript{2}\_Exch\_Plus scenario results in rising hydrogen prices for countries like France and Spain, which in the Central techno-economic scenario case had lower prices. This demonstrates that while hydrogen networks promote integration, they can also diminish the competitive advantage for countries with lower energy costs, potentially leading to increased costs for industrial consumers in these regions. 

Overall, the results indicate that the industrial sector's growing energy demands can be a crucial determinant in the success of Europe's energy transition. Achieving a sustainable, low-carbon energy system will require not only an expansion of renewable and hydrogen technologies but also a clear strategy for managing the energy-intensive demands of industries. The trade-offs between decarbonisation, affordability, and regional price disparities will need to be carefully balanced, particularly in regions with substantial industrial activity. Policymakers must consider these factors as they design strategies for the energy transition, ensuring that both the energy and industrial sectors can adapt to the challenges of a decarbonised future.

\subsection{Comparison with literature}

While our study gives an economically driven substantial role to renewable energy, accounting for at least 85\% of electricity production in 2050, \citet{Fleiter2023METISSystem.} finds a share of 87-88\%.

Regarding hydrogen demand, our results range from 22 to 30 Mt, depending on the scenario, which aligns with the lower end of the range reported by \citet{Tarvydas2022TheScenarios}, varying between 19 and 60 Mt. The difference with the upper range can be explained by the inclusion of hydrogen use in buildings in some scenarios, which we do not account for, as well as the fact that \citet{Tarvydas2022TheScenarios} considers the whole of Europe, while we focus on only 15 countries.

The inclusion of a hydrogen network in our analysis results in a reduction in weighted system costs of between 0.4\% and 1.9\%. In comparison, \citet{Neumann2023TheEurope} report a decrease in total system costs ranging from 1.6\% to 3.4\%. This difference could stem from several factors. Firstly, \citet{Neumann2023TheEurope} do not account for extra-European imports. Secondly, their hydrogen network design differs from ours; for instance while they allow for an interconnection between France and Italy, our model, based on TYNDP data, allows the construction of a pipeline between Spain and Italy, which \citet{Neumann2023TheEurope} do not consider. Finally, they exclude nuclear energy from their analysis, which we have seen substantially impacts electricity and hydrogen prices.

Similar to our study, \citet{Kountouris2024AProduction} emphasise the synergies between SMR/ATR with carbon capture and electrolysers. However, in their reference scenario H2E, electrolysers produce the majority of hydrogen, nearly 60\%, whereas our study finds the opposite. This discrepancy is due to different techno-economic assumptions. Specifically, \citet{Kountouris2024AProduction} use higher electrolyser efficiency assumptions (74\% in 2050) compared to ours (65\%). It appears they employed higher heating value (HHV) efficiency, while we used lower heating value (LHV). Moreover, their 2020 efficiency figure of 65.6\% differs from their source's 57.7\%, suggesting they might not have included Balance of Plant consumption. 

Additionally, the efficiencies for fossil-based hydrogen in \citet{Kountouris2024AProduction} are closer to our assumptions for SMRs than for ATRs. However, ATRs with carbon capture are ultimately less costly than their SMR counterparts.
Despite these differences, both studies highlight the presence of synergies between electrolysers and SMRs/ATRs.

Another result that can be linked to existing literature is the impact of a hydrogen network on the spatial distribution of hydrogen production facilities, as illustrated in \Cref{fig:map_ref_h2_network}. Similarly to the findings of \citet{Fleiter2023METISSystem.}, our study shows Germany and Belgium importing all of their hydrogen, significantly affecting their neighbouring countries. The main difference between \citet{Fleiter2023METISSystem.} and our study lies in the broader range of countries included in their analysis, as well as their assumption that hydrogen production is entirely based on electrolysis. In contrast, we allow for blue hydrogen production, and even grey hydrogen if economically viable.

The case study by \citet{Fleiter2023METISSystem.} aligns with our "No Fossil 2050" policy scenario in its H2\_Exch\_Plus variation. Although \citet{Fleiter2023METISSystem.} does not provide an economic analysis, their scenario leads, in our study, to an increase of 2.9\% in weighted system costs, 3.5\% in electricity prices, and 12.1\% in hydrogen prices compared to the Reference scenario, which permits the use of fossil fuels.

\subsection{Limitations and perspectives}
A limitation of this study pertains to the POMMES model, for which only a single weather time series (2018) was utilised. While incorporating multiple weather time series through a stochastic approach would offer more robust estimations, it would significantly increase computational demands and time. Future work could explore such an approach. Additionally, as a perspective, a comparison between the IND-OPT-POMMES model results, which integrate both industrial and energy systems, and a full POMMES model could be conducted. The full POMMES model would remove the intermediary layer that links electricity and hydrogen prices between the energy system and industry, allowing industry to be modelled as an integrated component of the overall system. This contrasts with the current approach, where industry is optimised for its own benefit rather than as part of a holistic energy system, and would require further modifications to POMMES.

Regarding the hydrogen network and imports from the MENA region, the imports are treated as exogenous, without considering ramping or capacity constraints. In this model, a country like Poland is assumed to import hydrogen directly from MENA, while in reality, such imports would likely flow through France, Spain, and Italy via pipelines. A more accurate representation would model MENA as a node that produces and transmits hydrogen to Europe, respecting pipeline capacity constraints.

Another limitation of this study is that the POMMES model only allows for the consumption of natural gas in SMR/ATR and even CCGT/TAC processes. This choice stems from the default allocation of biogas production to the industrial sector. Consequently, the IND-OPT model allows the use of biogas in its industrial processes and for the portion of hydrogen production that it manages internally. Future research could explore how the results evolve, particularly in the No Fossil 2050 policy scenario, when the use of biogas is permitted within the POMMES framework.

A potential improvement in modelling could involve modifying POMMES to allow for the flexibility of demand from future electric vehicles. Indeed, the integration of electric vehicles into the power system appears to be a well-researched topic \cite{Taljegard2019ElectricEurope,Loschan2023FlexibilityAustria,Colmenar-Santos2019ElectricScenario,Canigueral2021FlexibilityStudy,Nikoobakht2019ElectricSystems,Zhao2019QuantifyingResponse,Gerritsma2019FlexibilityFuture}, with TSO reports, such as those from RTE \cite{RTE2021Futurs2050,RTE2024Bilan2023-2035}, even incorporating Vehicle-to-Grid (V2G) technology in their prospective studies.

Finally, this study does not account for the potential negative impact of decarbonising heavy industry on competitiveness, particularly with regard to the associated costs. \citet{Cooper2024MeetingUK} demonstrated that decarbonisation in the UK could raise the prices of manufactured goods by 10-15\%. In such a scenario, two outcomes are possible: either industries relocate, leading to a reduction in domestic energy consumption, or competitiveness is maintained through some form of external support \cite{Cooper2024MeetingUK}. However, this issue lies beyond the scope of the present study.

\section{Conclusion}
\label{sec:conclusion}

A model of key heavy industry sectors in six major European countries was coupled with a model of the European energy system. The aim of this coupling was to assess the impact of electricity and hydrogen consumption by these key industrial sectors on the energy system under various policy and techno-economic scenarios. The findings show that policy scenarios lead to the greatest differences in the results. The study highlights synergies between energy system planning and heavy industry. Specifically, the inclusion of industrial electricity and hydrogen consumption leads to an increase in electricity and hydrogen prices, with significant disparities between countries. Additionally, phasing out fossil fuels results in greater price increases and a higher reliance on hydrogen and methanol imports.

While scenarios with lower carbon taxes lead to lower electricity and hydrogen prices, these reductions are not sufficient to increase industrial consumption. In fact, lower carbon taxes make continued use of fossil fuels more attractive to industries. 

The results also show that introducing a hydrogen network helps to reduce price disparities between countries but diminishes the competitiveness of countries that would otherwise benefit from lower prices in the absence of such a network. 

As well, the additional electricity and hydrogen consumption in the energy system is primarily met by increased renewable energy installations in the power sector. In the hydrogen sector, except in scenarios involving the complete phase-out of fossil fuels, most of the additional consumption is met through fossil-based hydrogen production with carbon capture.
The study thus demonstrates synergies between electrolytic hydrogen and blue hydrogen, which help to reduce overall hydrogen prices. These synergies appear as essential for achieving competitive hydrogen production prices.

\section*{CRediT author statement}
\textbf{Quentin Raillard{-}{-}Cazanove:} Conceptualization, Methodology, Software, Writing – original draft. \textbf{Robin Girard:} Supervision, Conceptualization, Writing - Review \& Editing. \textbf{Thibaut Knibiehly:} Software, Writing - Review \& Editing.

\section*{Declaration of generative AI and AI-assisted technologies in the writing process}
During the preparation of this work the authors used ChatGPT-4 in order to improve readability and language. After using this tool/service, the authors reviewed and edited the content as needed and take full responsibility for the content of the publication.

\appendix
\section{Supplementary results}
\label{appendix:sup_results}
\subsection{Industry consumption results}\label{appendix:indus_consum_results}

\begin{figure}[H]
    \centering
    \includegraphics[width=\textwidth]{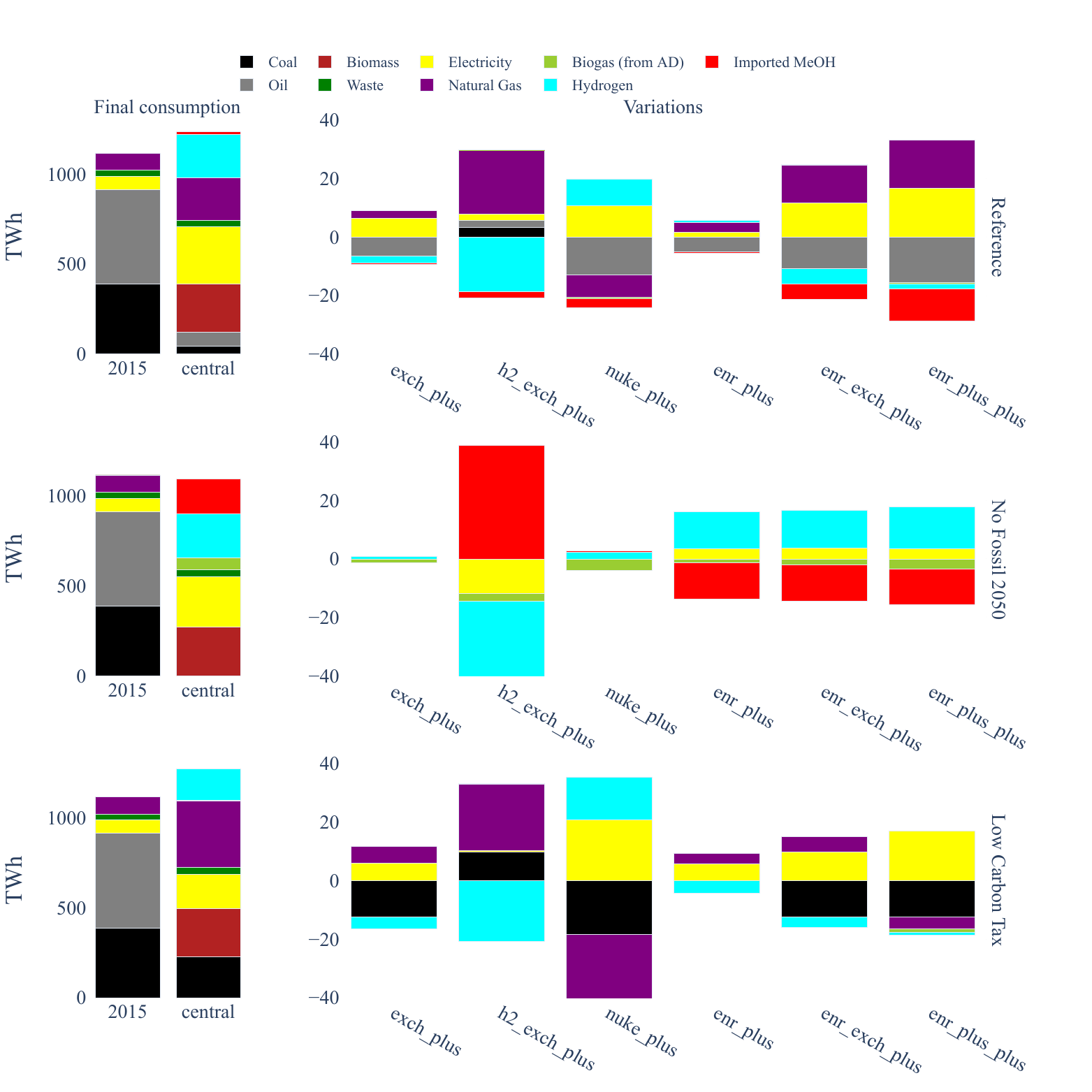}
    \caption{Heavy industry consumption and variations in EU-5+1 from IND-OPT in 2050 | Electricity and hydrogen demands are met by POMMES}\label{fig:full_coupling_indus_consum_appendix}
\end{figure}

\Cref{fig:full_coupling_indus_consum_appendix} shows the impact of the techno-economic scenarios on the central scenarios presented in \Cref{fig:full_coupling_indus_consum}. As stated in \Cref{subsec:coupled_model_int_res}, the variations due to the techno-economic scenarios on the final consumption of the sectors modelled by IND-OPT are marginal, of the order of 3\% maximum. The policy scenarios, on the other hand, have a huge impact on technological choices and therefore on consumption. 

\begin{figure}[H]
    \centering
    \includegraphics[width=\textwidth]{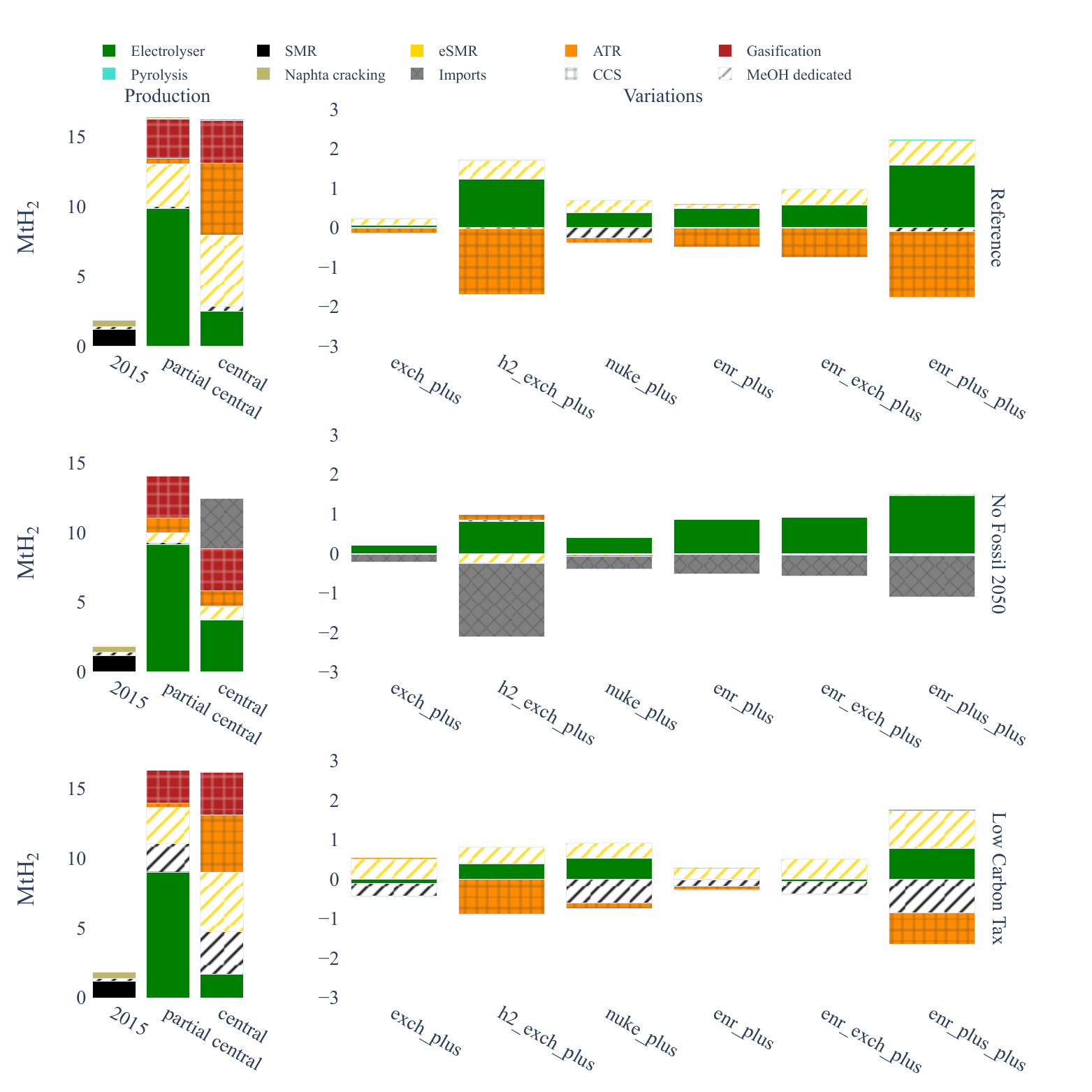}
    \caption{Heavy industry hydrogen origin in EU-5+1 from IND-OPT in 2050}\label{fig:full_coupling_h2_indus}
\end{figure}

\Cref{fig:full_coupling_h2_indus} illustrates the origin of hydrogen consumed in the sectors modelled by IND-OPT. To construct this figure, data from hydrogen production in POMMES and production managed by IND-OPT (notably for gasification or MeOH production) are cross-referenced. Additionally, IND-OPT can operate some ATR+CCS plants, which can also consume biogas produced from biomass gasification. In the No Fossil 2050 scenario, the ATR+CCS units operate entirely on biogas. The figure compares results in a scenario where IND-OPT entirely manages hydrogen production, based on electricity prices derived from the results presented in \Cref{subsec:energy_system_standalone} (partial coupling). 

Similar to \Cref{fig:full_coupling_h2_prod_var}, under coupled models, nearly all scenarios lead to a reduction in ATR+CCS production as well as a decrease in imports. In the EU-5+1 case, hydrogen imports, which typically come from the MENA region, can also come from neighbouring countries like Norway in the H\textsubscript{2}\_Exch\_Plus scenario. Likewise, except for the No Fossil 2050 scenario, the techno-economic variations largely replace SMRs for MeOH production with eSMRs that consume electricity rather than gas for heat production. This is due to the significant impact on electricity prices of the techno-economic scenarios. 

The difference between partial and full coupling is explained by the fact that IND-OPT, with its annual time step, does not manage supply-demand balance at the hourly level, a function that POMMES handles. This leads to higher costs in POMMES due to the inclusion of storage solutions. As a result, when operated independently, IND-OPT overestimates the competitiveness of electrolysers. This is a known limitation of IND-OPT, as identified in \citet{Raillard--Cazanove2024IndustryTrajectory}, further emphasising the benefits of coupling IND-OPT with POMMES, as showed in this study.

\subsection{Hydrogen network impact}\label{appendix:h2_network_impact}
\begin{figure}[H]
\begin{tikzpicture}
\centering
\node[above left] (a) at (-8,5) {\Large (a)};
\node[above left,inner sep=0] (image) at (0,0) {\underlay{4in}{250pt}{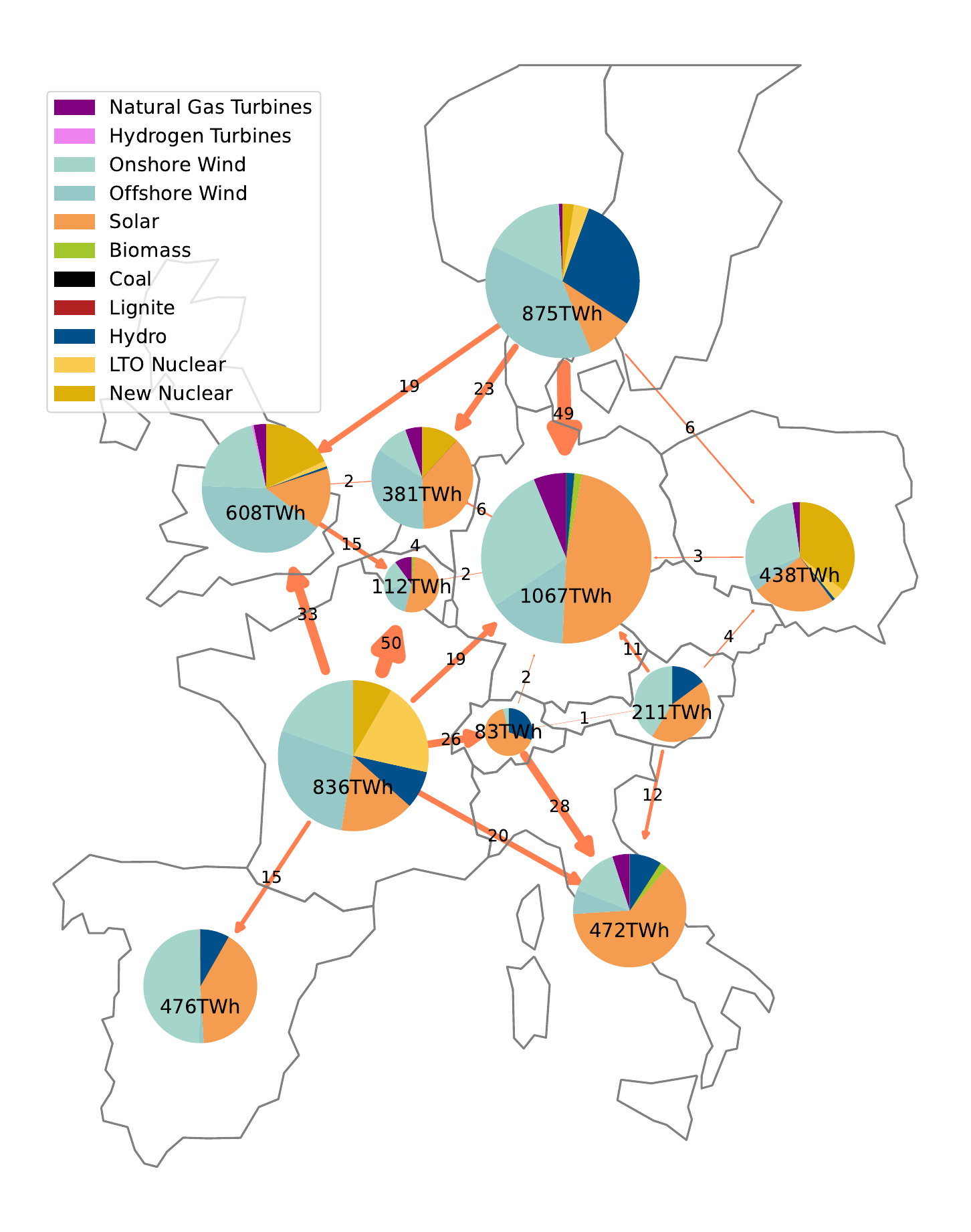}{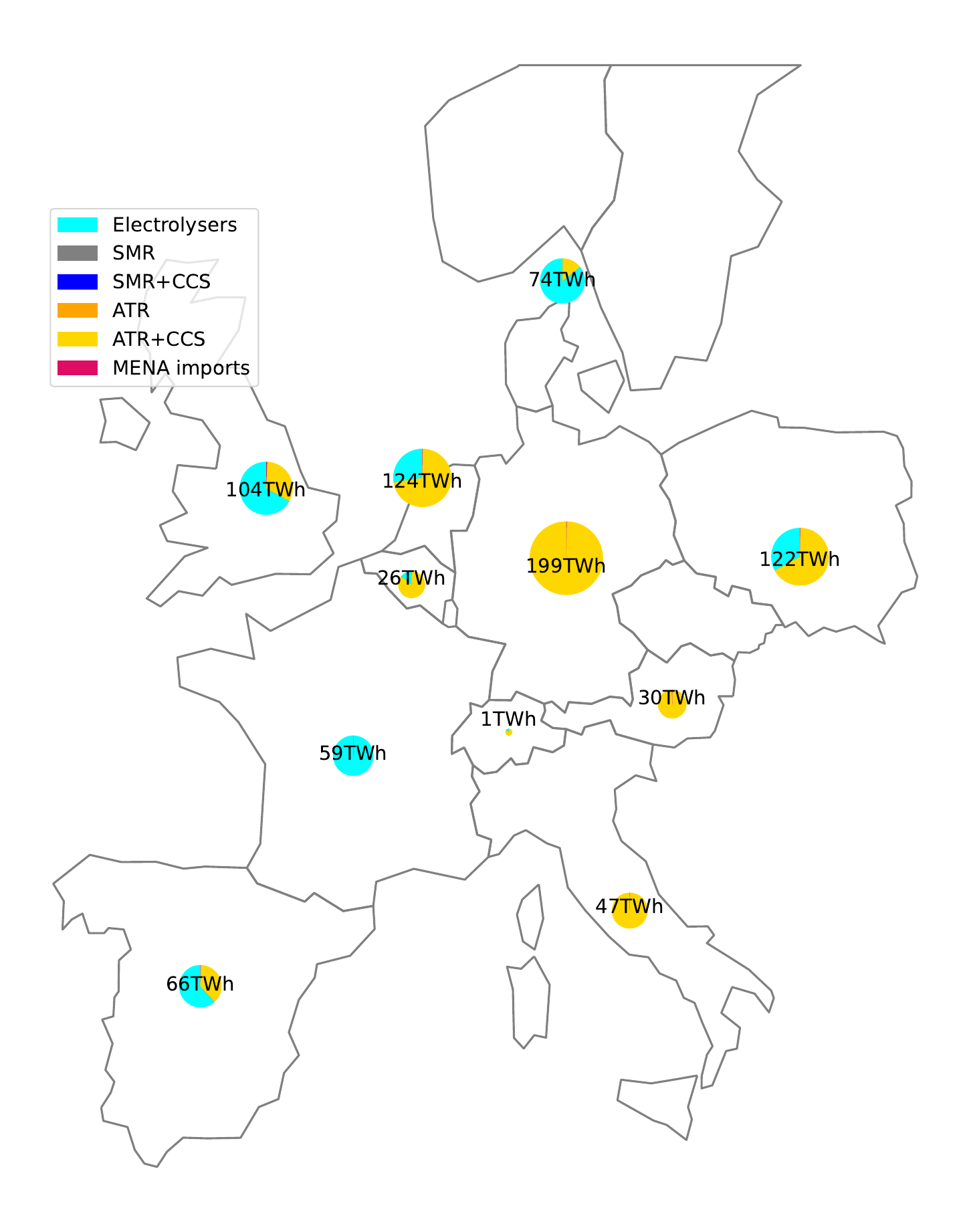}};
\node[below left] (b) at (-8,-5) {\Large (b)};
\node[below left,inner sep=0] (image2) at (0,0) {\underlay{4in}{250pt}{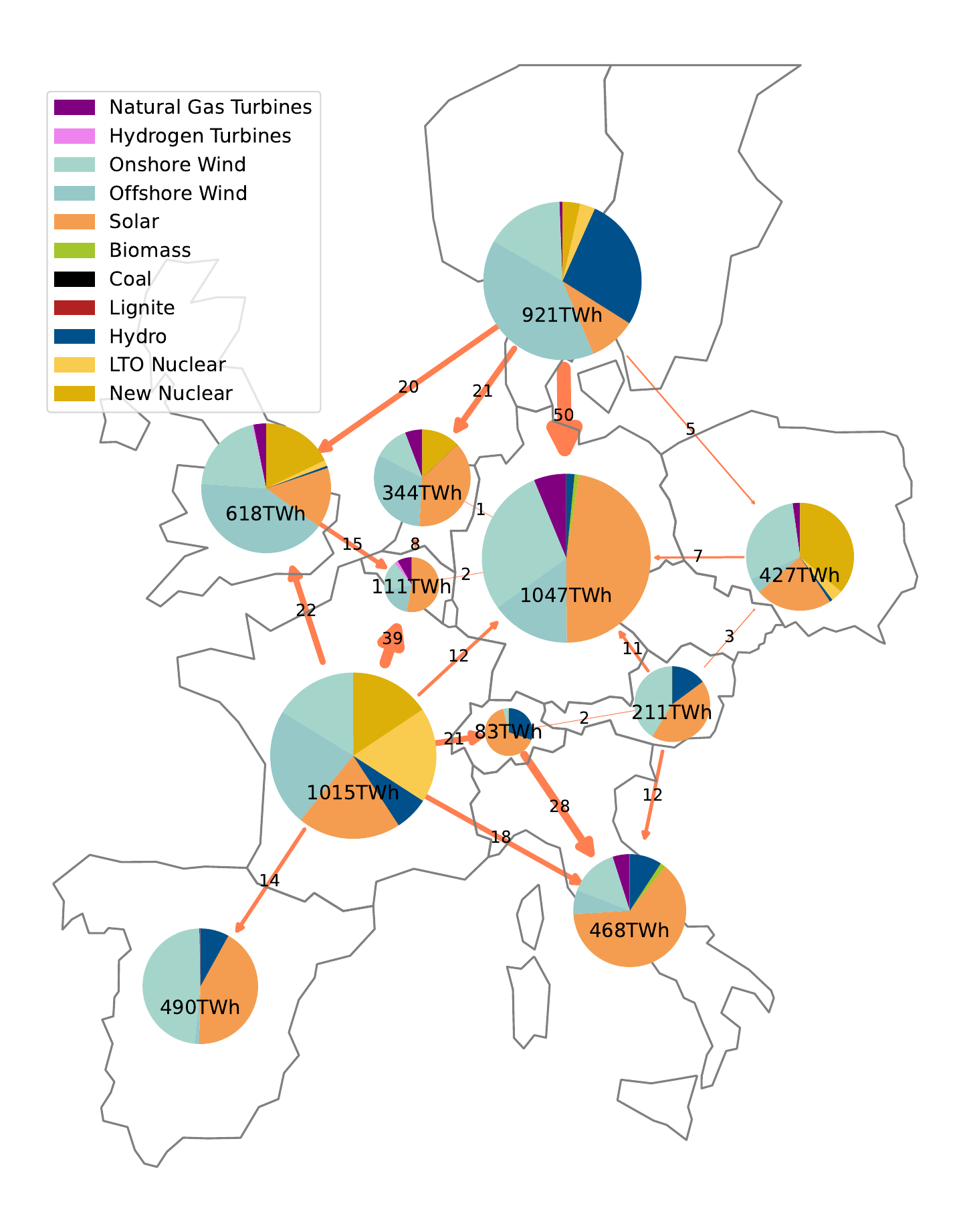}{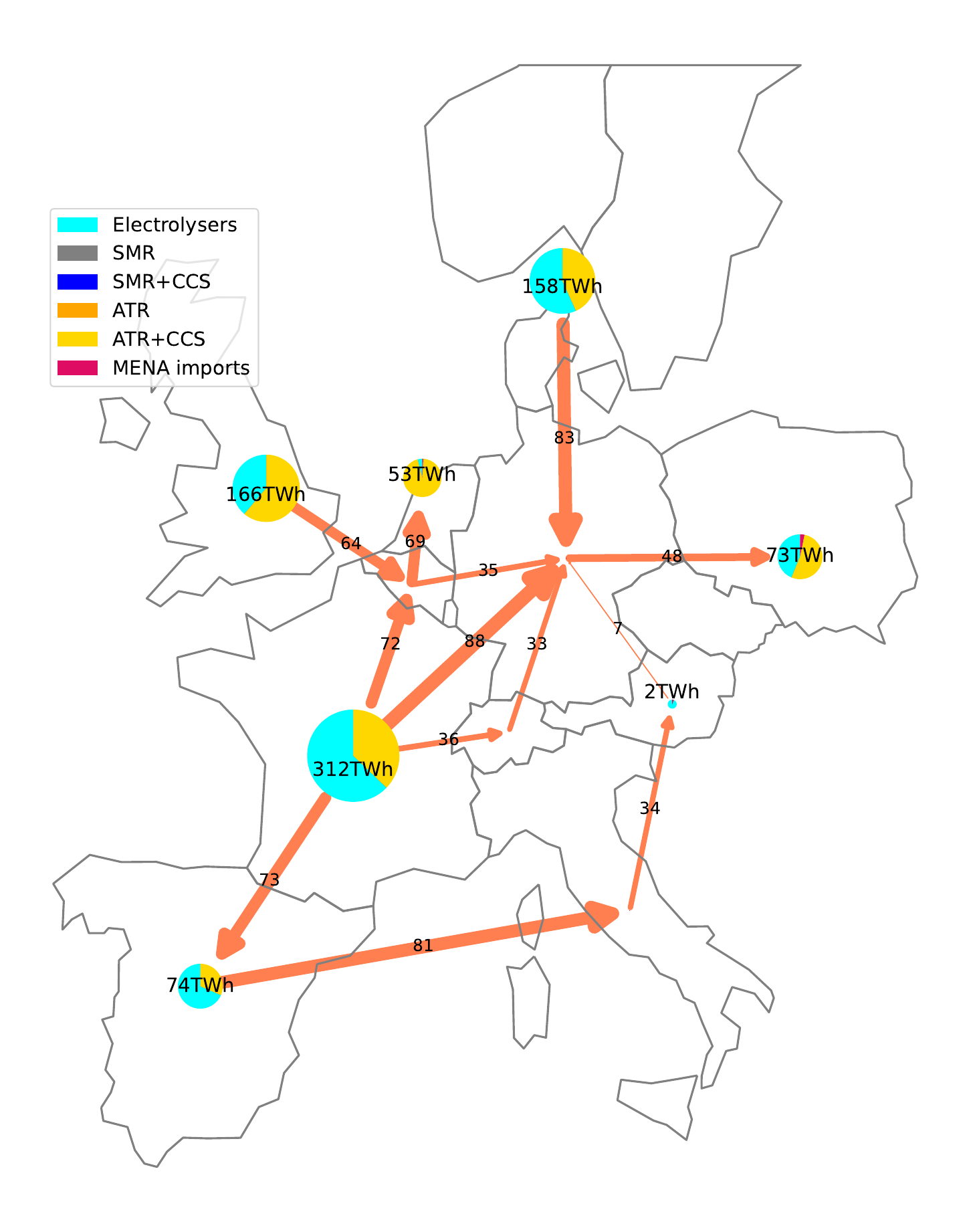}};
\end{tikzpicture}
\caption{Reference Central (a) and H\textsubscript{2}\_Exch\_Plus (b) scenarios electricity and hydrogen systems in 2050}
\label{fig:map_ref_h2_network}
\end{figure}

\subsection{Model coupling impact}\label{appendix:model_coupling_impact}
\begin{figure}[H]
\begin{tikzpicture}
\centering
\node[above left] (a) at (-8,5) {\Large (a)};
\node[above left,inner sep=0] (image) at (0,0) {\underlay{4in}{250pt}{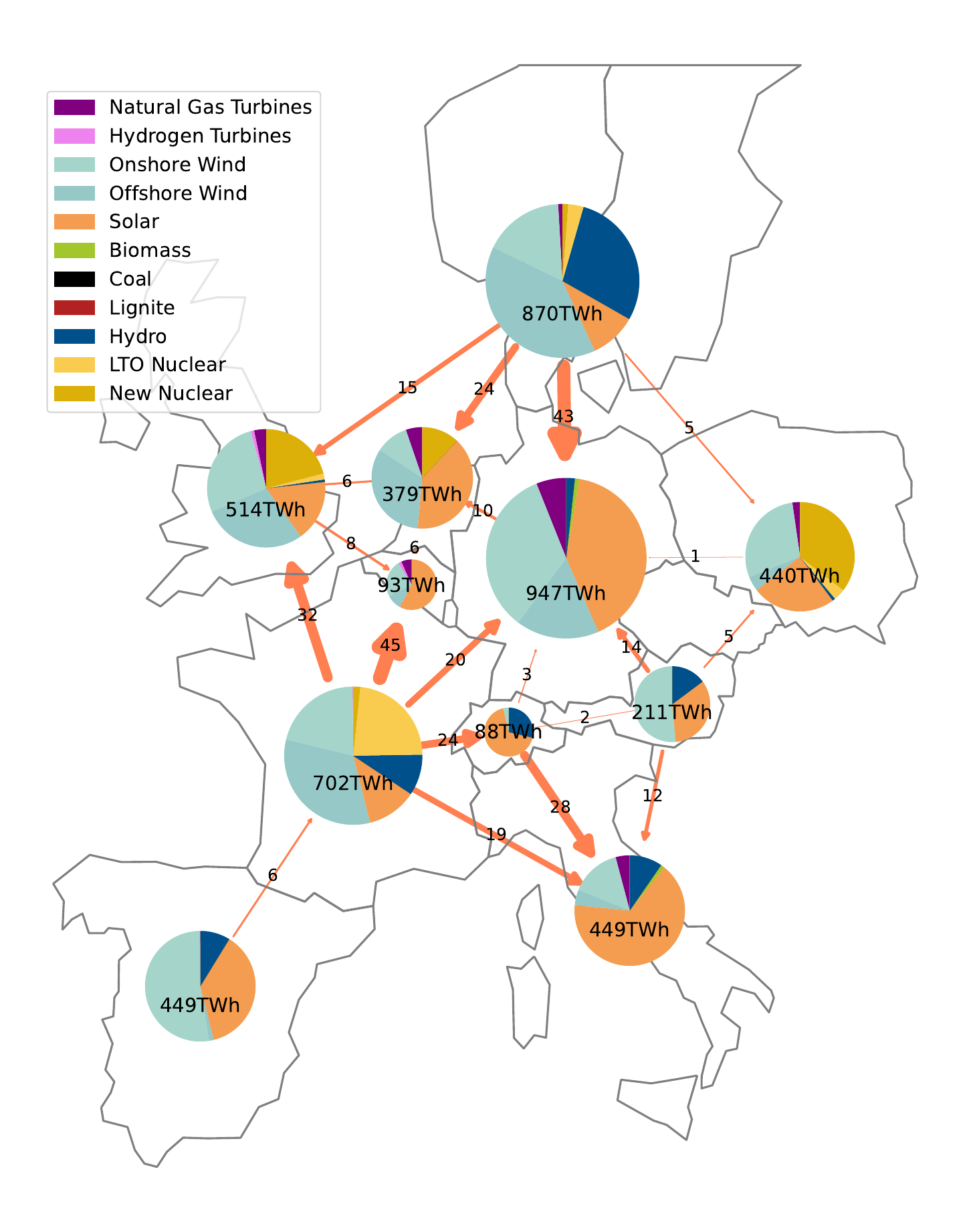}{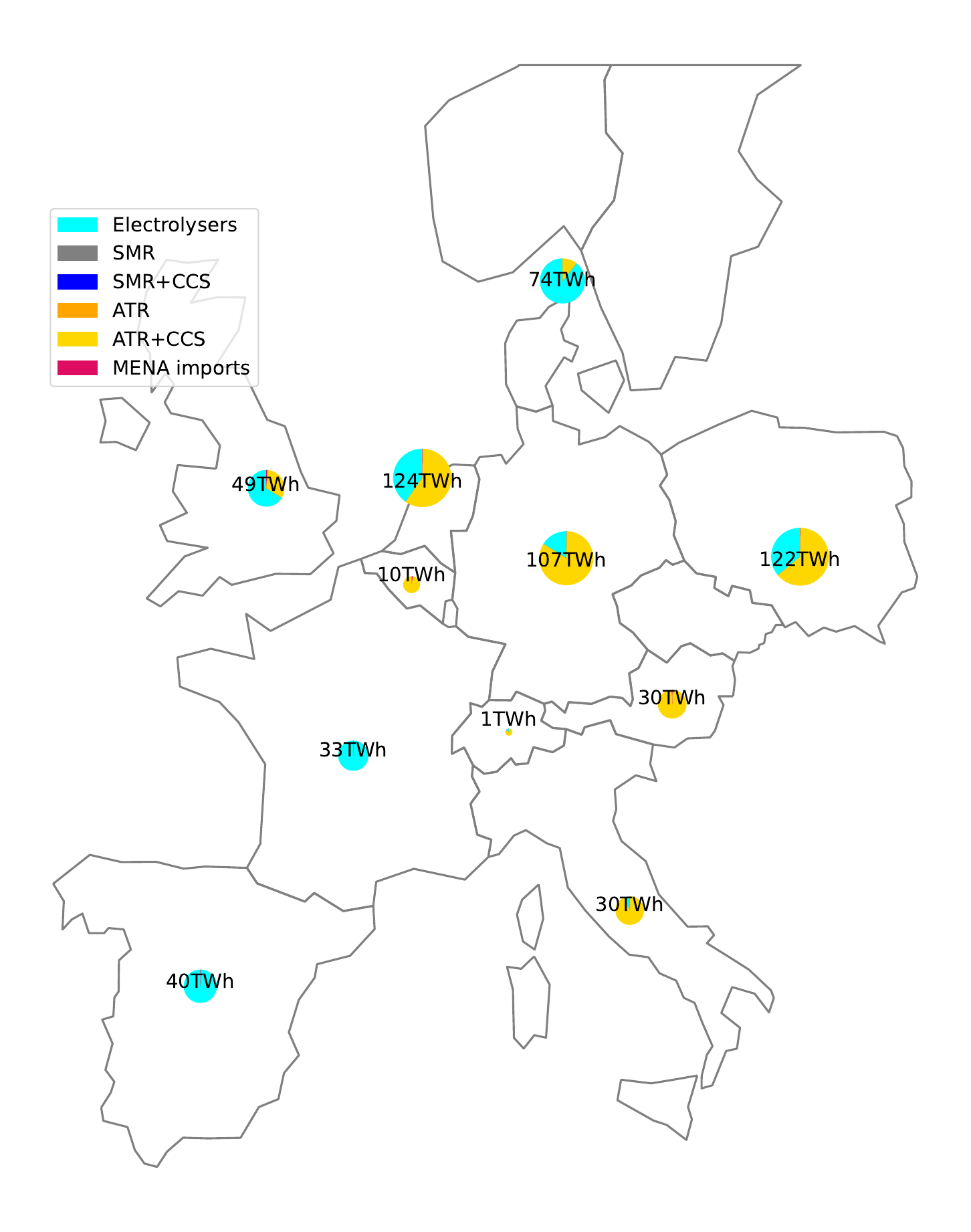}};
\node[below left] (b) at (-8,-5) {\Large (b)};
\node[below left,inner sep=0] (image2) at (0,0) {\underlay{4in}{250pt}{Figures/Maps/elec_map_reference_central}{Figures/Maps/h2_map_reference_central}};
\end{tikzpicture}
\caption{Reference Central scenario electricity and hydrogen systems in 2050 for POMMES standalone (a) and model coupling (b)}
\label{fig:map_coupling_vs_standalone}
\end{figure}

\begin{figure}[H]
    \centering
    \includegraphics[width=\textwidth]{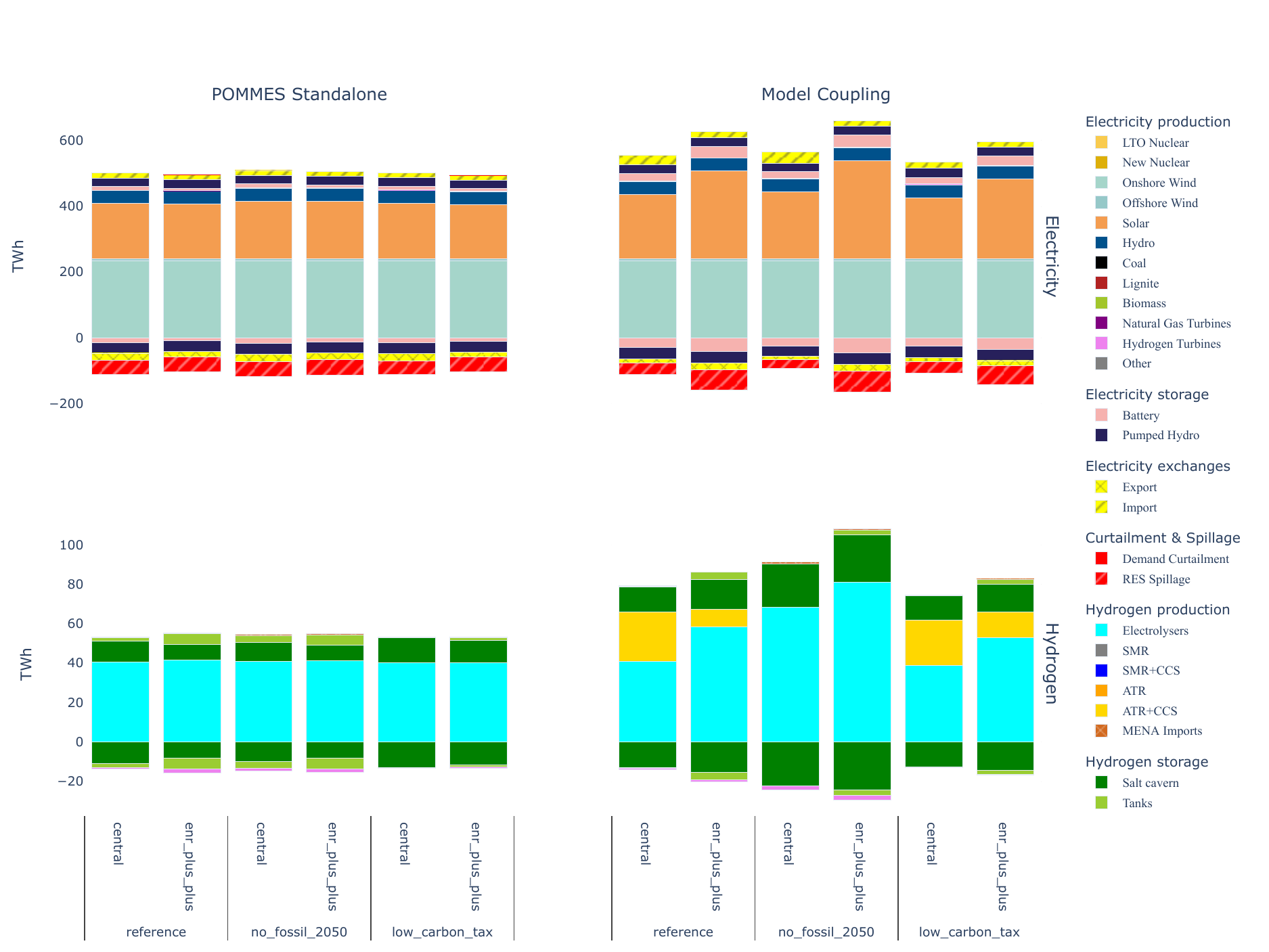}
    \caption{Electricity and hydrogen production in Spain | POMMES standalone vs model coupling comparison}\label{fig:comp_spain}
\end{figure}

\begin{figure}[H]
    \centering
    \includegraphics[width=\textwidth]{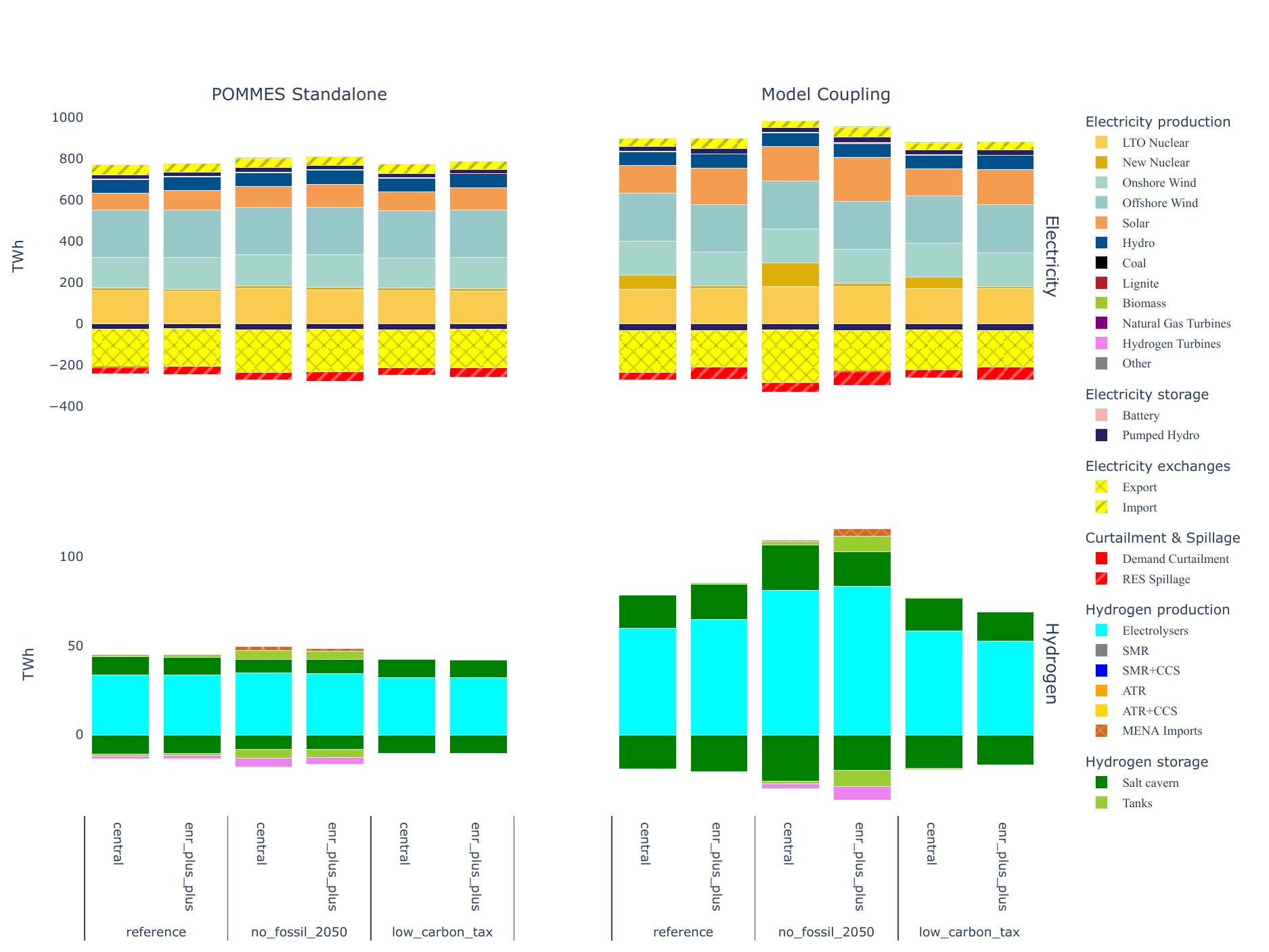}
    \caption{Electricity and hydrogen production in France | POMMES standalone vs model coupling comparison}\label{fig:comp_france}
\end{figure}

\section{Techno-economic data for POMMES}
\label{appendix:techno_economic_pommes}

\subsection{Prices and tax}
\begin{table}[H]
\centering
\begin{tabular}{lllll}
\hline
 & Policy scenario & 2030 & 2040 & 2050 \\ \hline
Carbon tax {[}€/tCO2{]} & Reference & 100 & 200 & 300 \\
 & No Fossil 2050 & 100 & 200 & 300 \\
 & Low Carbon Tax & 50 & 100 & 150 \\
Natural gas {[}€/MWh{]} &  & 35 & 35 & 35 \\
Hydrogen imports {[}€/MWh{]} &  & 170 & 140 & 110 \\ \hline
\end{tabular}
\caption{Carbon tax and import prices}
\label{tab:ctax_import_prices}
\end{table}

\begin{table}[H]
\centering
\begin{tabularx}{\textwidth}{llllllllllll}
\hline
Zone & FR & DE & ES & IT & BE & GB & CH & NL & AT & Nordic & PL\_CZ \\
Network tax (€/MWh) & 8.9 & 15.3 & 7 & 7.5 & 7.8 & 10.3 & 11.2 & 10.6 & 16.7 & 9.87 & 15.2 \\ \hline
\end{tabularx}
\caption{Electricity network tax per area}
\label{tab:network_cost}
\end{table}

\subsection{Electricity production}
The characteristics of electricity generation technologies are presented in \Cref{tab:tec-ec-elec}. The data for wind and solar are sourced from \citet{ENTSOE2024TYNDPReport}, while the data for nuclear, hydro, and gas and hydrogen power plants are drawn from \citet{RTE2021Futurs2050}. Information for coal, lignite, and biomass power plants is obtained from \citet{Kost2021LevelizedTechnologies}. 

Based on RTE BP2050, all WACCs have been established at 4\%, with the exception of new nuclear, which has been set at 8\%. These WACCs are the focus of sensitivity analyses in the study, particularly for the ENR\_Plus\_Plus and Nuke\_Plus scenarios.

\begin{table}[H]
\centering
\begin{tabularx}{\textwidth}{lcllllllllll}
\hline
 & \multicolumn{1}{l}{Direct emissions} &  & \multicolumn{5}{l}{Overnight Invesment Costs {[}€/kW{]}} & Fixed Costs &  & Lifetime &  \\ \cline{4-8}
Technologies & \multicolumn{1}{l}{{[}tCO\textsubscript{2}/MWh{]}} &  & 2030 &  & 2040 &  & 2050 & {[}€/kW/yr{]} &  & {[}yr{]} &  \\ \hline
Wind Onshore &  &  & 1040 &  & 990 &  & 970 & 1.2\% CAPEX &  & 30 &  \\
Wind Offshore &  &  & 1800 &  & 1650 &  & 1640 & 3\% CAPEX &  & 30 &  \\
Solar PV &  &  & 380 &  & 330 &  & 290 & 2.5\% CAPEX &  & 40 &  \\
Biomass & \multicolumn{1}{l}{0.150} &  & 3000 &  & 3000 &  & 3000 & 33 &  & 20 &  \\
LTO Nuclear &  &  & \multicolumn{1}{c}{} &  & \multicolumn{1}{c}{} &  & \multicolumn{1}{c}{} & 186 &  & 10 &  \\
New Nuclear &  &  & 11900 &  & 5035 &  & 4505 & 100 &  & 60 &  \\
Hydro River &  &  & \multicolumn{1}{c}{} &  & \multicolumn{1}{c}{} &  & \multicolumn{1}{c}{} & 121 &  & 70 &  \\
Hydro Lake &  &  & \multicolumn{1}{c}{} &  & \multicolumn{1}{c}{} &  & \multicolumn{1}{c}{} & 121 &  & 70 &  \\
Coal & \multicolumn{1}{l}{0.855} &  & 1500 &  & 1500 &  & 1500 & 22 &  & 30 &  \\
Lignite & \multicolumn{1}{l}{0.933} &  & 1800 &  & 1800 &  & 1800 & 22 &  & 30 &  \\
OCGT & \multicolumn{1}{l}{0.534} &  & 600 &  & 600 &  & 600 & 20 &  & 30 &  \\
CCGT & \multicolumn{1}{l}{0.356} &  & 900 &  & 900 &  & 900 & 40 &  & 40 &  \\
OCGT-H\textsubscript{2} &  &  & 800 &  & 800 &  & 800 & 20 &  & 30 &  \\
CCGT-H\textsubscript{2} &  &  & 1100 &  & 1100 &  & 1100 & 40 &  & 40 &  \\ \hline
\end{tabularx}
\caption{Economic characteristics in POMMES for electricity production technologies}
\label{tab:tec-ec-elec}
\end{table}

\subsection{Hydrogen production}
The modelled hydrogen production technologies characteristics are described \Cref{tab:tec-ec-h2} and \Cref{tab:tec-h2}. Steam Methane Reforming (SMR) and Autothermal Reforming (ATR) reactors data were adapted from \citet{Oni2022ComparativeRegions} and \citet{Raillard--Cazanove2024IndustryTrajectory}. Electrolyser costs were taken from \citet{RTE2021Futurs2050} with an efficiency and lifetime adapted from \citet{Brissaud2024LessonsPower-to-Gas}, assuming alkaline electrolysis. The initial electrolyser's efficiency (including balance of plant) is 54\%\cite{Brissaud2024LessonsPower-to-Gas} which we assume to increase linearly up to 65\% in 2050. 

Ramp-up/down characteristics are taken from \citet{Jodry2023IndustrialOptimisation} and electrolysis minimum power rate is as reported by \citet{Brissaud2024LessonsPower-to-Gas}. Regarding SMR/ATR, minimum power rates assumptions were made following discussions with industry stakeholders such as Air Liquide. 

\begin{table}[H]
\centering
\begin{tabularx}{\textwidth}{lllllllll}
\hline
 & Direct emissions & \multicolumn{5}{l}{Overnight Invesment Costs {[}€/kW{]}} & Fixed Costs & Lifetime \\ \cline{3-7}
Technologies & {[}tCO\textsubscript{2}/MWh{]} & 2030 &  & 2040 &  & 2050 & {[}€/kW/yr{]} & {[}yr{]} \\ \hline
Electrolysis &  & 641 &  & 574 &  & 507 & 12 & 10 \\
SMR & 0.28 & 850 &  & 850 &  & 850 & 144 & 25 \\
SMR + partial CCS & 0.169 & 950 &  & 950 &  & 950 & 199 & 25 \\
SMR + CCS & 0.059 & 1150 &  & 1150 &  & 1150 & 302 & 25 \\
ATR & 0.25 & 1300 &  & 1300 &  & 1300 & 89 & 25 \\
ATR + CCS & 0.023 & 1600 &  & 1600 &  & 1600 & 108 & 25 \\ \hline
\end{tabularx}
\caption{Economic characteristics in POMMES for hydrogen production technologies}
\label{tab:tec-ec-h2}
\end{table}

Although some prospective studies model the ATR technology \cite{Seck2022HydrogenAnalysis}, it is still often not represented \cite{Manuel2022HighDecarbonisation,Neumann2023TheEurope,Fleiter2023METISSystem.,Brown2018SynergiesSystem,Jodry2023IndustrialOptimisation}. Autothermal reforming of methane is a variant of SMR in which the gas is burnt directly in the reformer to provide heat. Oxygen is injected into the reformer to partially oxidise the CH\textsubscript{4}, leading to greater reaction energy efficiency \cite{Oni2022ComparativeRegions}. Even though ATR is essentially more expensive than SMR (see \Cref{tab:tec-ec-h2}), thanks to the partial oxidation of methane, syngas and flue gases are not mixed with N\textsubscript{2}. So capturing CO\textsubscript{2} is easier (and cheaper) with ATR than with SMR \cite{KhojastehSalkuyeh2017Techno-economicTechnologies}. One key characteristic is the substantially lower energy consumption for ATR+CCS compared to SMR+CCS as depicted \Cref{tab:tec-h2}.

\begin{table}[H]
\centering
\begin{tabularx}{\textwidth}{llllllll}
\hline
 &  & \multicolumn{4}{l}{Consumption   {[}kWh/kWh\textsubscript{H\textsubscript{2}}{]}} & Ramp up/down & Minimum power rate \\ \cline{3-6}
Technologies & Year & Electricity &  & Methane &  & {[}hr{]} & {[}\%{]} \\ \hline
Electrolysis & 2030 & 1.73 &  &  &  & \textless{}1 & 20 \\
 & 2040 & 1.63 &  &  &  & \textless{}1 & 20 \\
 & 2050 & 1.53 &  &  &  & \textless{}1 & 20 \\
SMR &  & 0.017 &  & 1.31 &  & 3.5 & 15 \\
SMR + partial CCS &  & 0.028 &  & 1.34 &  & 3.5 & 50 \\
SMR + CCS &  & 0.121 &  & 1.37 &  & 3.5 & 50 \\
ATR &  & 0.071 &  & 1.25 &  & 3.5 & 15 \\
ATR + CCS &  & 0.108 &  & 1.25 &  & 3.5 & 50 \\ \hline
\end{tabularx}

\caption{Technical characteristics in POMMES for hydrogen production technologies}
\label{tab:tec-h2}
\end{table}

\subsection{Storage}
The storage technologies characteristics used in POMMES are described \Cref{tab:tec-ec-stor}. Pumped Hydro Storage (PHS) data was taken from \citet{Schmidt2019ProjectingTechnologies} with 80\% round trip efficiency (RTE). Battery costs were derived from \citet{Cole2023CostUpdate} and \citet{RTE2021Futurs2050} and a 0.04\%/hr self-discharge was assumed for a 85\% RTE\cite{Cole2023CostUpdate}. As for salt cavern and hydrogen tanks, the data was taken from \citet{Jodry2023IndustrialOptimisation} with a 98\% RTE.

\nomenclature[A]{PHS}{Pumped Hydro Storage}
\nomenclature[A]{RTE}{Round Trip Efficiency}

\begin{table}[H]
\centering
\begin{tabularx}{\textwidth}{llllllll}
\hline
 &  &  & \multicolumn{2}{l}{Overnight Invesment Costs} & Fixed Costs & Lifetime \\ \cline{4-6}
Technologies & Resource & Year & Power {[}€/kW{]} & Volume {[}€/kWh{]} & Power {[}€/kW/yr{]} & {[}yr{]} \\ \hline
Pumped Hydro Storage & electricity &  & 1130 & 80 & 8 & 50 \\
Battery & electricity & 2030 & 315 & 240 & 30 & 15 \\
 &  & 2040 & 300 & 200 & 30 & 15 \\
 &  & 2050 & 285 & 155 & 30 & 15 \\
Salt Cavern & hydrogen &  & 545 & 0.28 & 2 & 40 \\
Tank & hydrogen &  & 12.6 & 5.4 & 2 & 20 \\ \hline
\end{tabularx}
\caption{Economic characteristics in POMMES for storage technologies}
\label{tab:tec-ec-stor}
\end{table}




\bibliographystyle{elsarticle-num-names} 
\bibliography{references}

\end{document}


\maketitle

\renewcommand\nomgroup[1]{%
   \item[\bfseries
   \ifstrequal{#1}{A}{Indices and index sets}{%
   \ifstrequal{#1}{B}{General parameters}{%
   \ifstrequal{#1}{C}{Conversion parameters}{%
   \ifstrequal{#1}{D}{Storage parameters}{%
   \ifstrequal{#1}{E}{Transport parameters}{%
   \ifstrequal{#1}{F}{Exchange with ROW parameters}{%
   \ifstrequal{#1}{G}{Repurposing parameters}{%
   \ifstrequal{#1}{H}{Investment variables}{%
   \ifstrequal{#1}{I}{Operation Variables}{%
   \ifstrequal{#1}{J}{Intermediate cost variables}{%
  }}}}}}}}}}%
]}


\nomenclature[A, 01]{\(a \in \mathcal{A}\)}{areas}

\nomenclature[A, 02]{\(i \in \mathcal{Y}^{inv}\)}{investment years}
\nomenclature[A, 03]{\(y \in \mathcal{Y}^{op}\)}{operation years}
\nomenclature[A, 04]{\(y \in \mathcal{Y}^{dec}\)}{decommissioning years}
\nomenclature[A, 06]{\(t \in \mathcal{T}\)}{operation snapshots}
\nomenclature[A, 07]{\(r \in \mathcal{R}\)}{resources}
\nomenclature[A, 08]{\(c \in \mathcal{C}^{tech}\)}{conversion technologies}
\nomenclature[A, 09]{\(s \in \mathcal{S}^{tech}\)}{storage technologies}
\nomenclature[A, 10]{\(w \in \mathcal{W}^{tech}\)}{transport technologies}

\nomenclature[A, 11]{\(\mathcal{R}_{a}\)}{resources considered in area \(a\)}

\nomenclature[A, 12]{\(\mathcal{C}^{tech}_{a, i}\)}{conversion technologies which can be installed in area \(a\) for investment year \(i\)}
\nomenclature[A, 12]{\(\mathcal{S}^{tech}_{a, i}\)}{storage technologies which can be installed in area \(a\) for investment year \(i\)}
\nomenclature[A, 13]{\(\mathcal{W}^{tech}_{\afrom, \ato, i}\)}{transport technologies which can be installed from area \(\afrom\) to  area \(\ato\) for investment year \(i\)}
\nomenclature[A, 14]{\(\mathcal{W}^{tech}_{a, i}\)}{transport technologies which connect area \(a\), \(\mathcal{W}^{tech}_{a, i} = \bigcup_{b \in \mathcal{A}} \left( \mathcal{W}^{tech}_{b, a, i} \cup \mathcal{W}^{tech}_{a, b, i} \right) \)}




\nomenclature[B, 01]{\(d_{r, t, y}\)}{exogenous demand of \(r\) in \(y\) at \(t\) [MWh]}
\nomenclature[B, 02]{\(\lambda^{shed}_{r, y}\)}{load shedding cost [\euro/MWh]}
\nomenclature[B, 03]{\(\lambda^{spill}_{r, y}\)}{resource spillage cost [\euro/MWh]}
\nomenclature[B, 04]{\(g^{CO_2}_{a, y}\)}{maximum total \ce{CO2} emission in \(y\) [kg\ce{CO2}]}
\nomenclature[B, 05]{\(t^{CO_2}_{a, y}\)}{carbon tax value in \(y\) [\euro{}/kg\ce{CO2}]}

\nomenclature[B, 06]{\(\tau\)}{discount rate [\(\%\)]}

\nomenclature[B, 10]{\(y_0\)}{reference year for actualisation [year]}
\nomenclature[B, 11]{\(\phi_1(\tau,y, y_0)\)}{discount factor to bring back values from year \(y\) to actualised value in year \(y_0\) [\(\%\)]}
\nomenclature[B, 12]{\(\phi_2(\alpha, n)\)}{capital recovery factor [\(\%\)]}

\nomenclature[B, 13]{\(\Delta i = \Delta j = \Delta y\)}{investment/decommissioning/operation year time step [year]}
\nomenclature[B, 14]{\((\Delta T)_{t}\)}{snapshot length [hour]}

\nomenclature[C, 00]{\(\alpha_{c}\)}{conversion technologies finance rate [\(\%\)]}
\nomenclature[C, 01]{\(\beta_{c}\)}{specific overnight cost of \(c\) [\euro{}/MW]}
\nomenclature[C, 02]{\(\omega_{c}\)}{fixed OPEX of \(c\) [\euro{}/MW/yr]}
\nomenclature[C, 03]{\(\lambda_{c}\)}{variable OPEX of \(c\) [\euro{}/MWh]}
\nomenclature[C, 04]{\(a_{y, t, c}\)}{availability of \(c\) in operation year \(y\) at \(t\) [\(\in [0, 1]\)]}
\nomenclature[C, 05]{\(k_{c, r}\)}{conversion factor of \(c\), \(k_{c, r}\leq 0\; (\text{resp}. \geq 0)\) for consumption (resp. emission) of \(r\) [\%]}
\nomenclature[C, 06]{\(\varepsilon^{CO_2}_{c}\)}{\ce{CO2} emission rate for \(c\) invested for year \(i\) [kg\ce{CO2}/MWh]}
\nomenclature[C, 07]{\(l_{c}\)}{life length of \(c\) invested for year \(i\) [year]}
\nomenclature[C, 08]{\(p^{c, min}_{c}\)}{minimum capacity that must be invested in \(c\) [MW]}
\nomenclature[C, 09]{\(p^{c, max}_{c}\)}{maximum capacity that can be invested in \(c\) [MW]}


\nomenclature[D, 00]{\(\alpha_{s}\)}{storage technologies finance rate [\(\%\)]}
\nomenclature[D, 01]{\(\beta_{s}\)}{power capacity specific overnight cost of technology \(s\) [\euro{}/MW]}
\nomenclature[D, 02]{\(\sigma_{s}\)}{energy capacity specific overnight cost of technology \(s\) [\euro{}/MWh]}
\nomenclature[D, 03]{\(\omega_{s}\)}{fixed OPEX of technology \(s\) [\euro{}/MW/yr]}
\nomenclature[D, 04]{\(\eta_{s}\)}{Dissipation losses in \(s\) over 1 hour [\%]}

\nomenclature[D, 05]{\(k^{in}_{s, r}\)}{Charging factor to load 1 MWh into \(s\), \(k^{in}_{s, r} \leq 0\; (\text{resp}. \geq 0)\) for consumption (resp. emission) of \(r\) [\%]}
\nomenclature[D, 06]{\(k^{keep}_{s, r}\)}{Consumption factor to keep 1 MWh during 1 hour in \(s\) (not taking into account dissipation), \(k^{keep}_{s, r} \leq 0\; (\text{resp}. \geq 0)\) for consumption (resp. collecting or emission) of \(r\) [h\(^{-1}\)]}
\nomenclature[D, 06]{\(k^{out}_{s, r}\)}{Discharging factor, \(k^{out}_{s, r} \leq 0\; (\text{resp}. \geq 0)\) for consumption (resp. collecting or emission) of \(r\) [\%]}
\nomenclature[D, 07]{\(res_{s}\)}{main resource stored by \(s\)}
\nomenclature[D, 08]{\(l_{s}\)}{life length of \(s\) [year]}

\nomenclature[D, 09]{\(p^{s, min}_{s}\)}{minimum storage power capacity that must be invested in \(s\) [MW]}
\nomenclature[D, 10]{\(p^{s, max}_{s}\)}{maximum storage power capacity that can be invested in \(s\) [MW]}
\nomenclature[D, 11]{\(s^{s, min}_{s}\)}{minimum storage energy capacity that must be invested in \(s\) [MWh]}
\nomenclature[D, 12]{\(s^{s, max}_{s}\)}{maximum storage energy capacity that can be invested in \(s\) [MWh]}


\nomenclature[E, 01]{\(\beta_{w}\)}{specific overnight cost of \(w\) [\euro{}/MW]}
\nomenclature[E, 02]{\(\omega_{w}\)}{fixed OPEX of \(w\) [\euro{}/MW/yr]}
\nomenclature[E, 03]{\(\lambda_{w}\)}{variable OPEX of \(w\) [\euro{}/MWh]}
\nomenclature[E, 04]{\(a_{y, t, w}\)}{availability of \(w\) in operation year \(y\) at \(t\) [\(\in [0, 1]\)]}
\nomenclature[E, 05]{\(k_{w, r}\)}{conversion factor of \(w\), \(k_{w, r}\leq 0\; (\text{resp}. \geq 0)\) for consumption (resp. emission) of \(r\) [\%]}
\nomenclature[E, 06]{\(\varepsilon^{CO_2}_{w}\)}{\ce{CO2} emission rate for \(w\) [kg\ce{CO2}/MWh]}
\nomenclature[E, 07]{\(l_{w}\)}{life length of \(w\) [year]}
\nomenclature[E, 08]{\(p^{w, min}_{w}\)}{minimum capacity that must be invested in \(w\) [MW]}
\nomenclature[E, 09]{\(p^{w, max}_{w}\)}{maximum capacity that can be invested in  \(w\) [MW]}


\nomenclature[F, 01]{\(\gamma_{y, r, t}\)}{importation cost of \(r\) in \(y\) at \(t\) from ROW [\euro{}/MWh]}
\nomenclature[F, 01]{\(\varepsilon^{CO_2}_{y, r, t}\)}{importation emission rate of \(r\) in \(y\) at \(t\) from ROW [kg\ce/MWh]}
\nomenclature[F, 02]{\(\pi^{max, imp}_{y, r, t}\)}{max importation volume of \(r\) in \(y\) at \(t\) from ROW [MWh]}
\nomenclature[F, 03]{\(\pi^{max, exp}_{y, r, t}\)}{max exportation volume of \(r\) in \(y\) at \(t\)to ROW [MWh]}


\nomenclature[H, 01]{\(\mathbf P^{c, inv}_{c, d}\)}{capacity of \(c \in \mathcal{C}^{tech}_{a, i}\)  decommissioned for year \(d\) [MW]}

\nomenclature[H, 03]{\(\mathbf P^{s, inv}_{s, d}\)}{storage power capacity of \(s \in \mathcal{S}^{tech}_{a, i}\) decommissioned for year \(d\) [MW]}
\nomenclature[H, 04]{\(\mathbf S^{inv}_{s, d}\)}{storage energy capacity of \(s \in \mathcal{S}^{tech}_{a, i}\) decommissioned for year \(d\) [MWh]}
\nomenclature[H, 04]{\(\mathbf P^{w, inv}_{w, d}\)}{capacity of \(w \in \mathcal{W}^{tech}_{\afrom, \ato, i}\) decommissioned for year \(d\) [MWh]}

\nomenclature[H, 05]{\(\bar P^{c}_{y, c}\)}{total capacity of \(c\) in operation in year \(y\) [MW]}
\nomenclature[H, 06]{\(\bar P^{s}_{y, s}\)}{total storage power capacity of \(s\) in operation in year \(y\) [MW]}
\nomenclature[H, 07]{\(\bar S_{y, s}\)}{total storage energy capacity of \(s\) in operation in year \(y\) [MWh]}
\nomenclature[H, 08]{\(\bar P^{w}_{y, w}\)}{total capacity of \(w\) invested in operation in year \(y\) [MW]}


\nomenclature[I, 01]{\(\mathbf P^{c}_{y, t}\)}{power of \(c\) in \(y\) at \(t\) [MW]}
\nomenclature[I, 02]{\(U^{net, c}_{y, t, r}\)}{resource \(r\) net generation from conversion technologies in \(y\) at \(t\) [MWh]}

\nomenclature[I, 03]{\(\mathbf P^{s, in}_{y, t, s}\)}{storage charging power of \(s\) in \(y\) at \(t\) [MW]}
\nomenclature[I, 04]{\(\mathbf P^{s, out}_{y, t, s}\)}{storage discharging power of \(s\) in \(y\) at \(t\) [MW]}
\nomenclature[I, 05]{\(\mathbf S_{y, t, s}\)}{amount of energy in \(s\) in \(y\) at \(t\) [MWh]}
\nomenclature[I, 06]{\(U^{net, s}_{y, t, r}\)}{resource \(r\) net generation from storage technologies in \(y\) at \(t\) [MWh]}

\nomenclature[I, 07]{\(\mathbf P^{w}_{y, t}\)}{power of \(w\) in \(y\) at \(t\) [MW]}
\nomenclature[I, 08]{\(U^{net, w}_{y, t, r}\)}{resource \(r\) net generation from transport technologies in \(y\) at \(t\) [MWh]}

\nomenclature[I, 09]{\(\mathbf U^{shed}_{y, t, r}\)}{load shedding for resource \(r\) in\(y\) at \(t\) [MWh]}
\nomenclature[I, 10]{\(U^{spill}_{y, t, r}\)}{resource spillage for \(r\) in \(y\) at \(t\) [MWh]}

\nomenclature[I, 11]{\(\mathbf I_{y, t, r}\)}{importation of \(r\) in \(y\) at \(t\) from ROW [MWh]}
\nomenclature[I, 12]{\(\mathbf E_{y, t, r}\)}{exportation of \(r\) in \(y\) at \(t\) to ROW [MWh]}
\nomenclature[I, 13]{\(I^{net}_{y, t, r}\)}{net imports from ROW [MWh]}


\nomenclature[J, 01]{\(\mathcal{CAP}_{a, y}\)}{annualised cost of capital of the system in \(y\)}
\nomenclature[J, 02]{\(\mathcal{FIX}_{a, y}\)}{fixed operation costs of the system in \(y\)}
\nomenclature[J, 03]{\(\mathcal{VAR}_{a, y}\)}{variable (proportional) operation costs of the system in \(y\)}

\nomenclature[J, 04]{\(\mathcal A^{var}_{a, y}\)}{adequacy (load shedding and spillage) costs in \(y\)}

\nomenclature[J, 05]{\(\mathcal C^{cap}_{a, y}\)}{conversion annualised capital cost in \(y\)}
\nomenclature[J, 07]{\(\mathcal C^{fix}_{a, y}\)}{conversion fixed operation cost in \(y\)}
\nomenclature[J, 08]{\(\mathcal C^{var}_{a, y}\)}{conversion variable operation cost in \(y\)}

\nomenclature[J, 09]{\(\mathcal S^{cap}_{a, y}\)}{storage annualised capital cost in \(y\)}
\nomenclature[J, 10]{\(\mathcal S^{fix}_{a, y}\)}{storage fixed operation cost in \(y\)}

\nomenclature[J, 11]{\(\mathcal W^{cap}_{a, y}\)}{transport annualised capital cost in \(y\)}
\nomenclature[J, 12]{\(\mathcal W^{fix}_{a, y}\)}{transport fixed operation cost in \(y\)}
\nomenclature[J, 13]{\(\mathcal W^{var}_{a, y}\)}{transport variable operation cost in \(y\)}

\nomenclature[J, 14]{\(\mathcal I^{net, var}_{a, y}\)}{Net imports from ROW variable costs in \(y\)}

\printnomenclature

\section{POMMES equations}

\subsection{Objective}

The objective function of the problem is given by the minimisation of the actualised costs in equations \eqref{eq:objective}. All variables are continuous and positive.

\begin{equation}
\label{eq:objective}
    \min_{
        \mathbf P^{inv},     
        \mathbf P^,   
        \mathbf S^{inv},        
        \mathbf S,
        \mathbf I,
        \mathbf E,
        \mathbf U
    }
        \sum_{a, y}
            \left[
                \phi_1 \left(
                    \tau,y + \frac{\Delta y}{2}, y_0
                \right)
                \left(
                    \mathcal{CAP}_{a, y} + \mathcal{FIX}_{a, y} + \mathcal{VAR}_{a, y}
                \right)
            \right] 
\end{equation}

Where \(\phi_1(\tau, y, y_0)\) is the discount factor to actualise all the costs to reference year \(y_0\) \eqref{eq:discount_factor_def}.

\begin{equation}
    \label{eq:discount_factor_def}
    \phi_1(\tau, y, y_0) = (1 + \tau)^{-(y - y_0)}
\end{equation}




\subsection{Costs definition}

\subsubsection{Annualised capital costs}

Annualised capital costs are given in equation \eqref{eq:capex}, and details are given in \eqref{eq:capex_conv}, \eqref{eq:capex_storage}, and \eqref{eq:capex_transport}.

\begin{equation}
        \label{eq:capex}
        \mathcal{CAP}_{a, y} = 
            \mathcal C^{cap}_{a, y} +
            \mathcal S^{cap}_{a, y} +
            \mathcal W^{cap}_{a, y}
    \quad \forall \, a, y
\end{equation}

Where

\begin{equation}
    \label{eq:capex_conv}
    \mathcal C^{cap}_{a, y} =
        \sum_{i \leq y}
        \sum_{c \in \mathcal{C}^{tech}_{a, i}}
            \phi_2(\alpha_{c}, l_c) \,
            \beta_{c} \,
            \sum_{d}
                \mathbf P^{c, inv}_{c, d}
    \quad \forall \, a, y
\end{equation}

\begin{equation}
    \label{eq:capex_storage}
    \mathcal S^{cap}_{a, y} =
        \sum_{i \leq y} 
        \sum_{s \in \mathcal{S}^{tech}_{a, i}} 
            \phi_2(\alpha_{s}, l_s)
            \left(
                \beta_{s}\, 
                    \sum_{d}
                        \mathbf P^{s, inv}_{s, d} +
                \sigma_{s}\, 
                    \sum_{d}
                        \mathbf S^{inv}_{s, d}
            \right)
    \quad \forall \, a, y
\end{equation}

\begin{equation}
    \label{eq:capex_transport}
    \mathcal W^{cap}_{a, y} =
    \frac{1}{2}
        \sum_{i \leq y} 
        \sum_{w \in \mathcal{W}^{tech}_{a, i}} 
            \phi_2(\alpha_{w}, l_w) \,
            \beta_{w} \,
                \sum_{d}
                    \mathbf P^{w, inv}_{w, d}
    \quad \forall \, a, y
\end{equation}

The coefficient \(\phi_2(\alpha, n)\) represents the capital recovery factor in annualising the costs with a finance rate of \(\alpha\) during \(n\) years with one term per year (equation \eqref{eq:crf_def}). Payments occur at the end of the year.

\begin{equation}
    \label{eq:crf_def}
    \phi_2(\alpha, n) = 
        \frac{\alpha}{1 - (1 + \alpha)^{-n}}
\end{equation}

The annualisation of the CAPEX is done on the whole life length of the technologies, even if they can be early decommissioned.

\subsubsection{Fixed operation costs}

Fixed operation costs are given in equation \eqref{eq:fixed_opex} and details are provided in \eqref{eq:fixed_costs_conv}, \eqref{eq:fixed_costs_storage} and \eqref{eq:fixed_costs_transport}.

\begin{equation}
    \label{eq:fixed_opex}
    \mathcal{FIX}_{a, y} =
        \mathcal C^{fix}_{a, y} +
        \mathcal S^{fix}_{a, y} +
        \mathcal W^{fix}_{a, y}
    \quad \forall \, a, y
\end{equation}

Where

\begin{equation}
    \label{eq:fixed_costs_conv}
    \mathcal C^{fix}_{a, y} =
         \sum_{i \leq y}
         \sum_{c \in \mathcal{C}^{tech}_{a, i}}
            \omega_{c} \bar P^{c}_{y, c}
    \quad \forall \, a, y
\end{equation}

\begin{equation}
    \label{eq:fixed_costs_storage}
    \mathcal S^{fix}_{a, y} =
         \sum_{i \leq y}
         \sum_{s \in \mathcal{S}^{tech}_{a, i}}
            \omega_{s} \bar P^{s}_{y, s}
    \quad \forall \, a, y
\end{equation}

\begin{equation}
    \label{eq:fixed_costs_transport}
     \mathcal W^{fix}_{a, y} =
     \frac{1}{2}
         \sum_{i \leq y}
         \sum_{w \in \mathcal{W}^{tech}_{a, i}}
            \omega_{c} \bar P^{c}_{y, c, i}
    \quad \forall \, a ,y
\end{equation}

Moreover, storage fixed costs are only proportional to installed power capacity, not energy capacity.

\subsubsection{Variable operation costs}

Variable operation costs are given in equation \eqref{eq:var_opex} and details are given in \eqref{eq:var_opex_conv}, \eqref{eq:var_opex_net_imp}, \eqref{eq:var_opex_transport} and \eqref{eq:var_opex_adequacy}.

\begin{equation}
    \label{eq:var_opex}
    \mathcal{VAR}_{a, y} =
        \mathcal C^{var}_{a, y}
        + \mathcal I^{net, var}_{a, y}
        + \mathcal W^{var}_{a, y}
        + \mathcal A^{var}_{a, y}
    \quad \forall \, a, y
\end{equation}

Where

\begin{equation}
    \label{eq:var_opex_conv}
    \mathcal C^{var}_{a, y} =
        \sum_{i \leq y, t}
        \sum_{c \in \mathcal{C}^{tech}_{a, i}}
            (\Delta T)_{t} \times
                \left(
                    t^{CO_2}_{a, y} \varepsilon^{CO_2}_{c} + \lambda_{c}
                \right)
                \mathbf P^{c}_{y, t, c}
    \quad \forall \, a, y
\end{equation}

\begin{equation}
    \label{eq:var_opex_net_imp}
    \mathcal I^{net, var}_{a, y} =
        \sum_{t, r}
            \left(
                t^{CO_2}_{a, y} \varepsilon^{CO_2}_{y, r, t} + \gamma_{y, r, t}
            \right)
            I^{net}_{y, r, t}
    \quad \forall \, a, y
\end{equation}

\begin{equation}
    \label{eq:var_opex_transport}
    \mathcal W^{var}_{a, y} =
        \frac{1}{2}
        \sum_{i \leq y, t}
        \sum_{w \in \mathcal{W}^{tech}_{a, i}}
            (\Delta T)_{t} \times
                \left(
                    t^{CO_2}_{a, y} \varepsilon^{CO_2}_{w} + \lambda_{w}
                \right)
                \mathbf P^{w}_{y, t, w}
    \quad \forall \, a, y
\end{equation}

The model has no proportional power capacity or energy capacity costs for storage.

\begin{equation}
    \label{eq:var_opex_adequacy}
    \mathcal{A}^{var}_{a, y} = 
        \sum_{t, r \in \mathcal{R}_{a}}
            \left(
                \lambda^{shed}_{r, y}  \, \mathbf U^{shed}_{y, t, r}
                + \lambda^{spill}_{r, y}  \, U^{spill}_{y, t, r}
            \right)
    \quad \forall \, a, y
\end{equation}

\subsection{Adequacy constraint}

The adequacy is met for each resource at each operation time \eqref{eq:adequacy}. The net generation of the conversion (resp. storage) technologies are aggregated for each time step in the \(U^{net, c}_{y, t, r}\) (resp. \(U^{net, s}_{y, t, r}\)) variable defined in equation \eqref{eq:conv_net_gen_def} (resp. \eqref{eq:storage_net_gen_def}).

\begin{equation}
    \label{eq:adequacy}
    \boxed{
        d_{r, t, y}
        + U^{spill}_{y, t, r} =
            U^{net, c}_{y, t, r}
            + U^{net, s}_{y, t, r}
            + U^{net, w}_{y, t, r}
            + I^{net}_{y, t, r}
            + \mathbf U^{shed}_{y, t, r}
        \quad \forall \, a, y, t, \quad \forall\, r \in \mathcal{R}_{a}
    }
\end{equation}

\begin{equation}
    \label{eq:conv_net_gen_def}
    U^{net, c}_{y, t, r} = 
        (\Delta T)_{t} \times 
            \sum_{i \leq y}
            \sum_{c \in \mathcal{C}^{tech}_{a, i}}
                k_{c, r} \, \mathbf P^{c}_{y, t, c}
    \quad \forall \, a, y, t, \quad \forall\, r \in \mathcal{R}_{a}
\end{equation}

\begin{equation}
    \label{eq:storage_net_gen_def}
    U^{net, s}_{y, t, r} = 
        (\Delta T)_{t} \times
            \sum_{i \leq y}
            \sum_{s \in \mathcal{S}^{tech}_{a, i}}
                \left( 
                    k^{in}_{s, r} \, \mathbf P^{s, in}_{y, t, s} +
                    k^{keep}_{s, r} \, \mathbf S_{y, t, s} +
                    k^{out}_{s, r} \, \mathbf P^{s, out}_{y, t, s} 
                \right)
    \quad \forall \, a, y, t, \quad \forall\, r \in \mathcal{R}_{a}
\end{equation}

\begin{equation}
    \label{eq:transport_net_gen_def}
    U^{net, w}_{y, t, r} = 
        (\Delta T)_{t} \times 
            \sum_{i \leq y}
            \sum_{(\afrom, \ato) \in \mathcal{A}^{2}}
            \sum_{w \in \mathcal{W}^{tech}_{\afrom, \ato, i}}
                (\delta_{\afrom, a} - \delta_{a, \ato}) \times k_{w, r} \, \mathbf P^{w}_{y, t, w}
    \quad \forall \, a, y, t, \quad \forall\, r \in \mathcal{R}_{a}
\end{equation}

\paragraph*{Remark} 
One could be surprised by the addition of the \(\eta^{in}_{s, i, r} \, P^{s, in}_{y, t, s}\) term to the resource production as the resource \(r\) is supposed to be consumed if loaded into storage, or at least, that the signs before the \(out\) and \(in\) are the same in the storage part of the equation. However, this approach comes from single energy carrier modelling. The storage does not store electricity or gas. It stores energy in MWh. Let's consider low-temperature storage, for example. It consumes electricity when loading (thus \(\eta^{in}_{elec} < 0\)) and consumes electricity when discharging (\(\eta^{out}_{elec} < 0 \) with the heat pumps consumption) and discharge low-temperature heat (\(\eta^{out}_{heat} > 0 \)).

\subsection{Capacity constraints}

The instant power of the conversion technologies is lower than the available installed capacity \eqref{eq:conv_availability}.

\begin{equation}
    \label{eq:conv_availability}
    \mathbf P^{c}_{y, t, c} \leq a_{y, t, c} \, \bar P^{c}_{y, c} 
    \quad \forall\, a, y, t, \quad \forall i \leq y, \quad \forall\, c \in \mathcal{C}^{tech}_{a, i}
\end{equation}

Considering this, the total installed capacity of \(c\) for operation year \(y\) (\(\bar P^{c}_{y, c}\)) is defined as the total installed capacity of technology \(c\) that is not yet decommissioned in \(y\).

\begin{equation}
    \label{eq:pbar_def}
    \bar P^{c}_{y, c} =  
        \sum_{d > y}
            \mathbf P^{c, inv}_{c, d}
        \quad \forall \, a, y, \quad \forall\, i \leq y, \quad \forall\, c \in \mathcal{C}^{tech}_{a, i}
\end{equation}

Moreover, decommissioning must happen strictly after investment and before the end of the technology life length \eqref{eq:pinv_dec}.

\begin{equation}
    \label{eq:pinv_dec}
     P^{c, inv}_{c, d} = 0  \quad \forall \, a, i,\quad \forall\, c \in \mathcal{C}^{tech}_{a, i}, \quad \forall\, d \in \left\{d \mid d \leq i \right\} \cup \left\{d \mid d > i + l_{c}\right\} 
\end{equation}

The minimum bounds are the invested capacity and the maximum allowed each year by the decision maker \eqref{eq:conv_inv_bounds}. This constraint could be related to the deployment rate or to the limited space of the local area, for example.

\begin{equation}
    \label{eq:conv_inv_bounds}
    p^{c, min}_{c} \leq
    \sum_{d}
        \mathbf P^{c, inv}_{c, d} \leq 
    p^{c, max}_{c} 
        \quad \forall \, a, i, \quad \forall\, c \in \mathcal{C}^{tech}_{a, i}
\end{equation}

\subsection{Storage constraints}

The total storage power (resp. energy) capacity for operation year \(y\) is defined as the total power (resp. energy) capacity of  \((s, i)\) that is not yet decommissioned in \(y\) in equation \eqref{eq:stor_power_tot_def} (resp. \eqref{eq:stor_energy_tot_def}).

\begin{equation}
    \label{eq:stor_power_tot_def}
    \bar P^{s}_{y, s} =
        \sum_{d > y}
            \mathbf P^{s, inv}_{s, d}
    \quad \forall \, a, y, \quad \forall\, i \leq y, \quad \forall\, s \in \mathcal{S}^{tech}_{a, i}
\end{equation}

\begin{equation}
    \label{eq:stor_energy_tot_def}
     \bar S_{y, s} =
        \sum_{d > y}
            \mathbf S^{inv}_{s, d}
    \quad \forall \, a, y, \quad \forall\, i \leq y, \quad \forall\, s \in \mathcal{S}^{tech}_{a, i}
\end{equation}

Where as in \eqref{eq:pinv_dec} the decommissioning must happen strictly after investment and before end of life length \eqref{eq:stor_dec}.

\begin{equation}
    \label{eq:stor_dec}
     P^{s, inv}_{s, d} = S^{inv}_{s, d} = 0  \quad \forall \, a, i,\quad \forall\, s \in \mathcal{S}^{tech}_{a, i}, \quad \forall\, d \in \left\{d \mid d \leq i \right\} \cup \left\{d \mid d > i + l_{s}\right\} 
\end{equation}

Any resource of the model can be stored in the right storage technology \(s\) is invested in. For all time steps \(t\), the input rate of storage \(P^{s, in/out}_{y, t, s}\) is bounded by the total invested power capacity \(\bar P^{s}_{y, s}\):

\begin{equation}
    \label{eq:stor_power_capa}
    \mathbf P^{s, in/out}_{y, t, s} \leq \bar P^{s}_{y, s}
    \quad \forall \, a, y, \quad \forall\, i \leq y, \quad \forall\, s \in \mathcal{S}^{tech}_{a, i}
\end{equation}

The total amount of energy in storage \(S_{y, t, s}\) should remain lower than the total invested energy capacity \(\bar S_{y, s}\):

\begin{equation}
    \label{eq:stor_energy_capa}
    \mathbf S_{y, t, s} \leq \bar S_{y, s}
    \quad \forall \, a, y, \quad \forall\, i \leq y, \quad \forall\, s \in \mathcal{S}^{tech}_{a, i}
\end{equation}

Invested power and energy capacities are bounded by the minimum and maximum allowed values in equations \eqref{eq:stor_power_bounds} and \eqref{eq:stor_energy_bounds}.

\begin{equation}
    \label{eq:stor_power_bounds}
    p^{s, min}_{s} \leq 
        \sum_{d}
            \mathbf P^{s, inv}_{s, d}
        \leq  p^{s, max}_{s}
    \quad \forall \, a, i \quad \forall\, s \in \mathcal{S}^{tech}_{a, i}
\end{equation}

\begin{equation}
    \label{eq:stor_energy_bounds}
    s^{s, min}_{s} \leq
        \sum_{d}
            \mathbf S^{inv}_{s,d}
        \leq s^{s, max}_{s}
    \quad \forall \, a, i \quad \forall\, s \in \mathcal{S}^{tech}_{a, i}
\end{equation}

The total amount of energy in the storage at each time step \(S_{y, t, s}\) is defined as the total amount of energy in the storage at the previous time step minus the dissipation plus the loaded energy minus the discharged energy \eqref{eq:stor_level}. The constraint is cyclic to avoid side effects. 

Note that \(P^{s, in}_{y, t, s}\) is the power to load 1 MWh into the storage and  \(P^{s, out}_{y, t, s}\) is discharge power to lower the storage level of 1 MWh.

\begin{equation}
    \label{eq:stor_level}
    \mathbf S_{y, t, s} =
        \left( 
            1 - \eta_{s}
        \right)^{\frac{(\Delta T)_{t}}{1[h]}} \, \mathbf S_{y, t - 1, s} 
        + (\Delta T)_{t}
        \times (
            \mathbf P^{s, in}_{y, t, s}
            - \mathbf P^{s, out}_{y, t, s}
        )
    \quad \forall \, a, y, t, \quad \forall\, i \leq y, \quad \forall\, s \in \mathcal{S}^{tech}_{a, i}
\end{equation}

\subsection{Transport constraints}

The instant power of the transport technologies is lower than the available installed capacity \eqref{eq:transport_availability}.

\begin{equation}
    \label{eq:transport_availability}
    \mathbf P^{w}_{y, t, w} \leq a_{y, t, w} \, \bar P^{w}_{y, w} 
    \quad \forall\, \afrom, \ato, y, t, \quad \forall i \leq y, \quad \forall\, w \in \mathcal{W}^{tech}_{\afrom, \ato, i}
\end{equation}

The total installed capacity of \(w\) for operation year \(y\) (\(\bar P^{c}_{y, w}\)) is defined as the total installed capacity of technology \(w\) that is not yet decommissioned in \(y\) \eqref{eq:wbar_def}.

\begin{equation}
    \label{eq:wbar_def}
    \bar P^{w}_{y, w} =  
        \sum_{d > y}
            \mathbf P^{w, inv}_{w, d}
        \quad \forall \, \afrom, \ato, y, \quad \forall\, i \leq y, \quad \forall\, w \in \mathcal{W}^{tech}_{\afrom, \ato, i}
\end{equation}

Moreover, decommissioning must happen strictly after investment and before the end of the technology life length \eqref{eq:winv_dec}.

\begin{equation}
    \label{eq:winv_dec}
     P^{w, inv}_{w, d} = 0  \quad \forall \, \afrom, \ato, i,\quad \forall\, w \in \mathcal{W}^{tech}_{\afrom, \ato, i}, \quad \forall\, d \in \left\{d \mid d \leq i \right\} \cup \left\{d \mid d > i + l_{w}\right\} 
\end{equation}

The minimum bounds are the invested capacity and the maximum allowed each year by the decision maker \eqref{eq:transport_inv_bounds}.

\begin{equation}
    \label{eq:transport_inv_bounds}
    p^{w, min}_{w} \leq
    \sum_{d}
        \mathbf P^{w, inv}_{w, d} \leq 
    p^{w, max}_{w} 
        \quad \forall \, \afrom, \ato, y, \quad \forall\, i \leq y, \quad \forall\, w \in \mathcal{W}^{tech}_{\afrom, \ato, i}
\end{equation}

\subsection{Carbon related constraints}

Equation \eqref{eq:emission_def} defines the operation emission variable.

\begin{equation}
    \label{eq:emission_def}
    E^{\ce{CO2}}_{a, y, t} =
        (\Delta T)_{t} \times
        \sum_{i \leq y}
        \sum_{c \in \mathcal{C}^{tech}_{a, i}}
            \varepsilon^{\ce{CO2}}_{c} \, \mathbf P^{c}_{y, t, c} 
        + \sum_{r \in \mathcal{R}_{a}}
            \varepsilon^{\ce{CO2}}_{y, r, t} \, I^{net}_{y, r, t}
    \quad \forall \, a, y, t
\end{equation}

The total emission constraint is in equation \eqref{eq:emission_max}.

\begin{equation}
    \label{eq:emission_max}
    \sum_{t}
        E^{\ce{CO2}}_{a, y, t}
    \leq g^{\ce{CO2}}_{a, y}
    \quad \forall \ a, y
\end{equation}

\subsection{Imports constraints}

Imports bound are defined in equation \eqref{eq:imports}. Variables are in energy units (MWh).

\begin{equation}
    \label{eq:net_imports_def}
    I^{net}_{y, r, t} =
        \mathbf I_{y, r, t} - \mathbf E_{y, r, t}
    \quad \forall \, y, r, t
\end{equation}

\begin{equation}
    \label{eq:imports}
    \mathbf I_{y, r, t} \leq \pi^{max, imp}_{y, r, t} \quad \forall \, y, r, t
\end{equation}

\begin{equation}
    \label{eq:exports}
    \mathbf E_{y, r, t} \leq \pi^{max, exp}_{y, r, t} \quad \forall \, y, r, t
\end{equation}

\section{Supplementary results}

\begin{figure}[H]
\centering
\underlay{4.2in}{250pt}{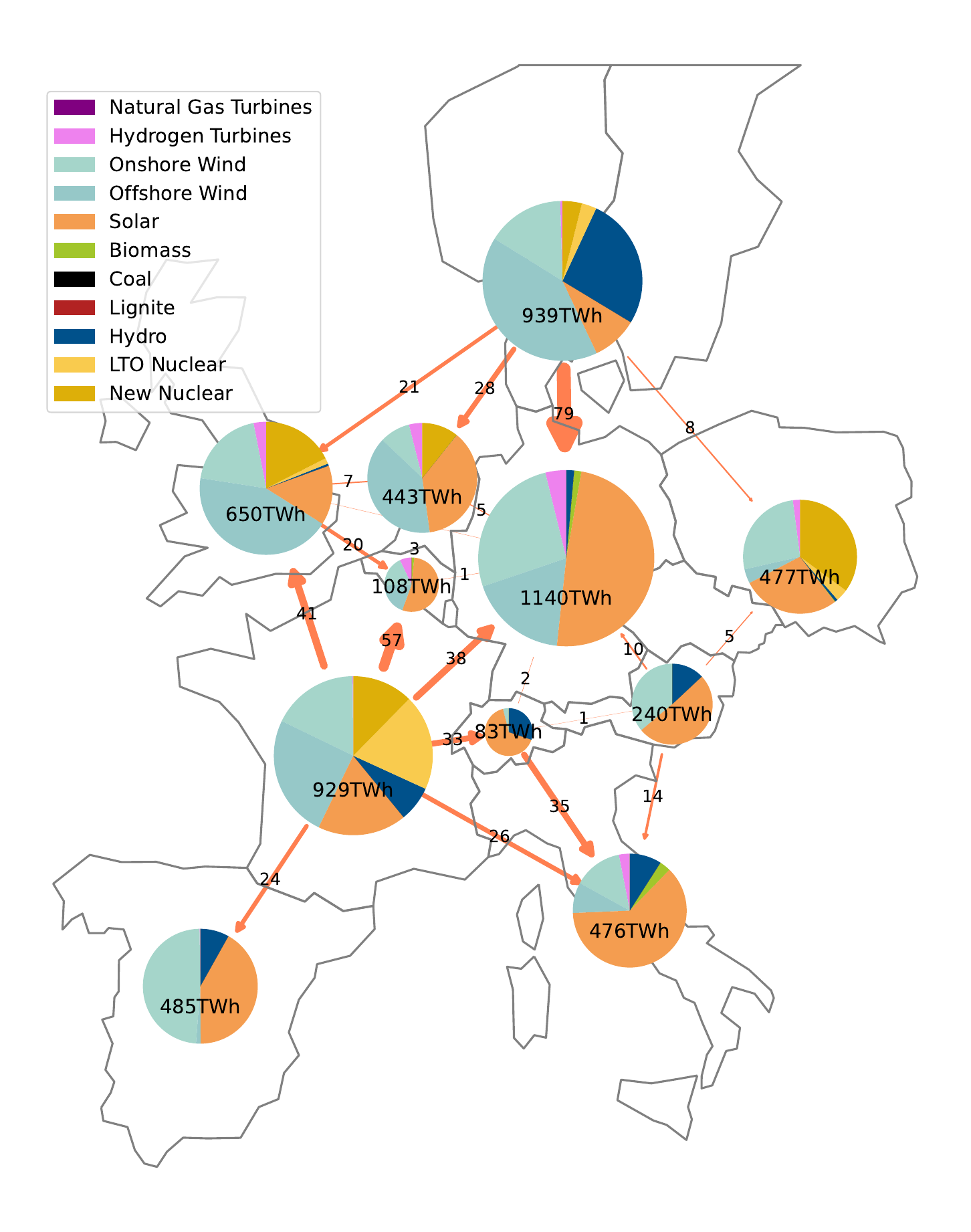}{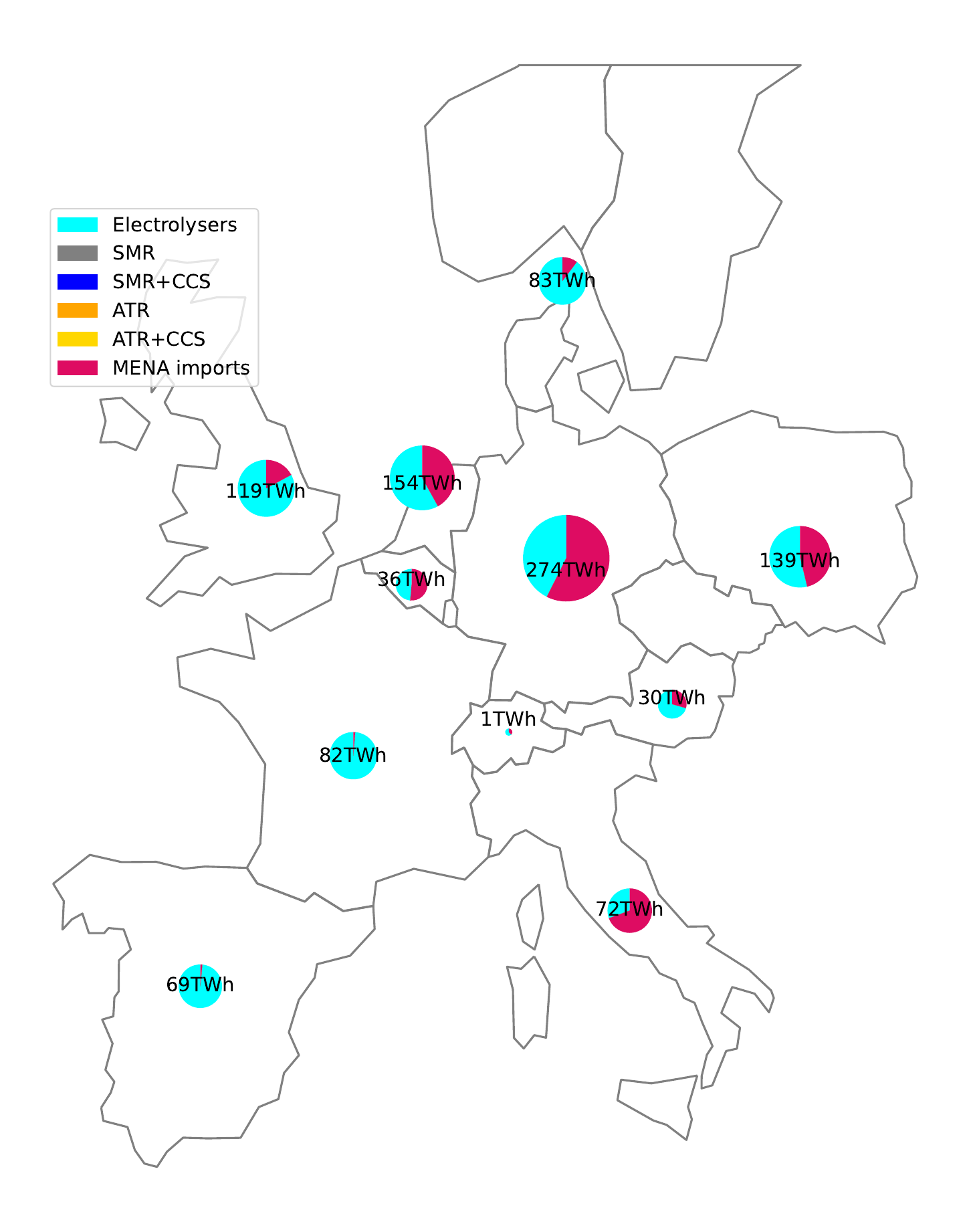}
\caption{No Fossil 2050 Central scenario electricity and hydrogen system in 2050 | from POMMES}
\end{figure}

\begin{figure}[H]
\centering
\underlay{4.2in}{250pt}{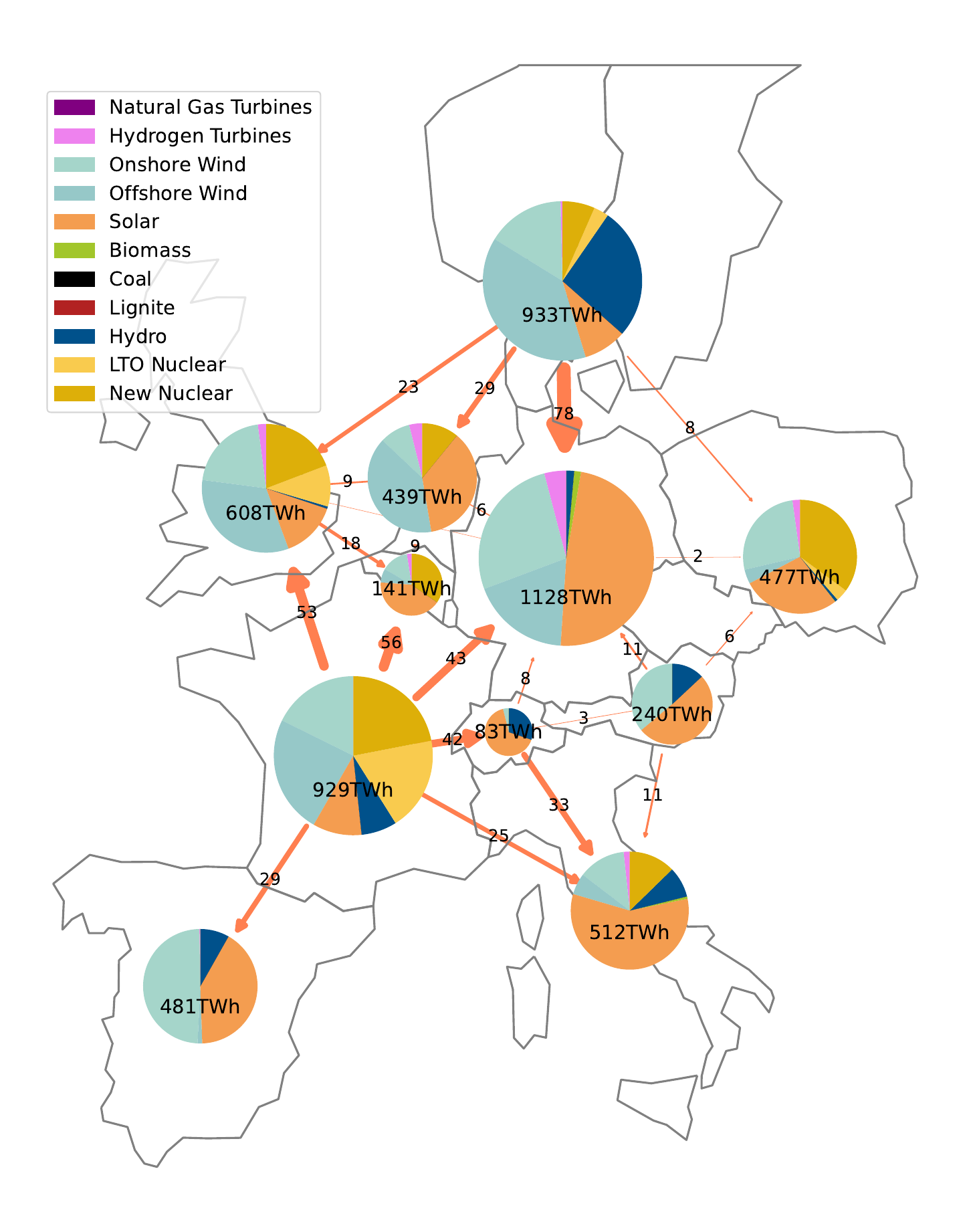}{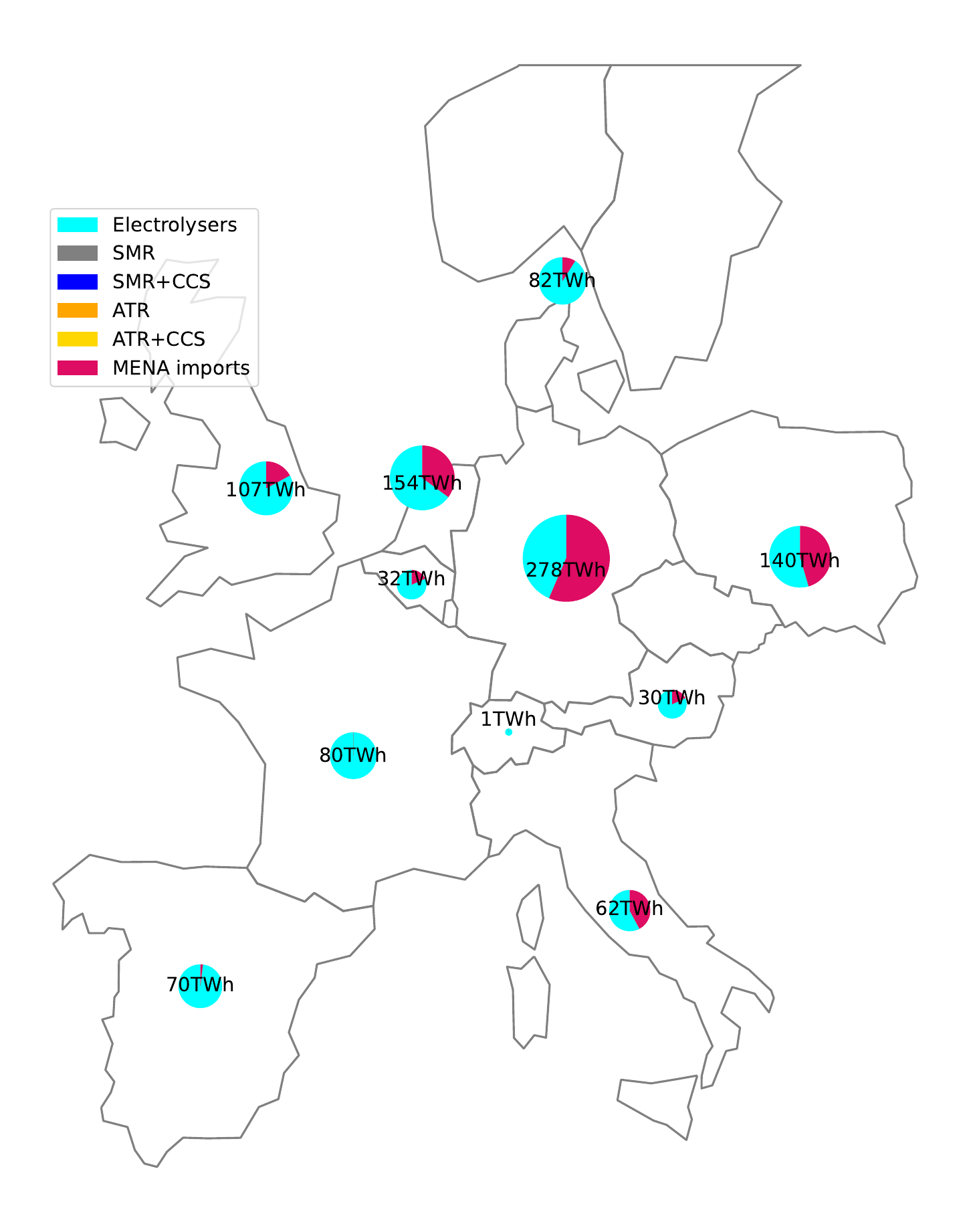}
\caption{No Fossil 2050 Nuke\_Plus scenario electricity and hydrogen system in 2050 | from POMMES}
\end{figure}

\begin{figure}[H]
\centering
\underlay{4.2in}{250pt}{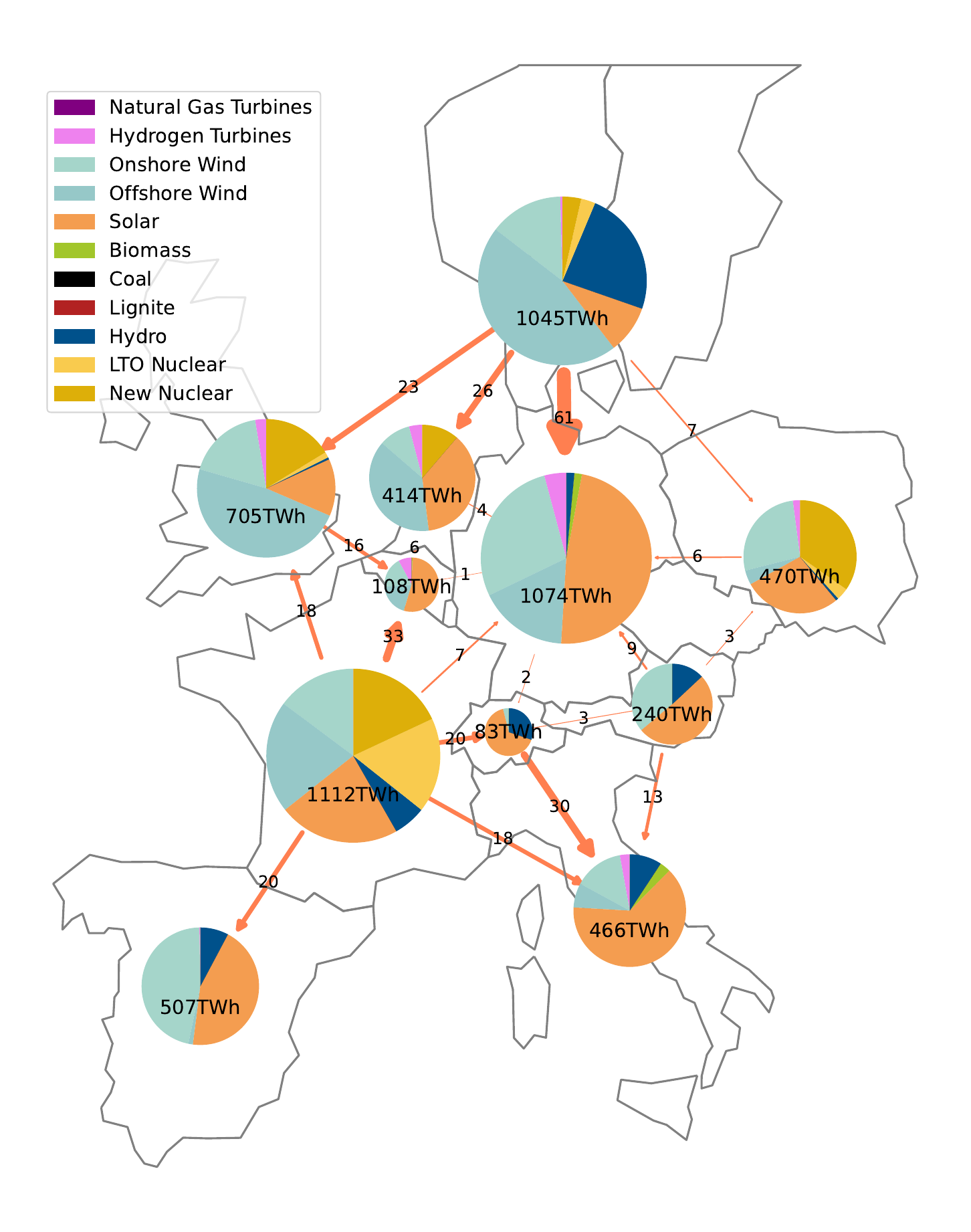}{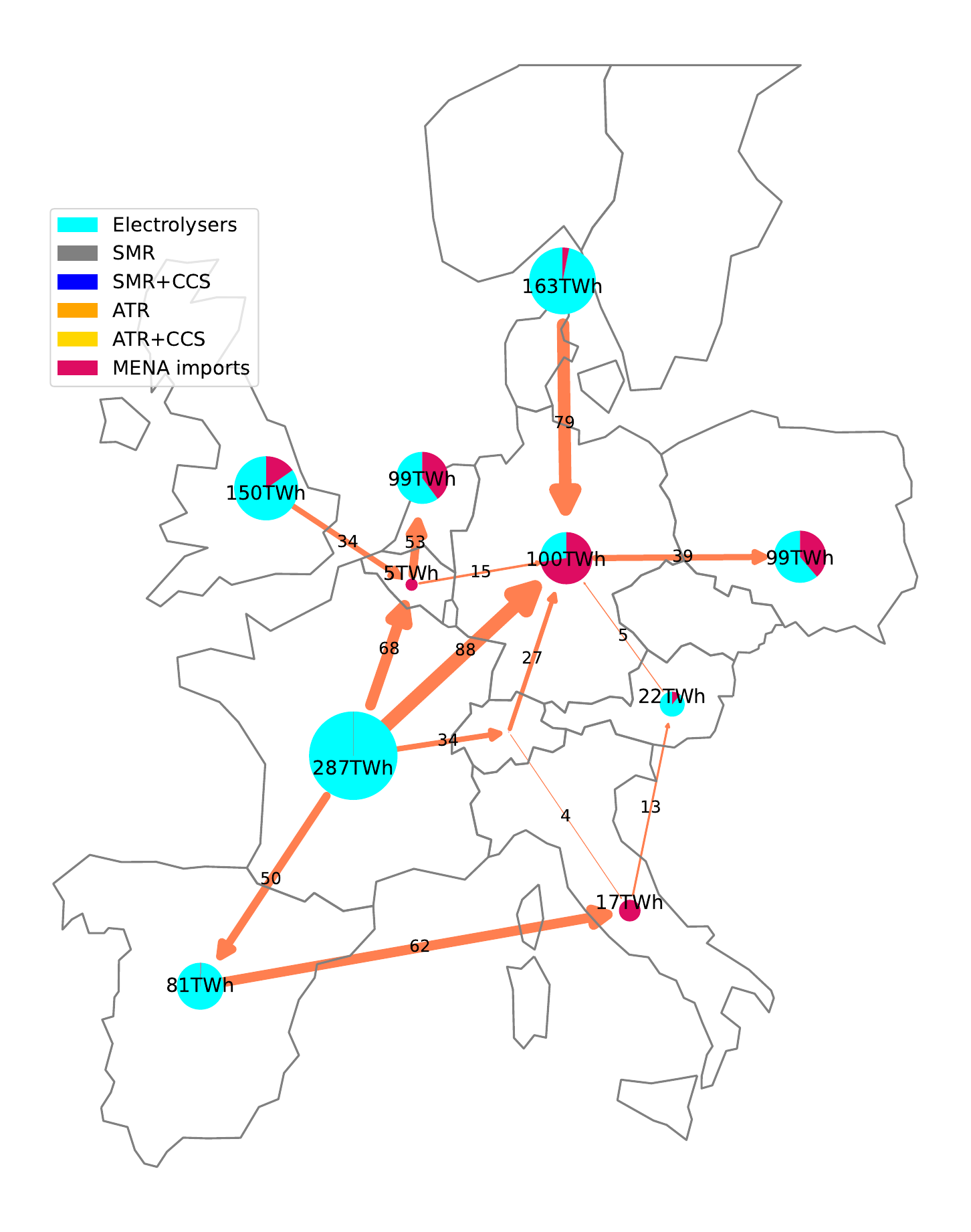}
\caption{No Fossil 2050 H\textsubscript{2}\_Exch\_Plus scenario electricity and hydrogen system in 2050 | from POMMES}
\end{figure}

\begin{figure}[H]
\centering
\underlay{4.2in}{250pt}{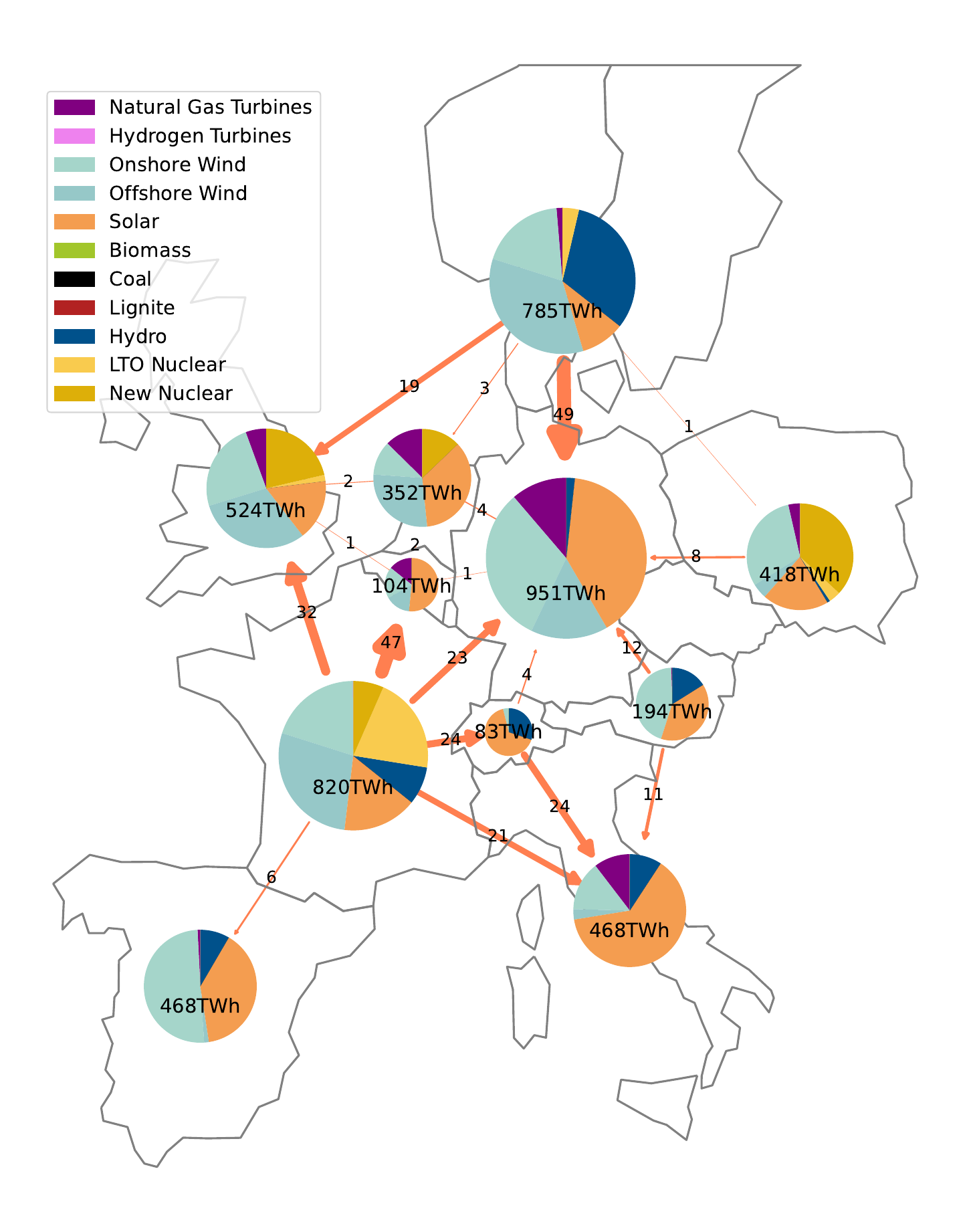}{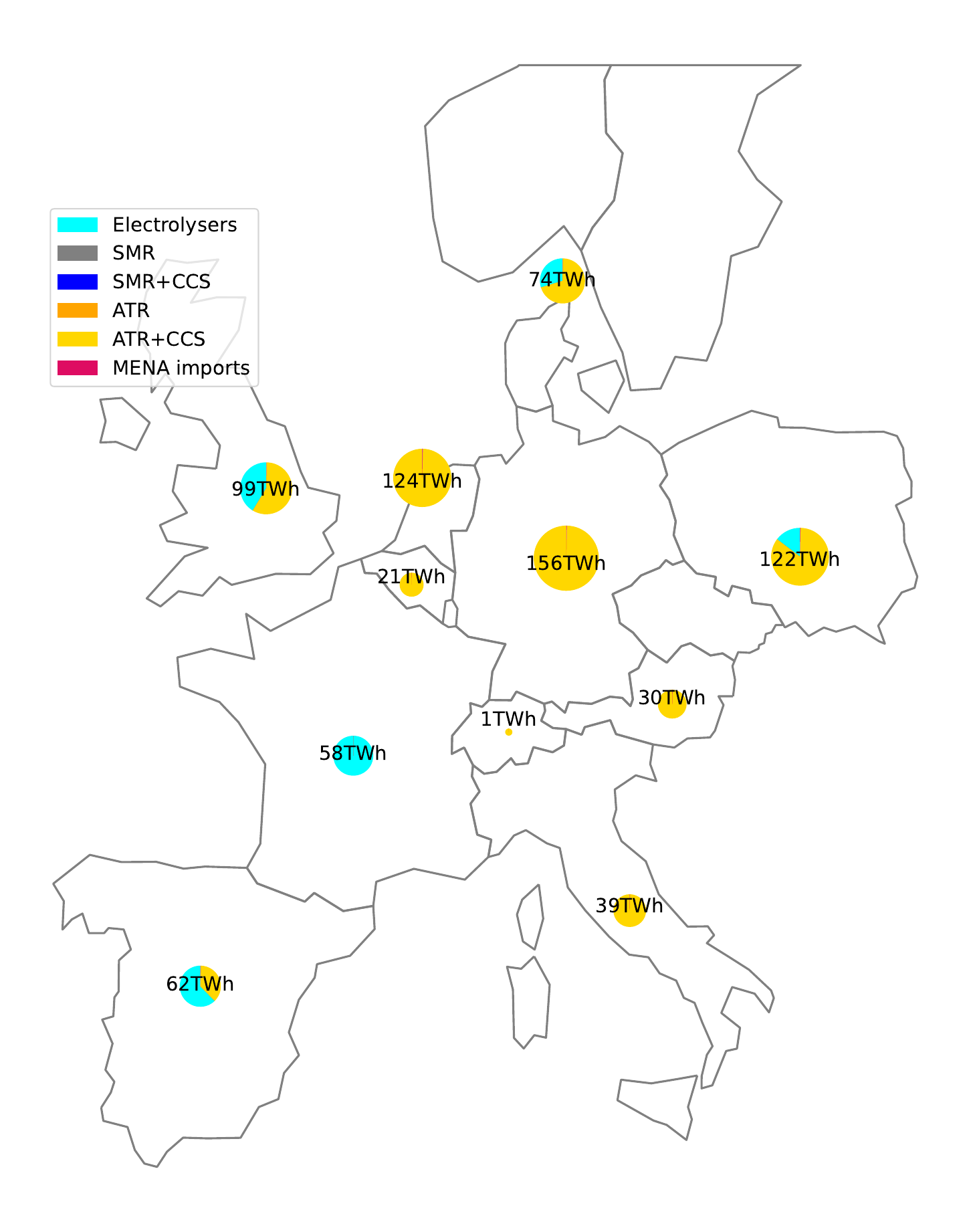}
\caption{Low Carbon Tax Central scenario electricity and hydrogen system in 2050 | from POMMES}
\end{figure}

\begin{figure}[H]
\centering
\underlay{4.2in}{250pt}{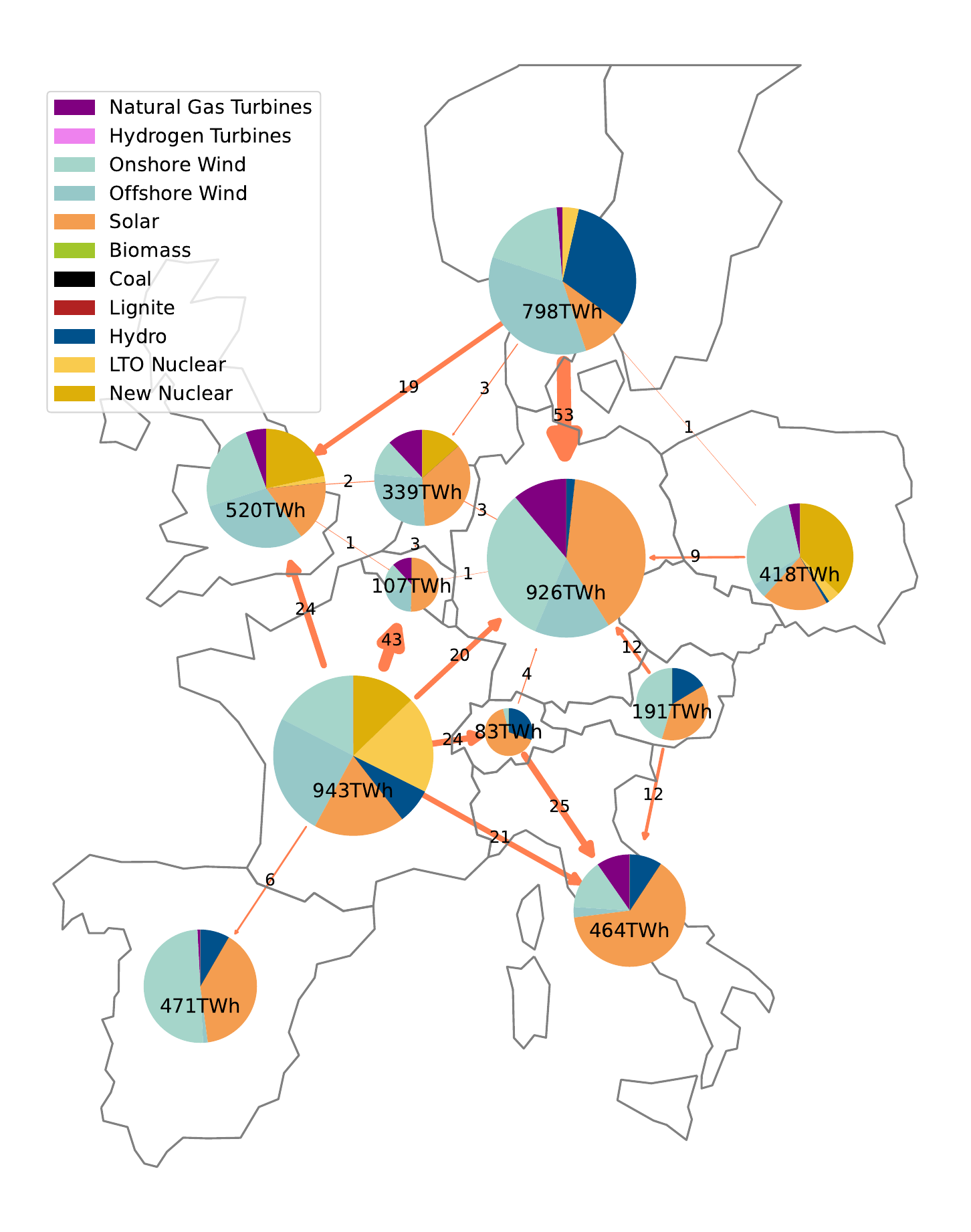}{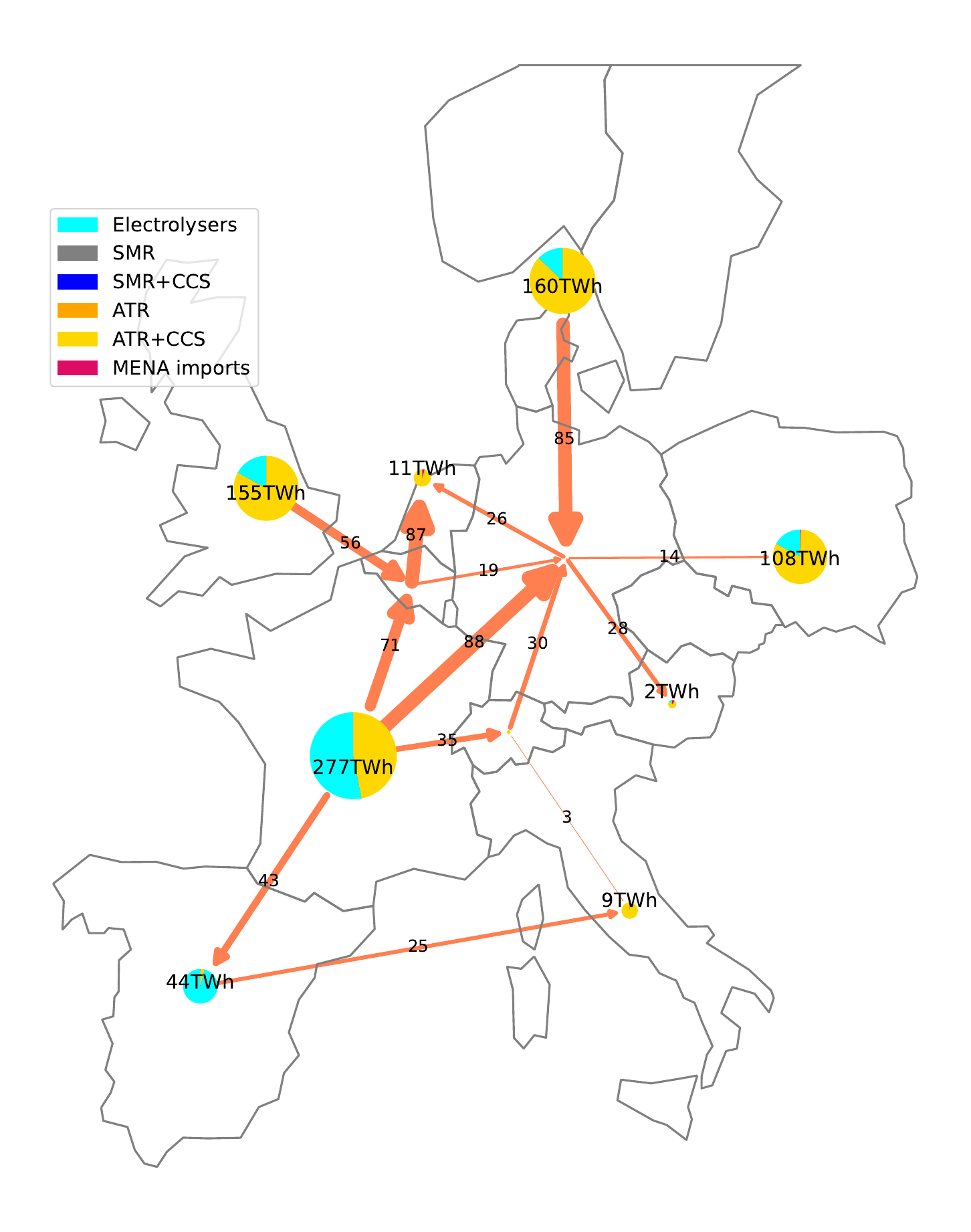}
\caption{Low Carbon Tax H\textsubscript{2}\_Exch\_Plus scenario electricity and hydrogen system in 2050 | from POMMES}
\end{figure}